\documentclass[fleqn,usenatbib]{mnras}

\usepackage[T1]{fontenc}

\DeclareRobustCommand{\VAN}[3]{#2}
\let\VANthebibliography\thebibliography
\def\thebibliography{\DeclareRobustCommand{\VAN}[3]{##3}\VANthebibliography}

\usepackage{graphicx}	
\usepackage{amsmath}	
\usepackage{amssymb}	

\usepackage{multicol}

\newcommand{\mcl}{M$_\mathrm{cl}$}

\title[Stochastic sampling with BPASS]{Exploring the impact of IMF and binary parameter stochasticity with a binary population synthesis code.}

\author[Stanway \& Eldridge]{
Elizabeth R. Stanway$^{1}$\thanks{E-mail: e.r.stanway@warwick.ac.uk}
and J. J. Eldridge$^{2}$\thanks{E-mail: j.eldridge@auckland.ac.nz}
\\
$^{1}$Department of Physics, University of Warwick, Gibbet Hill Road, Coventry, CV4 7AL, UK\\
$^{2}$Department of Physics, University of Auckland, Private Bag 92019, Auckland, New Zealand
}

\date{Accepted 2023 April 13. Received 2023 April 6; in original form 2022 August 22}

\pubyear{2023}

\begin{document}
\label{firstpage}
\pagerange{\pageref{firstpage}--\pageref{lastpage}}
\maketitle

\begin{abstract}
Low mass star formation regions are unlikely to fully populate their initial mass functions, leading to a deficit of massive stars. In binary stellar populations, the full range of binary separations and mass ratios will also be underpopulated. To  explore the effects of stochastic sampling in the integrated light of stellar clusters, we calculate models at a broad range of cluster masses, from $10^2$ to $10^7$\,M$_\odot$, using a binary stellar population synthesis code. For clusters with stellar masses less than $10^5\,M_\odot$, observable quantities show substantial scatter and their mean properties reflect the expected deficit of massive stars. In common with previous work, we find that purely stochastic sampling of the initial mass function appears to underestimate the mass of the most massive star in known clusters. However, even with this constraint, the majority of clusters likely inject sufficient kinetic energy to clear their birth clusters of gas. For quantities which directly measure the impact of the most massive stars, such as $N_\mathrm{ion}$,  $\xi_\mathrm{ion}$ and $\beta_\mathrm{UV}$, uncertainties due to stochastic sampling dominate over those from the IMF shape or distribution of binary parameters, while stochastic sampling has a negligible effect on the stellar continuum luminosity density. \end{abstract}

\begin{keywords}
methods: numerical -- binaries: general -- stars: luminosity function, mass function -- galaxies: stellar content
\end{keywords}



\section{Introduction}\label{sec:intro}

Observations of stellar populations are invariably interpreted in the context of models which predict the properties of individual stars in those populations. Stellar evolution models with a known distribution in age, mass and composition can be combined with matching stellar atmosphere models in order to predict the observable features of the population as a whole.  The matching of such stellar population synthesis (SPS) models  to observational data is invaluable for interpreting the origin and impact of star-forming galaxies throughout cosmic time.

However the use of SPS models presupposes that the uncertainties in such models are well understood and (typically) that they are small compared to the uncertainties on observational constraints. As our understanding of stellar model uncertainties evolves and observations become more sensitive, it is becoming clear that this supposition may not be secure, particularly in young, low metallicity environments such the distant Universe. In particular, the impact of binary star evolution pathways has risen to prominence as a significant source of both variation between SPS models and uncertainty within model grids. The effects of binary interactions are most dramatic in massive stars and at early stellar population ages, and so their impact on the integrated light of populations also depends on populating the upper end of the stellar initial mass function (IMF).

In recent work, we have explored the impact of varying the shape of the initial mass function \citep{2019A&A...621A.105S}, and resampling uncertainties on the initial assumptions regarding mass dependence of binary fraction, mass ratio and period distribution \citep{2020MNRAS.495.4605S}, on observable properties of a binary stellar population synthesis model. In both cases, the initial parameter distributions for the synthetic populations were statistically sampled: i.e. each stellar model in a grid was deemed to contribute to the output spectrum, even when the fractional contribution of that model was significantly less than one star.  While this is true of most SPS algorithms, the occurrence of low probability models is more common in a binary population synthesis due to the significantly larger number of models required in a grid. Where a single star population synthesis might have one stellar model at a given stellar mass, a binary synthesis requires that the same stellar mass is represented by a grid of models with different initial binary separations and mass ratios. This grid dilutes the initial mass function weighting of the primary star over several hundred potential secondary models, and statistical sampling effectively averages these to give a `typical' evolution model for binary stars at that primary mass. 

The stochastic sampling approach instead models stellar populations as forming from a finite gas reservoir. Mass is drawn from this reservoir one model at a time, randomly selecting stellar models from underlying distributions defined by their statistical weighting. This process continues until the mass reservoir is exhausted.

The arguments in favour of such an approach, and its impact in a single star evolution context have long been established. Much of the focus in this area has naturally been on the effect of such sampling on the most massive (and rarest) stars in a stellar population \citep[e.g.][]{2010MNRAS.401..275W}, and hence on their ionizing photon production rate \citep[e.g.][]{2012MNRAS.422..794E}. In low mass clusters, such as might be expected in dwarf galaxies, stochastic sampling will lead to an absence of massive stars and a deficit of ionizing photon production relative to a statistically sampled IMF. This observation has led to proposals for an integrated galactic initial mass function (IGIMF), in which the  upper stellar initial mass limit in large stellar populations depends on the upper limit of star forming cluster mass, which in turn depends on the galaxy-wide average star formation rate \citep{2003ApJ...598.1076K,2005ApJ...625..754W}. Such stochastic sampling approaches have also been applied to stellar population synthesis modelling. \citet{2013ApJ...778..138A}, for example, presented a large grid of stochastically-sampled stellar cluster models and considered their impact on the integrated light photometry with a varying cluster stellar mass, \mcl. Unsurprisingly, the cluster-to-cluster variation in photometry translated to a small but significant uncertainty in population parameters such as age and mass when these were recovered with a statistically-sampled model fit to the spectral energy distribution. Another notable effort is that of the Stochastically Lighting Up Galaxies (SLUG) project which has explored the impact of not only stochastically sampling the IMF as a function of cluster mass, but also of stochastically sampling the cluster mass function to generate galaxy models \citep{2011ApJ...741L..26F,2012ApJ...745..145D}. 

However, while stochastic sampling may provide a more physical picture of the stellar population in any one star formation region,  the random element means that many realisations are required to fully probe the range of possible cluster properties and evaluate the uncertainties in any one model. Repeated random sampling of this kind is computationally expensive, particularly in the context of population synthesis codes built on binary stellar evolution models. Early work in this area included the analysis of \citet{2012MNRAS.422..794E}, who considered the impact of stochastic IMF sampling on the ratio of star formation rate indicators with different timescales (a proxy for the ratio of very massive to mid-sized stars) with a relatively-small BPASS v1 binary model grid. More recently, work has focused on stochastic elements of the evolution of binary compact objects as gravitational wave transient progenitors, in particular on sampling plausible supernova kicks and their potential to disrupt or harden compact binaries \citep{2021MNRAS.500.1380M,2019MNRAS.490.5228B}. Indeed stochastic sampling of both these kicks and the initial mass and binary parameter distributions has recently been added to the population synthesis code COMPAS \citep{2021arXiv210910352T}. In common with many other rapid population synthesis models which aim to explore large parameter spaces, COMPAS is built upon the stellar evolution prescriptions of \citet{2002MNRAS.329..897H}. These are highly computationally efficient and accurately describe most stellar evolution pathways, but do not model the structure and evolution of individual stars in detail, and were initially calibrated on stellar models in the mass range $1<M<50\,$M$_\odot$.

In this paper, we continue our efforts to evaluate the relative impact of the most important sources of uncertainty in binary stellar populations, constructed from a large grid of detailed binary stellar evolution models. Here we focus on the effects of stochastic sampling on uncertainties in the integrated light of the population. We generate a large grid of stellar population realisations at three metallicities, in which the initial mass function, initial binary fraction, initial binary mass ratio, initial separation and supernova kick magnitude and direction are all stochastically sampled. The structure of the paper is as follows: in Section \ref{sec:method} we describe the stellar population models and implementation of stochastic sampling; in Section \ref{sec:results} we evaluate the impact of stochastic sampling on key observable parameters; in Section \ref{sec:disc} we first consider the scale of resulting uncertainties in the context of previous work, and then explore the impact of supernova rate uncertainties on mechanical feedback to clear their birth clouds; in Section \ref{sec:conc} we briefly summarise our main results and present our conclusions.

\section{Methodology}\label{sec:method}

 \subsection{The Binary Population and Spectral Synthesis (BPASS) project}

The Binary Population and Spectral Synthesis (BPASS) project was designed to calculate the properties and spectral energy distributions of synthetic stellar populations which incorporate the impact of binary star interactions and evolution processes \citep{2009MNRAS.400.1019E,2012MNRAS.419..479E}. Each synthetic population samples a grid of some 250,000 individual detailed stellar evolution models, including the effects of binary interactions and a small grid of low metallicity models which undergo chemically homogeneous evolution due to rotational mixing. These are combined with publicly available stellar atmosphere spectra to calculate observable properties of stellar populations. BPASS models are used to interpret populations in contexts ranging from young galaxies in the very distant Universe to proto-white dwarfs within our own galaxy \citep[see e.g.][]{2017PASA...34...58E,2016MNRAS.456..485S,2021MNRAS.507..621B,2022arXiv220201413E}.

In common with other SPS codes, the primary outputs of BPASS are the properties of simple stellar populations. These represent the products of a single episode of star formation, in which all the stars have the same age and metallicity. Models are generated at log(age/years)=6.0-11.0 at 0.1\,dex intervals, and at 13 metallicities between $Z=10^{-5}$ and $Z=0.040$. The distribution of initial masses, and initial binary periods, mass ratios and separations in standard BPASS models are described by well-defined analytic functions. These functions are assumed to be fully statistically sampled, and clusters are generated with a total stellar mass $M_\ast=10^6$\,M$_\odot$. The fiducial BPASS initial mass function is a broken power law of the form $N(M)\propto M^{-2.35}$ between 0.5 and 300\,M$_\odot$ and $N(M)\propto M^{-1.3}$ for $0.1<M<0.5$\,M$_\odot$. We do not include brown dwarfs. Mass-dependent binary parameters are selected to match the empirically determined functions of \citet{2017ApJS..230...15M}. The construction of BPASS models is described in \citet{2017PASA...34...58E} and \citet{2018MNRAS.479...75S}. This study makes use of the BPASS v2.2.1 population and spectral synthesis code.

\subsection{Stochastic sampling with BPASS}

In the fiducial, statistically-sampled BPASS models, the probability weighting associated with stars of a given initial mass (i.e. the IMF weighting) is further divided between models with ten possible binary mass ratios and 21 initial binary periods. As a result, for high mass stars, each stellar evolution model may have a final weighting of considerably less than 1 star in $10^6$\,M$_\odot$, but the cumulative total probability for models with that primary mass may well exceed 1 star.

Here we use the statistical weightings of BPASS v2.2.1 to define a probability of occurrence for each stellar model. The total probability of all models is normalised to unity and individual stars are then drawn at random from the joint IMF and binary parameter probability distribution. As each stellar model is drawn, its mass and that of its companion are subtracted from a total stellar mass budget for the cluster being modelled, M$_\mathrm{cl}$. The drawing process stops when the mass budget becomes negative, with the last star drawn included in this sample. Technically this allows each cluster to marginally exceed its target mass, in common with a number of other stochastic sampling studies \citep[e.g.][]{2013ApJ...778..138A}. A range of alternate approaches have been explored in the literature, including whether to keep the last selected star only if brings the total mass closer to the target or whether to apply analytic constraints on the maximum stellar mass in a cluster \citep[e.g.][]{2012MNRAS.422..794E,2020MNRAS.492....8A}. The sampling approach is purely stochastic and since the average individual model stellar mass is small ($\sim0.3$\,M$_\odot$ for this IMF), the perturbation on the total cluster mass is typically $<<1$\,per cent, and does not significantly affect any of the results presented here.

The drawing process is used to define a population in which each single star and binary model has an integer weighting (and, indeed, many of the available input models have a weighting of zero). These new primary model occurrence rates are used to determine the appropriate weighting for secondary models. These continue the evolution of the binary system after a merger, the first supernova or collapse into a white dwarf. Usually these are also populated statistically in BPASS, with a probability distribution sampled many times at any supernova in order to determine both the fraction of instances in which a binary remains bound and the distribution of possible orbits for the new compact remnant binary. The result is that usually many different secondary models can lead from the same initial primary model and vice versa. 
In the case of the stochastic sampling undertaken here, each star can undergo supernova precisely once, so a single supernova kick outcome (determined by a single random assignment of velocity and direction, rather than its distribution), is randomly sampled in each case. 

With a full set of resultant model weightings, the population statistics (in terms of star numbers by type and age, supernova numbers etc) and the output synthetic spectra are calculated.  
Inevitably, stochastically sampling stars for a high mass cluster takes substantially longer than doing so for a low mass cluster.  The number of individual stellar models which contribute to the final population is also larger in massive clusters, significantly slowing the population and spectral synthesis stages. While a BPASS synthesis of a \mcl$=10^2$\,M$_\odot$ population may be complete in under an hour, running a $10^7$\,M$_\odot$ population takes days. As a result, we scale the number of realisations of random sampling by the total stellar mass of the system generated.

In Table \ref{tab:maxmasses}, we list total cluster stellar masses explored in this work, which range from 100\,M$_\odot$ to 10 million\,M$_\odot$. We also show the number of random sampling iterations undertaken for each cluster mass, and the mean number of independent stellar evolution models which contribute to the output populations at a metallicity $Z=0.020$. We note that this latter number is marginally metallicity-dependent, since the secondary model selection depends on the primary star's end state and thus on stellar winds and prior interactions.  We undertake stochastic samplings at three representative metal mass fractions, $Z=0.002$, 0.008 and 0.020 (where $Z_\odot=0.020$ is canonically considered solar metallicity). Our canonical models incorporate binary populations. For comparison with earlier work, we also calculate models incorporating only single star models, although we do not recommend the use of these, given that observational evidence strongly favours a high binary contribution at all redshifts \citep[e.g.][]{2013A&A...550A.107S,2017ApJS..230...15M,2021arXiv210708304M}. The scale of uncertainties arising from stochastic sampling are similar in the binary and single star cases (see appendix).

  \begin{figure*}
      \centering
      \hspace{-0.9cm}
      \includegraphics[width=0.95\columnwidth]{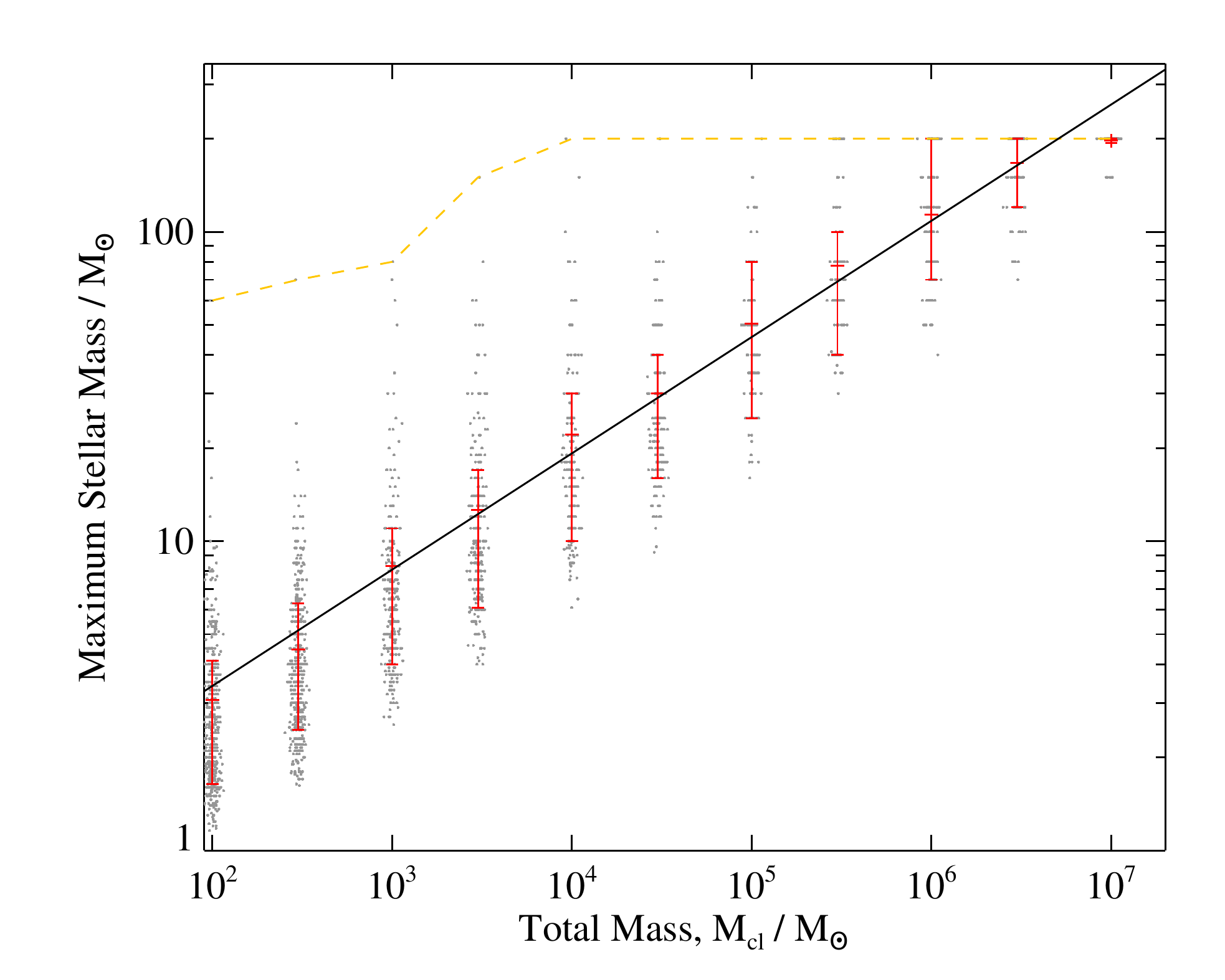}
      \includegraphics[width=0.95\columnwidth]{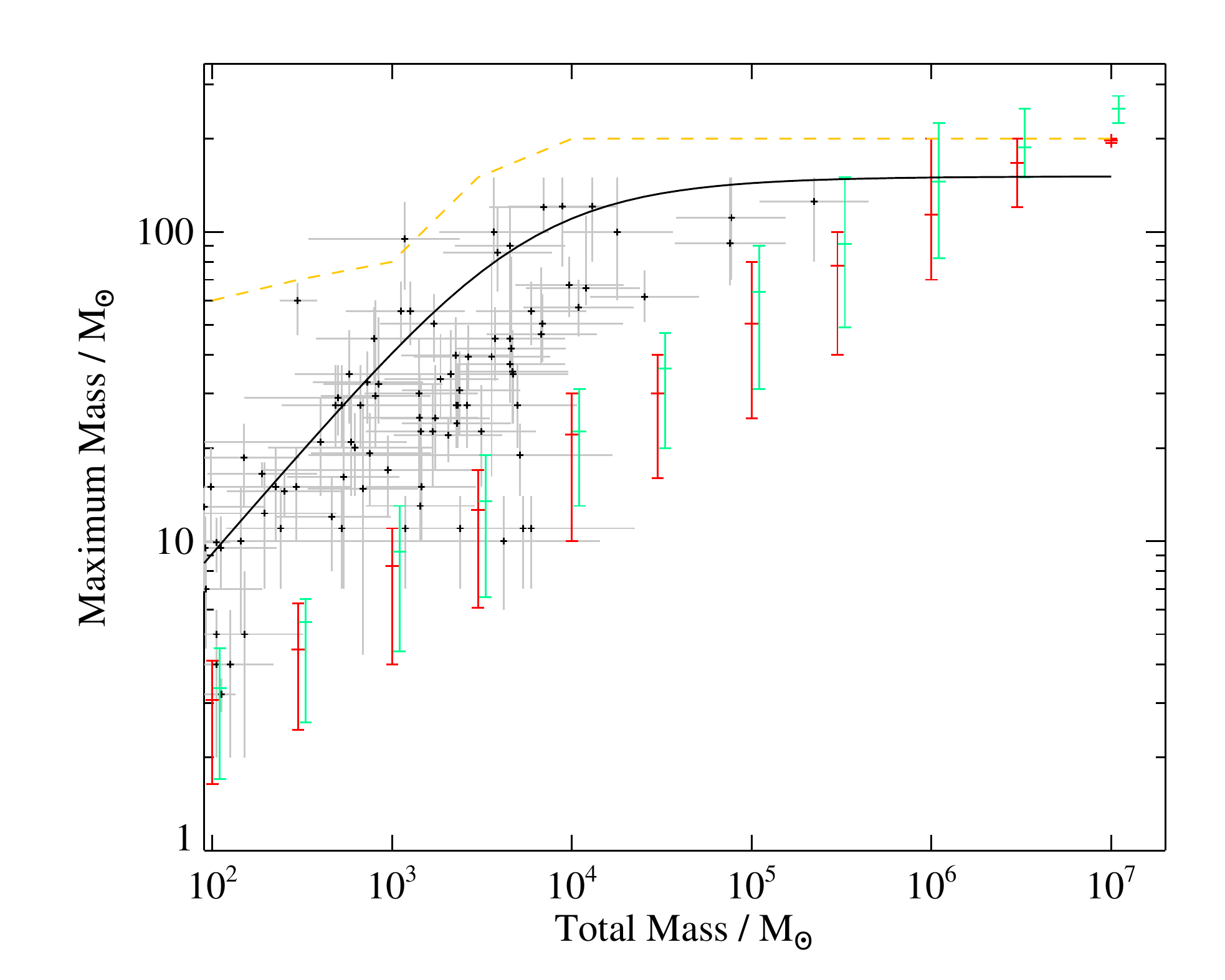}
      \caption{The highest mass stellar model in each stochastic sample realisation, shown as a function of cluster stellar mass. (Left) Cluster masses are plotted with a small random offset for clarity. Red symbols indicate the mean of the maximum masses at each cluster mass, and 16th and 84th percentile range. The solid line indicates a linear fit to the means as described in the text. The dashed line indicates the maximum stellar mass in any realization at that cluster mass. (Right) Comparison to the data compilation of \citet{2013MNRAS.434...84W}. The solid black line indicates an analytic estimate of the theoretical maximum stellar mass as a function of cluster size \citep{2007ApJ...671.1550P}. Red points indicate the typical maximum masses in a binary stochastic sample, as in the left hand plot. Pale green points indicate the same statistic for a stochastically-sampled single star stellar population, slightly offset in mass for clarity. }
      \label{fig:max_mass}
  \end{figure*}

\subsection{Comparison with Eldridge (2012)}

We note that \citet{2012MNRAS.422..794E} previously undertook a stochastic sampling exercise with an early version of BPASS (v1.1), specifically considering its impact on star formation rate indicators at solar metallicity. They found that the effects of binary interactions reduced the impact of stochastic IMF sampling, particularly at ages where stars rejuvenated or accreted mass, since interactions can imitate the impact of massive stars. 

The \citet{2012MNRAS.422..794E} work explored the random selection of stars from a power law initial mass function that extended either to the cluster mass or the cluster-mass-dependant stellar mass upper limit proposed by \citet{2007ApJ...671.1550P}. All clusters were deemed to comprise either entirely single stars or entirely binaries. The distribution of binary periods and initial mass ratios was described by a simple power law. 

The work presented here builds on this early work in several important ways. Since 2012, the BPASS stellar library has grown from $\sim$15,000 detailed stellar structure and evolution models to $\sim$250,000 (i.e. from around 3000 to 20,000 models per metallicity) allowing for far better sampling of the stellar mass and binary parameter distributions. Binary fractions and initial parameters are now assumed to be strongly mass-dependent, with populations including a mix of single and binary stars to match empirically-derived estimates. 

In the following, we make use of the much larger model grid, sample empirical distributions of initial parameters and consider a wider variety of observational predictions from the models. We also consider models at sub-solar metallicity and present results at three metallicities, including at $Z=0.002$ which lies below the threshold for accretion-induced spin-up and subsequent chemically homogeneous evolution.

However we note that \citet{2012MNRAS.422..794E} also considered the impact of a cluster mass function and age distribution in predicting the properties of galaxies. A key result was that binary populations showed less variation under the influence of stochastic sampling than single star populations, due to the effects of interactions. We opt not to explore that area further in this work, but defer it for later study.

\section{Results}\label{sec:results}

  \begin{figure*}
      \centering
      \includegraphics[width=0.32\textwidth]{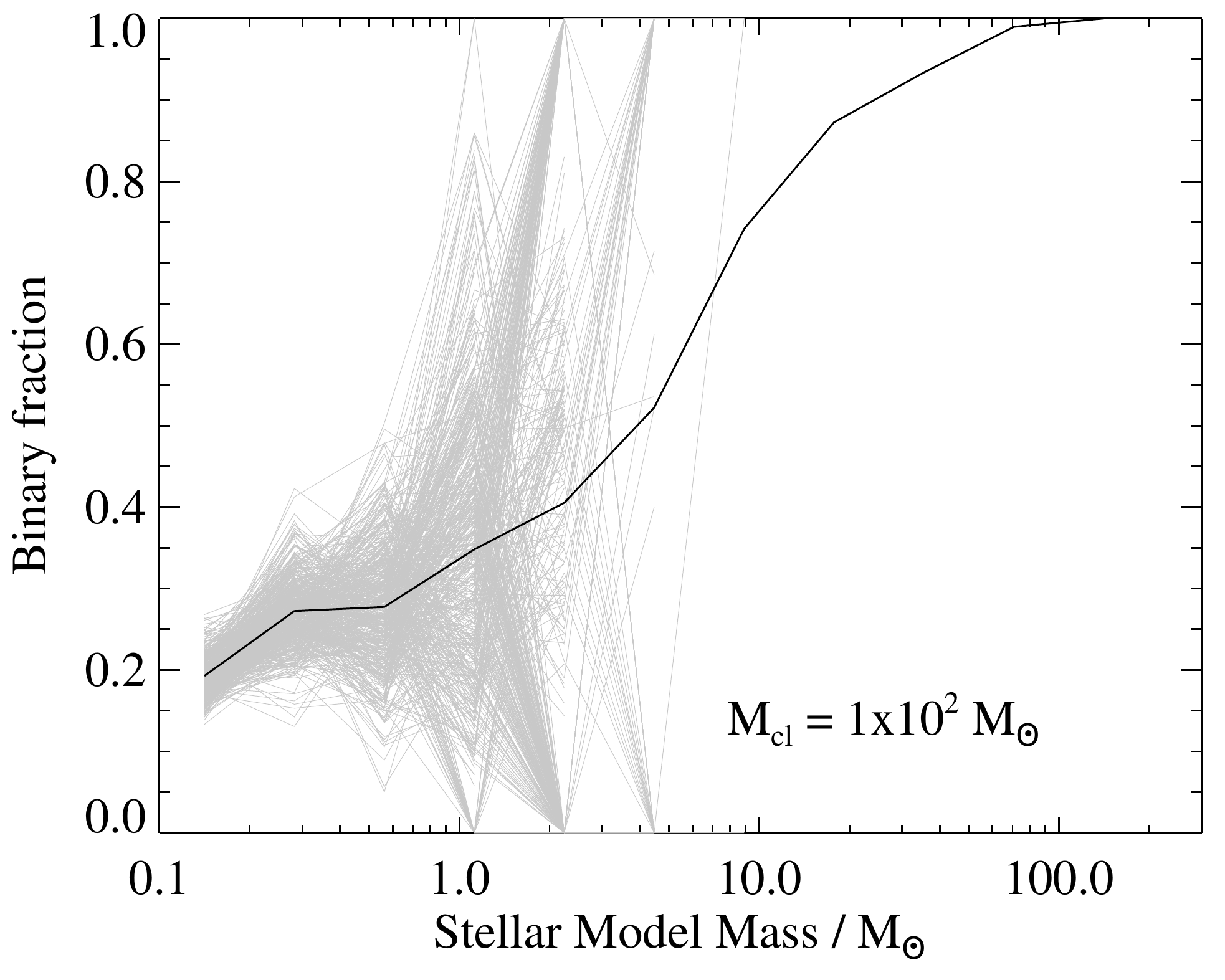}
      \includegraphics[width=0.32\textwidth]{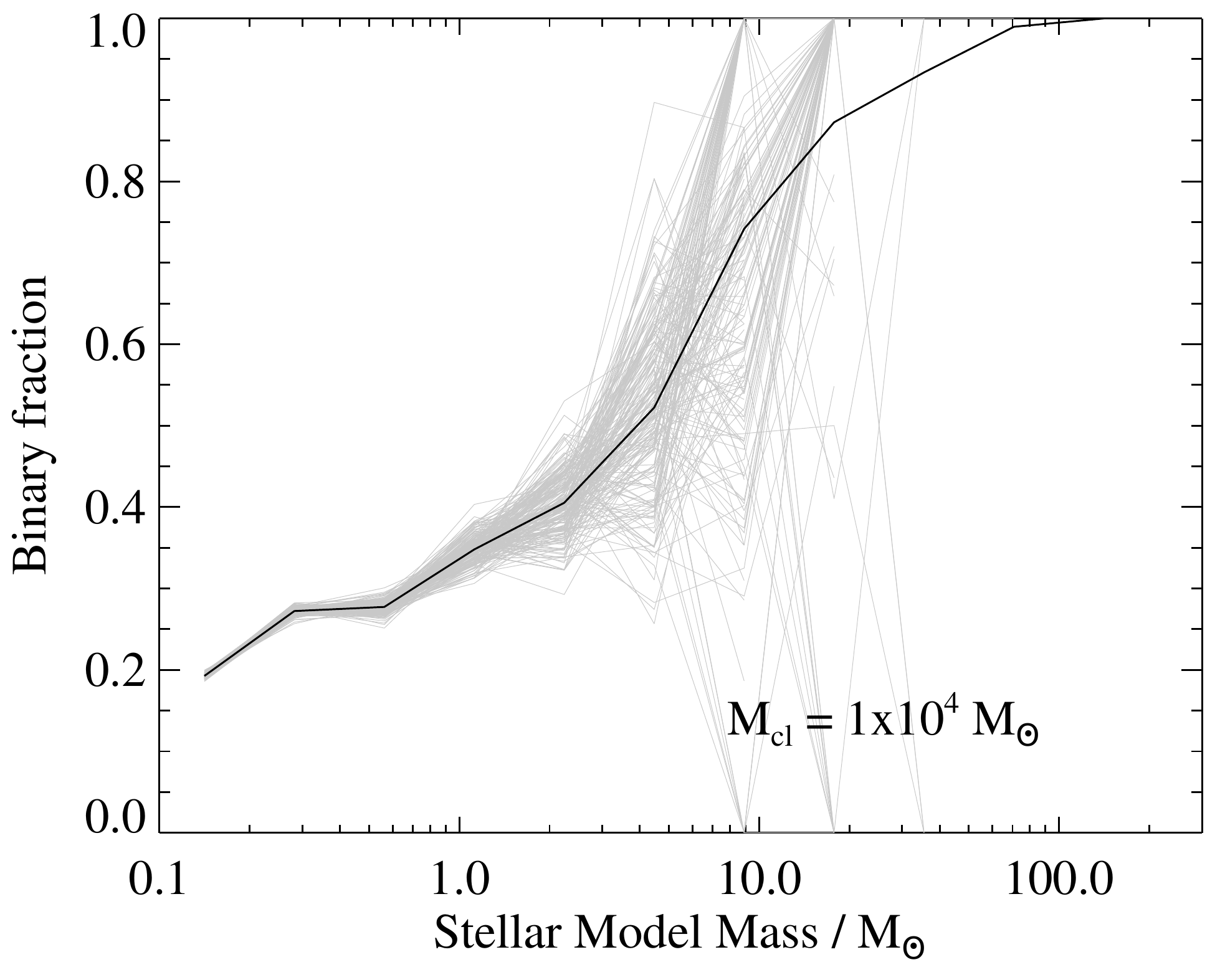}
      \includegraphics[width=0.32\textwidth]{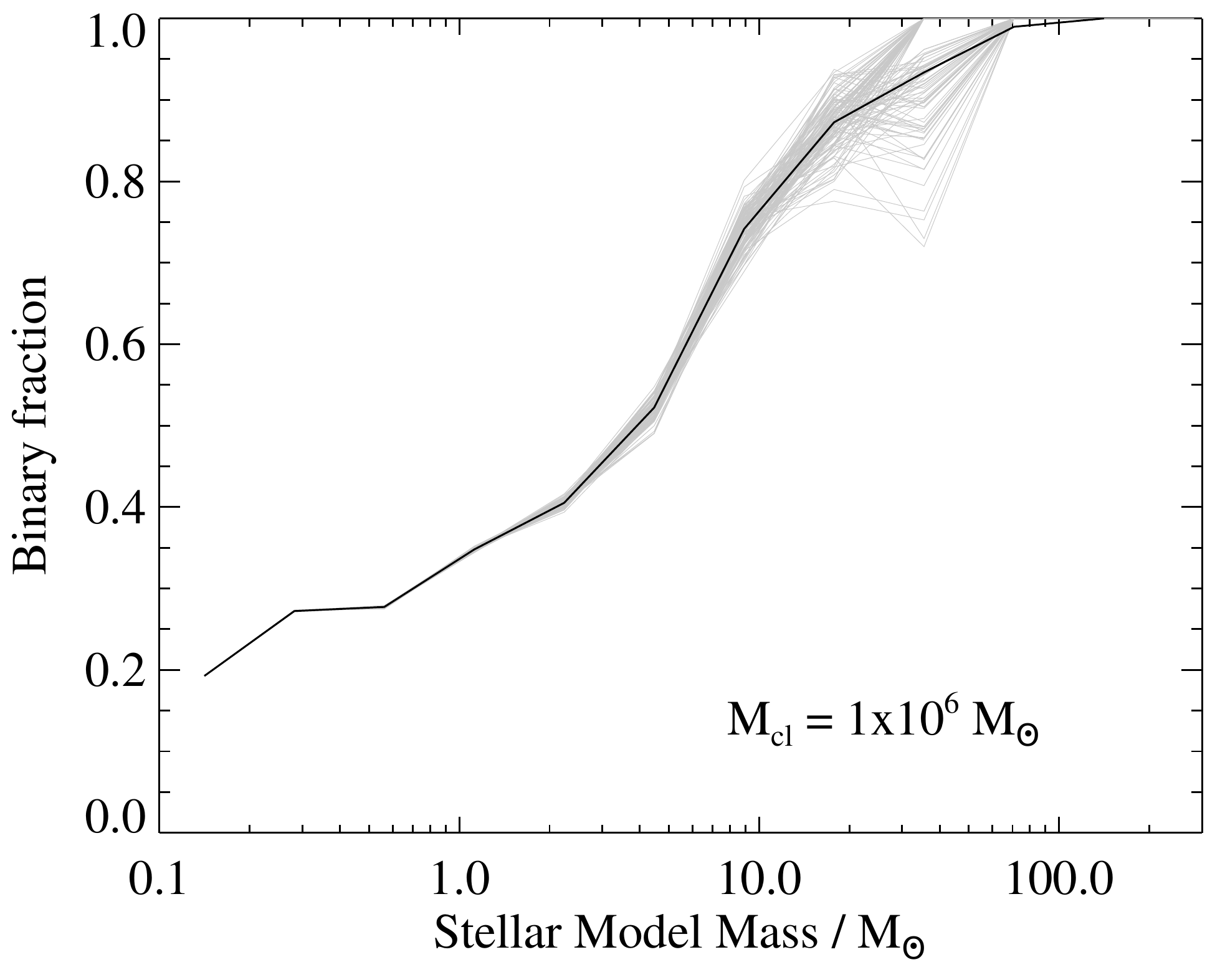}
      \caption{Examples of the binary fraction as a function of stellar mass for simple stellar populations with three different cluster masses. Thin grey lines indicate individual stochastic sampling realisations of the cluster stellar population, while the solid black line indicates the binary fraction in the statistically sampled population.}
      \label{fig:bfrac}
  \end{figure*}

\subsection{Maximum Masses}

Perhaps the most obvious property of a low mass cluster is its inability to routinely populate the high mass end of the stellar mass function.  If the total stellar mass of a system is 100\,M$_\odot$, it is highly unlikely that this will be found in a single massive star, and instead the initial mass function makes a large population of low mass stars more probable. This can have a significant impact on the star forming environment, since the most massive stars dominate the ionizing radiation and mechanical feedback from a cluster. The relationship of the most massive star to the mass of the cluster is also a key ingredient in integrated galaxy-wide initial mass function (IGIMF) estimates \citep[e.g.][]{2013MNRAS.434...84W}.

In the left panel of Figure \ref{fig:max_mass}, we illustrate the maximum stellar mass of models selected by our random sampling  realisations for each total stellar mass. Each of the hundreds of stochastic realisations is shown as a small black point, with small horizontal offsets applied at random to each point in order to improve clarity. For each total cluster mass, we also calculate the mean value for the maximum stellar mass. These are shown on the figure in red, and also listed in Table \ref{tab:maxmasses}. 

The maximum stellar mass $M_{\ast}\mathrm{(max)}$ is well described by a power-law function of the total stellar mass of the modelled cluster, $M_\mathrm{cl}$, with a linear fit to the data indicating: 
\begin{equation}
    M_{\ast}\mathrm{(max)} = 0.8\,M_\mathrm{cl}^{(0.35\pm0.02)}
\end{equation}
where $M_{\ast}\mathrm{(max)}$ and \mcl\ are in M$_\odot$.

There is, however, considerable scatter amongst the stochastic realisations at each M$_\mathrm{cl}$. Some 100\,M$_\odot$ clusters are comprised entirely of sub-M$_\odot$ dwarf stars. Others may host one or more stars with masses exceeding 20\,$M_\odot$. As would be expected, the maximum mass of the models shows less scatter at high \mcl, although we note that the relatively sparse underlying grid of individual stellar evolution models at high masses is visible in the figure. 

Realisations at the canonical BPASS model mass of M$_\mathrm{cl}=10^6$\,M$_\odot$ have a most massive star exceeding 100\,$M_\odot$ more than half the time.  We note, however, that in no realisation undertaken was a model from BPASS's grid of 300\,M$_\odot$ primary stars selected.

In the right hand panel of Figure \ref{fig:max_mass} we compare the results of our random sampling to the data compilation of \citet{2013MNRAS.434...84W}. These data are for dynamically-young clusters, which are still embedded in their birth clouds. The most massive star mass is usually determined by a fit to its spectral type (and is thus subject to evolutionary and binary-interaction uncertainties), or based on its ionizing photon output (and so subject to assumptions in far-ultraviolet stellar atmospheres), while cluster masses are derived from a star-counting approach, with a statistical correction for the contribution of low-mass stars below the observational limit \citep{2010MNRAS.401..275W}. 

Such approaches are inevitably subject to substantial uncertainties. They tend to be biased towards clusters which show an excess of massive stars, since these are easier to count and more compelling to report. Such systems will also be easier to identify as clusters in the optical since the presence of more massive stars will clear out some fraction of the natal cloud and reduce extinction. Observational estimates may also overestimate the mass of a star if it has a close binary companion. Nonetheless, it is notable that the mean properties of a pure stochastic sampling of the initial mass function differs significantly from the relationship suggested by the observed data. This is consistent with the similar conclusions of  \citet{2010MNRAS.401..275W,2013MNRAS.434...84W}, and may suggest that clusters form in a top-down process. Rather than fragmenting into a purely-stochastically-sampled IMF, clouds may collapse to form the highest mass star possible from the cluster before fragmenting further to populate the lower mass end of the IMF. However this conclusion is not universally accepted; other work has argued that the presence of more massive stars than expected in relatively low mass clusters is an indicator in favour of purely stochastic sampling \citep[since these would not occur in any cluster if an analytic constraint on the upper mass limit is assumed, ][]{2014ApJ...793....4A}.  

The highest stellar masses observed in any stochastic BPASS realisation at a given cluster mass (shown as a dashed line) does indeed provide an upper envelope to the empirical data sample and is similar in form to previous analytic estimates for the maximum mass-cluster mass relation. For comparison the right hand panel of Figure \ref{fig:max_mass} also shows the typical values of the most massive star selected in a single star population. This is slightly larger at each cluster mass than for the binary population, primarily because in the binary case the most massive star typically has a substantial companion whose mass must also be accounted for when sampling. The goal of the modelling in this work is to explore the uncertainties which arise from purely stochastic sampling. Given the biases and uncertainties associated with the observational data, we do not consider the mismatch here to necessarily rule out stochastically-constructed models, and note that regardless of the maximum mass, the models presented here provide information on the scatter in stellar properties.

\begin{table*}
    \centering
    \begin{tabular}{lccccccc}

    M$_\mathrm{cl}$/M$_\odot$ & N(iter) & N(mod,sin) & N(mod,bin) & log$_{10}($M$_{\ast}\mathrm{(max,sin)}$/M$_\odot$) & log$_{10}($M$_{\ast}\mathrm{(max,bin)}$/M$_\odot$) \\
    \hline\hline
    \smallskip
    $1\times10^2$  & 500    &  39 &  123 &  $0.52_{-0.29}^{+0.13}$  &  $0.49_{-0.27}^{+0.13}$ \\\smallskip
    $3\times10^2$  & 500    &  57 &  267 &  $0.74_{-0.32}^{+0.07}$  &  $0.65_{-0.26}^{+0.15}$\\\smallskip
    $1\times10^3$  & 300    &  84 &  541 &  $0.97_{-0.32}^{+0.15}$  &  $0.92_{-0.32}^{+0.12}$\\\smallskip
    $3\times10^3$  & 300    & 111 &  894 &  $1.13_{-0.31}^{+0.15}$  &  $1.10_{-0.32}^{+0.13}$\\\smallskip
    $1\times10^4$  & 200    & 138 & 1397 &  $1.35_{-0.24}^{+0.14}$  &  $1.35_{-0.35}^{+0.13}$ \\\smallskip
    $3\times10^4$  & 200    & 159 & 2007 &  $1.55_{-0.26}^{+0.11}$  &  $1.48_{-0.27}^{+0.12}$ \\\smallskip
    $1\times10^5$  & 100    & 184 & 2986 &  $1.81_{-0.31}^{+0.15}$  &  $1.70_{-0.31}^{+0.20}$ \\\smallskip
    $3\times10^5$  & 100    & 203 & 4185 &  $1.96_{-0.27}^{+0.22}$  &  $1.89_{-0.29}^{+0.11}$ \\\smallskip
    $1\times10^6$  & 100    & 218 & 5603 &  $2.16_{-0.25}^{+0.19}$  &  $2.06_{-0.21}^{+0.24}$  \\\smallskip
    $3\times10^6$  & 100    & 234 & 6753 &  $2.27_{-0.10}^{+0.12}$  &  $2.22_{-0.14}^{+0.08}$ \\\smallskip
    $1\times10^7$  & 100    & 255 & 7580 &  $2.40_{-0.05}^{+0.04}$  &  $2.29_{-0.07}^{+0.07}$\\
    \hline\smallskip
    Statistical    &        & 284 & 19089 &  2.48            & 2.48 \\
        \end{tabular}
    \caption{Typical properties of the cluster models as a function of  cluster stellar mass. The second column reports the number of random stochastic realisations undertaken per metallicity. The third and fourth columns gives the typical number of different stellar evolution models selected from the BPASS  grid in the case of an entirely single star population or a population including binaries, the fifth and sixth columns give the mean mass and standard deviation of the most massive star in each cluster in these two cases.}
    \label{tab:maxmasses}
\end{table*}

\subsection{Binary Fraction}

In all realisations, the overall fraction of the total cluster mass which is found in binary stars remains constant at $b_\mathrm{frac}=0.24$, with this fraction mostly determined by the low mass stars which dominate the weight budget. In Figure \ref{fig:bfrac} we show how the fraction of stellar mass in binaries depends on the primary mass of the stellar model, giving an indication of the uncertainty that stochastic sampling introduces on this most fundamental of binary population parameters. 

Unsurprisingly for the most massive stars in each realisation, the binary fraction becomes very erratic, since it must be either zero and one when each individual star is drawn from the underlying statistical distribution and it is not uncommon for the uppermost mass bins to be populated by only one or a few stars in any given realisation. For clusters with \mcl$=10^2$\,M$_\odot$, the binary fraction shows considerable scatter at all stellar masses. The scatter between stochastic realisations is still substantial for \mcl$=10^4$\,M$_\odot$, although by this cluster mass, stellar models below 3\,M$_\odot$ follow a well-defined sequence in binary fraction. Clusters with masses \mcl$=10^6$\,M$_\odot$ appear to track the underlying binary fraction distribution well at all stellar masses up to 20\,M$_\odot$, but demonstrate realisation-to-realisation variation in the binary fraction for stellar models at masses above this. This fraction is above 70\,per\,cent in all realisations.

It is important to note that this variation illustrates the challenge in determining an underlying distribution of binary parameters. Any observed cluster represents a single realisation which samples the physics of cluster fragmentation. Even if that physics leads to a fundamental binary fraction distribution, the empirical fraction measured in any given cluster may show large variations around this, particularly for massive stars.

 \subsection{Ionizing Photons}

  Photons with energies $h\nu>13.6$\,eV, i.e. those capable of ionizing hydrogen, are generated primarily by young (age$<10$\,Myr), massive ($M>10$\,M$_\odot$) stars. When binary population synthesis is included, further ionizing radiation is generated at moderate ages (10-100\,Myr) by stars which have been rejuvenated by mass transfer, or in which envelope-stripping has left an exposed helium core. At very late times (e.g. $>$1\,Gyr), white dwarfs and accreting compact objects take over as the dominant source of a stellar population's diffuse ionizing radiation field.
  
\begin{figure*}
      \centering
      \includegraphics[width=0.32\textwidth]{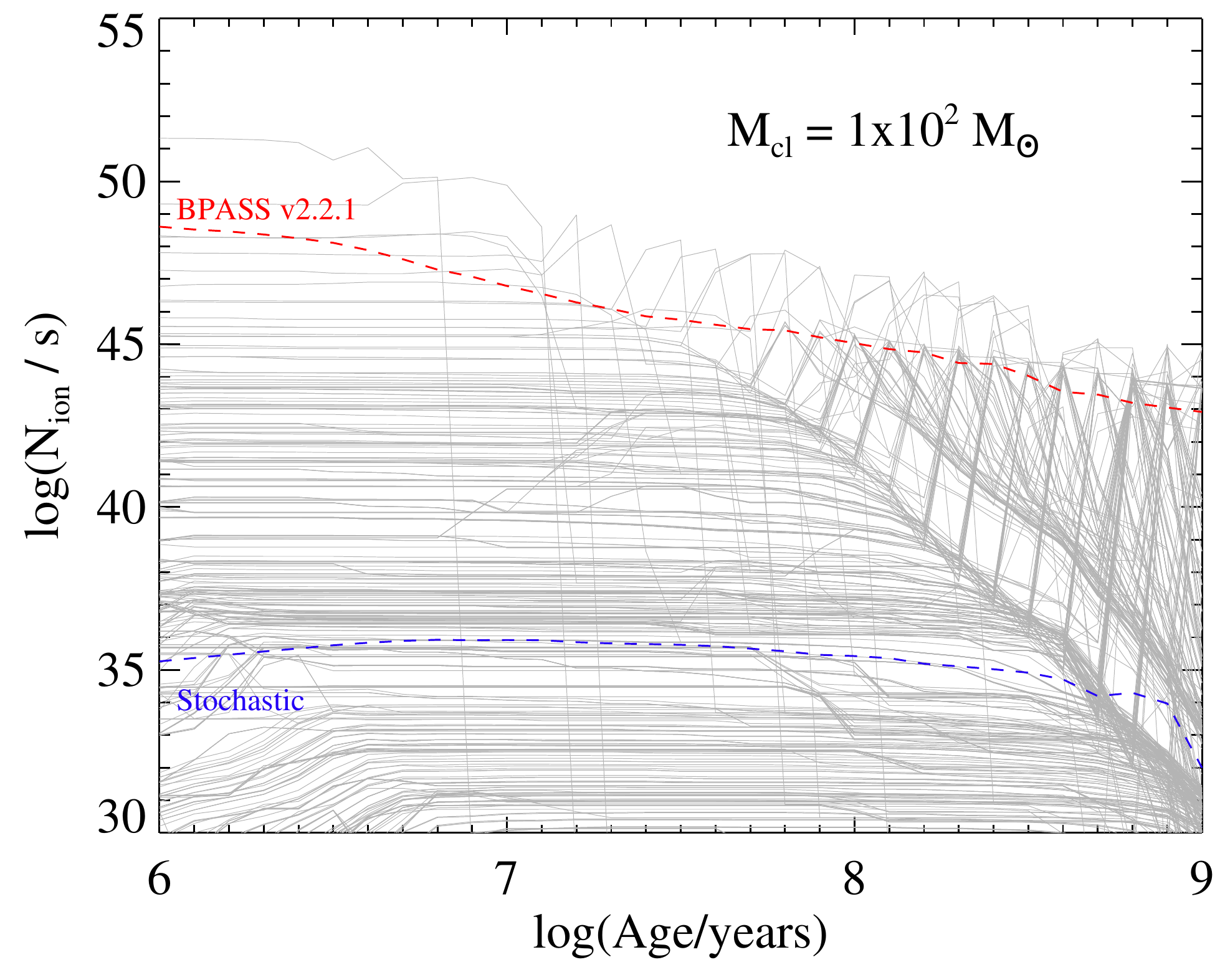}
      \includegraphics[width=0.32\textwidth]{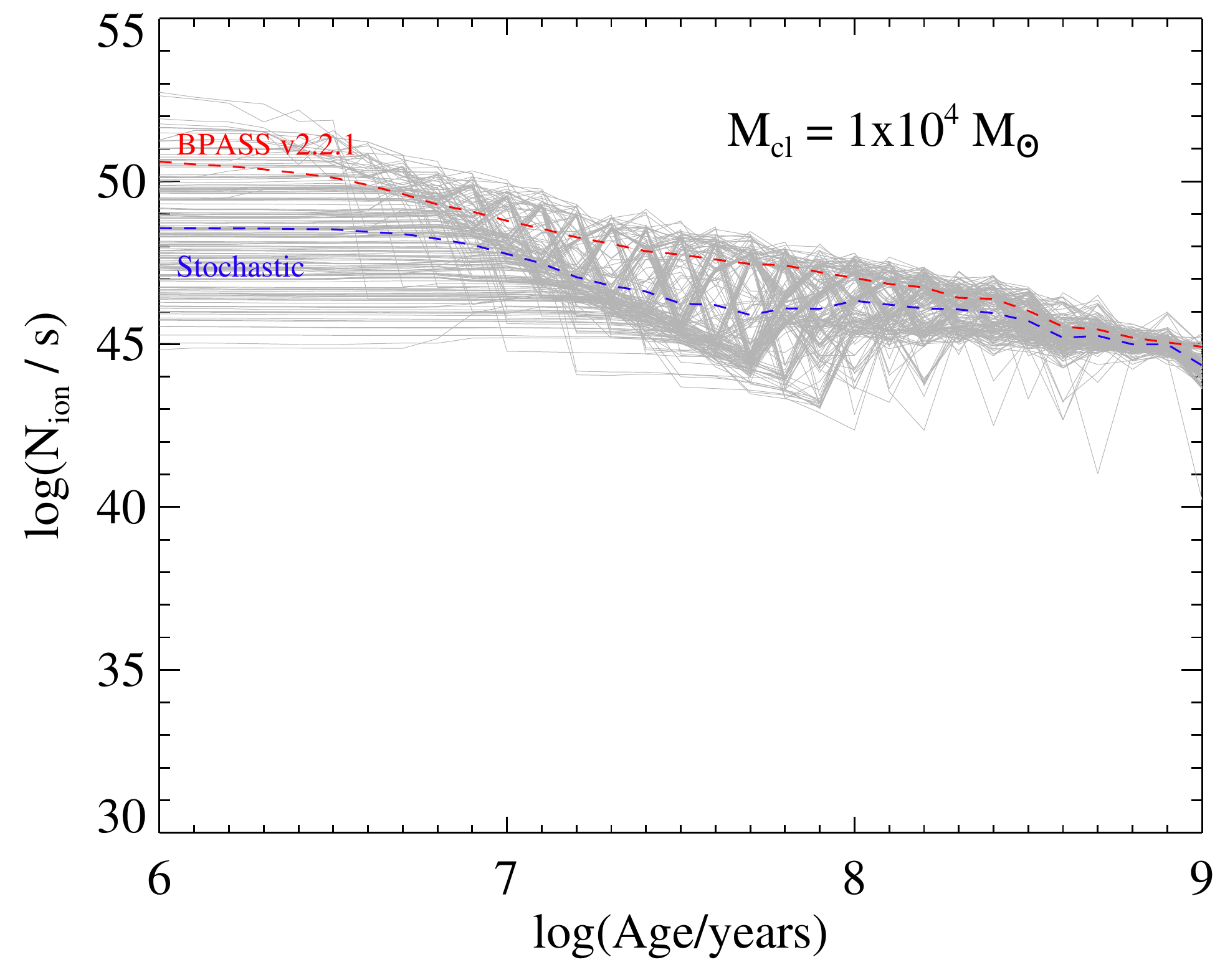}
      \includegraphics[width=0.32\textwidth]{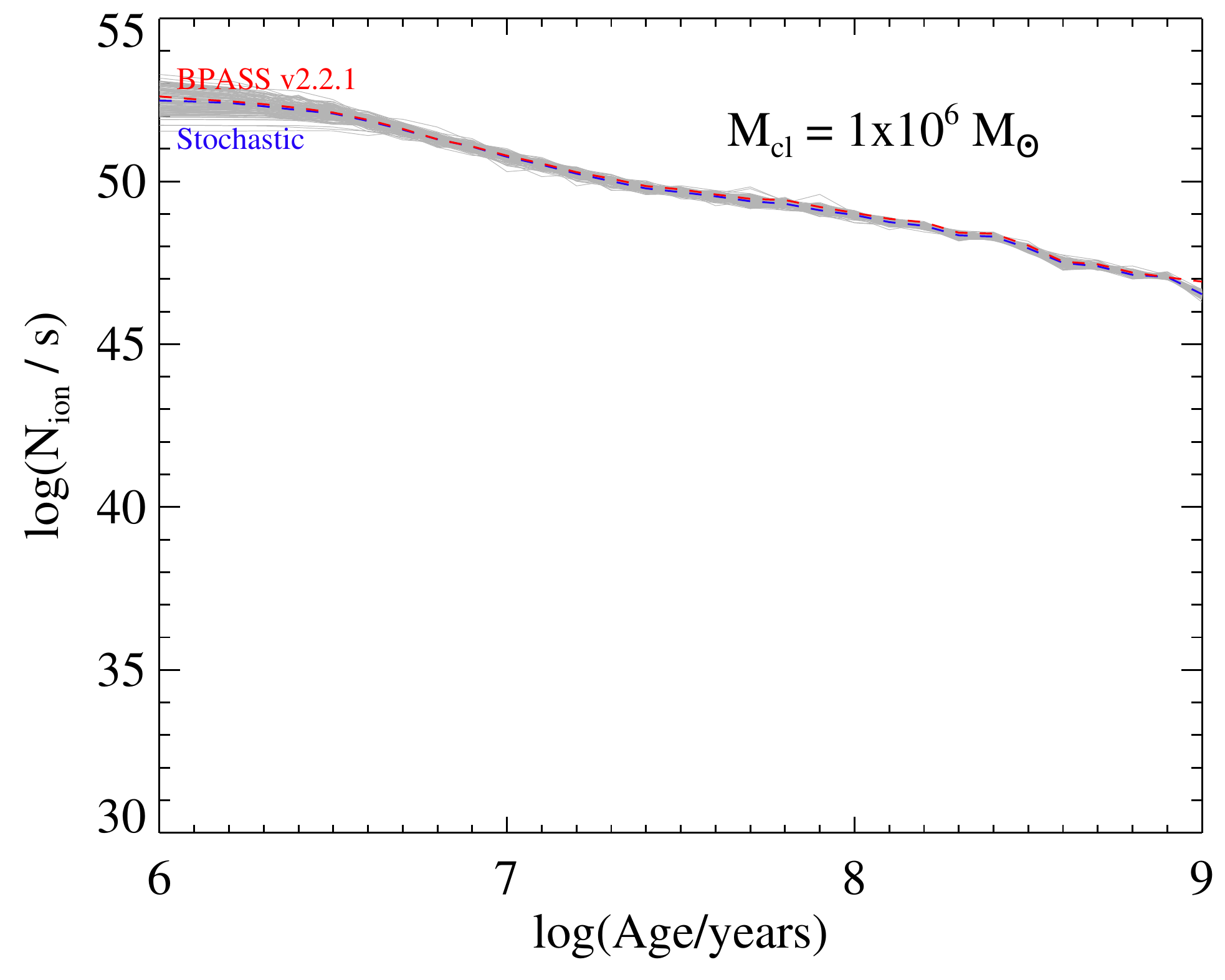}
      \caption{Examples of the evolution of ionizing flux for simple stellar populations with three different cluster masses. Thin grey lines indicate individual stochastic sampling realisations of the cluster stellar population, while blue dashed line indicates the mean for stochastic realisation. Red dashed line indicates the prediction from statistical sampling of parameter distributions.}
      \label{fig:ion_ind}
  \end{figure*}

  Given the importance of massive stars, it is unsurprising that the ionizing photon production rate is highly sensitive to the uncertainties of stochastic sampling. In Figure \ref{fig:ion_ind}, we show examples of the ionizing photon production histories from our models at $Z=0.020$ (the full set of histories, and a comparison of the time evolution of single and binary populations is presented in the Appendix).  Individual stochastic realisations are shown as thin grey lines. For clarity, a representative sample of individual realisations is shown at low cluster masses, rather than the full set. For each M$_\mathrm{cl}$, the mean ionizing photon production rate as a function of age is shown by the dashed blue line, while the ionizing photon production rate of the fiducial (statistically sampled) BPASS v2.2.1 model, scaled to the same total stellar mass, is shown as a red line. 
  
  For low mass clusters, the statistical sampling overpredicts the mean ionizing photon production rate, but any one cluster may show substantial variation in its photon production with time.  Jumps in the distribution are clearly visible at ages above 10\,Myr and are caused by individual stellar models with discrete masses undergoing binary interactions in different time bins. In low mass clusters, these individual models can contribute a substantial fraction of the total ionizing flux.  In higher mass clusters, the larger number of models, and better IMF sampling, tends to cancel out the largest variations, leading to a better defined relationship between ionizing photon production rate and stellar population age, but clusters with \mcl=$10^4$\,M$_\odot$ still span six dex in possible ionizing photon output at early times. Clusters of \mcl$=10^6$\,M$_\odot$, are well described by the fiducial BPASS v2.2.1 models, independent of stochastic sampling.
  
  To better quantify the uncertainty on the ionizing photon production rate, we instead consider a continously star forming population. This essentially provides a summation of the ionizing photons from a population with age, provided the ongoing star formation has been constant for an interval of at least $\sim$100\,Myr. For each \mcl, and each of three metallicities, we calculate the mean and standard deviation of the ionizing photon production rate, assuming a star formation rate of 1\,M$_\odot$\,yr$^{-1}$. This scenario is, of course, problematic for low mass \mcl\ cases, in which we would have to imagine many individual clusters forming sequentially within some larger star formation complex, but these photon rates can be scaled linearly with star formation rate and are shown here scaled to the same constant star formation rate for comparison purposes. 
  
\begin{figure}
      \centering
      \includegraphics[width=\columnwidth]{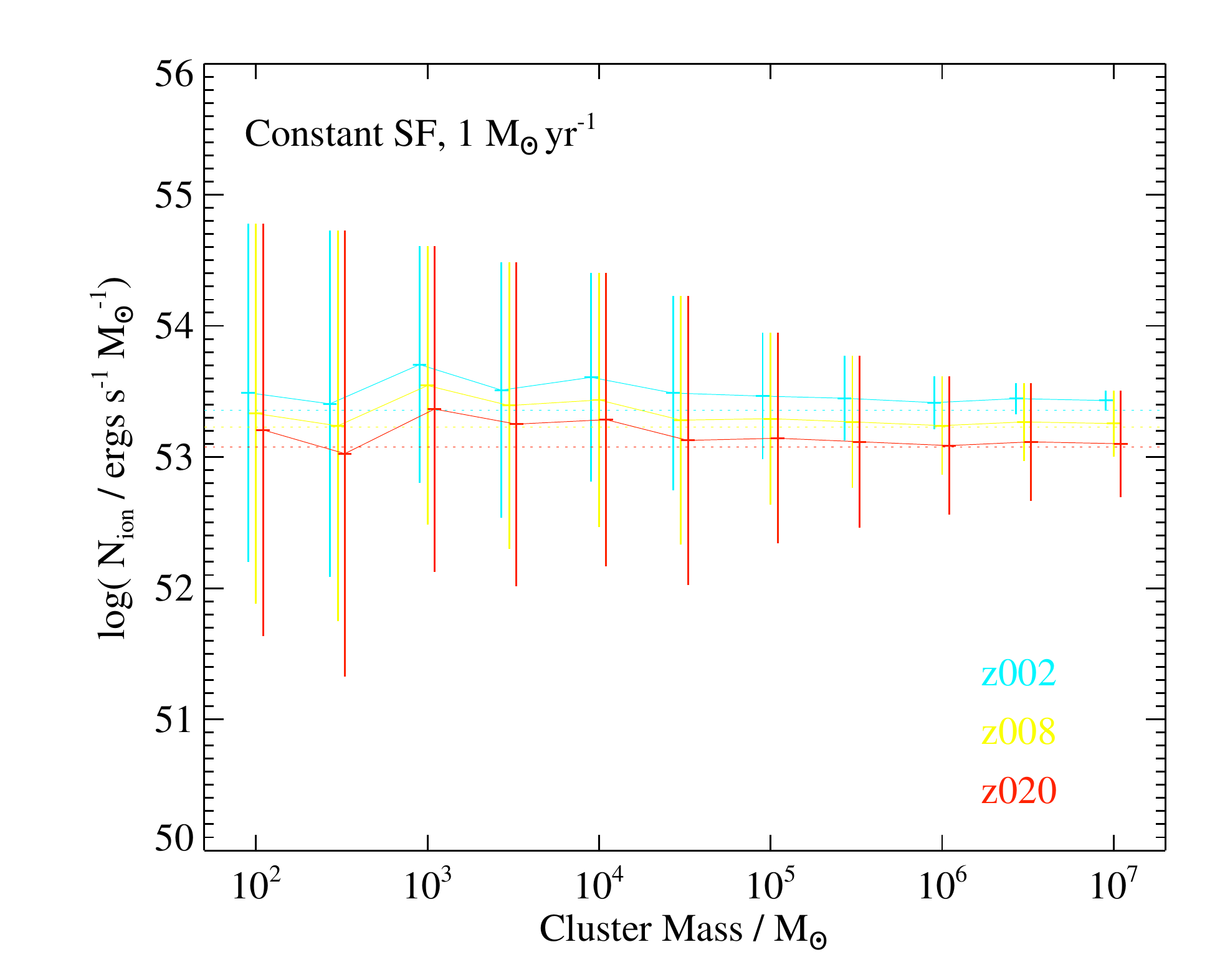}
      \caption{Average and standard deviation of ionizing flux from a continuous starburst as a function of metallicity and cluster mass. Horizontal lines indicate the expectation from statistical sampling.}
      \label{fig:ion_av}
  \end{figure}

  As Figure \ref{fig:ion_av} illustrates, the mean ionizing photon production per solar mass of ongoing constant star formation is dependent on cluster mass.  The mean lifetime-integrated ionizing photon rate is well described by the BPASS v2.2.1 value at  \mcl$>10^5$\,M$_\odot$, at all three metallicities investigated. However at low cluster masses, the population shows a deficit in ionizing photon output, and the scatter, both of the means and between stochastic realisations, is also heavily dependent on cluster mass. Low mass clusters show the largest scatter, but the uncertainty on log(N$_\mathrm{ion}$) is still 0.20, 0.38 and 0.53\,dex around the fiducial value for clusters with $10^6$\,M$_\odot$ at metallicities $Z=0.002$, 0.008 and 0.020 respectively. The lower scatter at low metallicities is likely a result of the higher overall ionizing photon productions. At low metallicities, stellar winds are weaker and stars retain more of their mass. As a result, stars with a wider range of initial masses contribute to the total ionizing flux, and the impact of any one model being selected or deselected is smaller. The dependence of $\sigma$(log(N$_\mathrm{ion}$/s)) on \mcl\ and $Z$ is tabulated in the Appendix.
  
\begin{figure}
      \centering
      \includegraphics[width=\columnwidth]{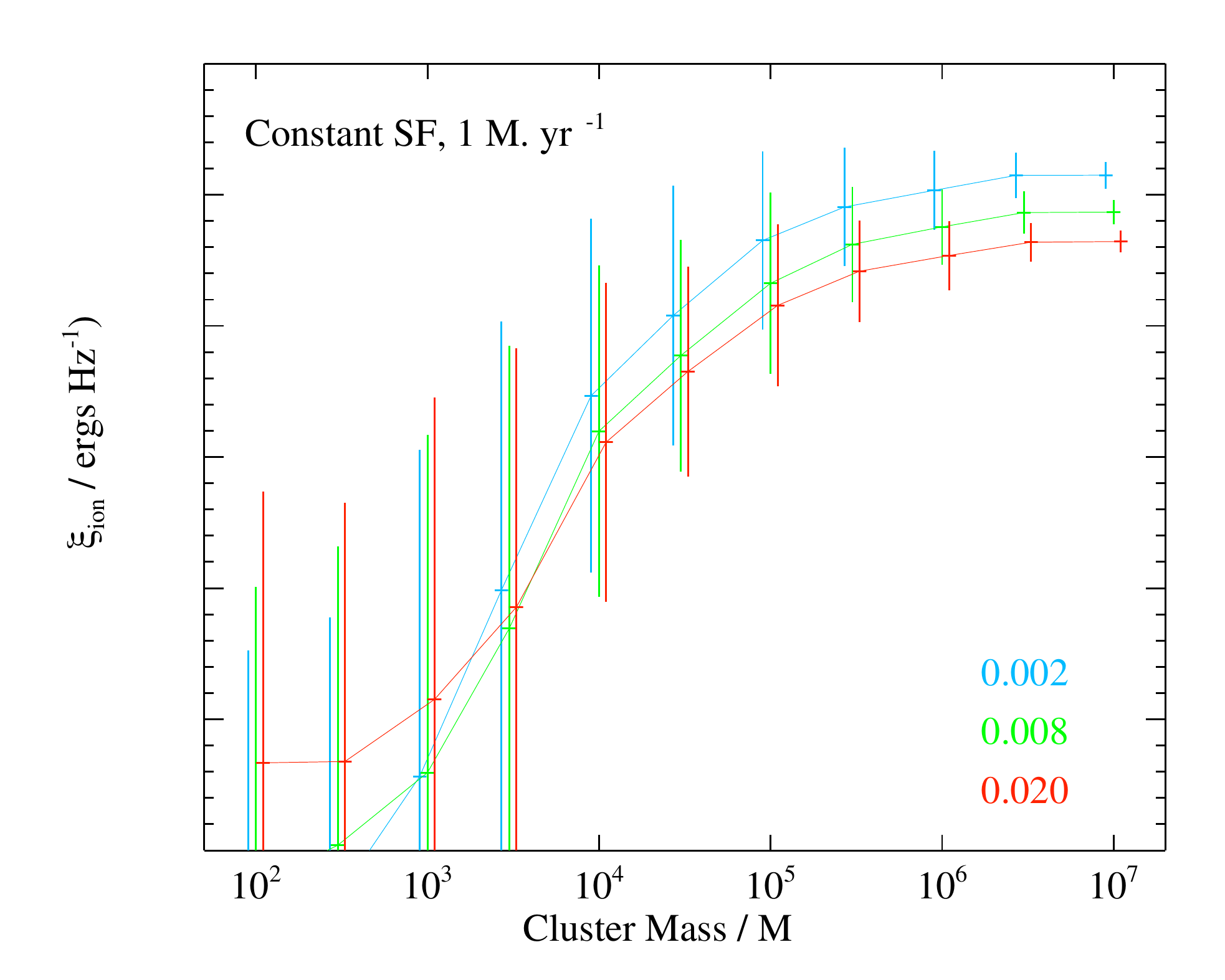}
      \caption{Average and standard deviation of ionizing flux production efficiency, $\xi_\mathrm{ion}$ from a continuous starburst as a function of metallicity and cluster mass.}
      \label{fig:xiion}
  \end{figure}

  We also consider the effect of stochastic sampling on the ionizing photon production efficiency, $\xi_\mathrm{ion} = N_\mathrm{ion}/L_\mathrm{1500\AA}$. The far-ultraviolet continuum luminosity $L_\mathrm{1500A}$ is generated by stars with a far wider range of masses and lifetimes than produce ionizing photons. As a result, clusters with relatively low \mcl\ and $M_\ast$(max) can still be luminous in the far-ultraviolet, but have a low $N_\mathrm{ion}$. The result is shown in Figure \ref{fig:xiion}. Star formation events comprised of large numbers of independently forming low mass clusters will have a very low ionizing photon production efficiency. As clusters exceed \mcl$=10^5$\,M$_\odot$, the production efficiencies approach the cannonical BPASS values, but the scatter remains large, particularly at solar metallicities.  For clusters with \mcl$=10^6$\,M$_\odot$, the uncertainty on log($\xi_\mathrm{ion}$) is still 0.16, 0.30 and 0.1\,dex around the fiducial value at metallicities $Z=0.002$, 0.008 and 0.020 respectively.

 \subsection{Colours and spectral features}
 
 \begin{figure}
      \centering
      \includegraphics[width=\columnwidth]{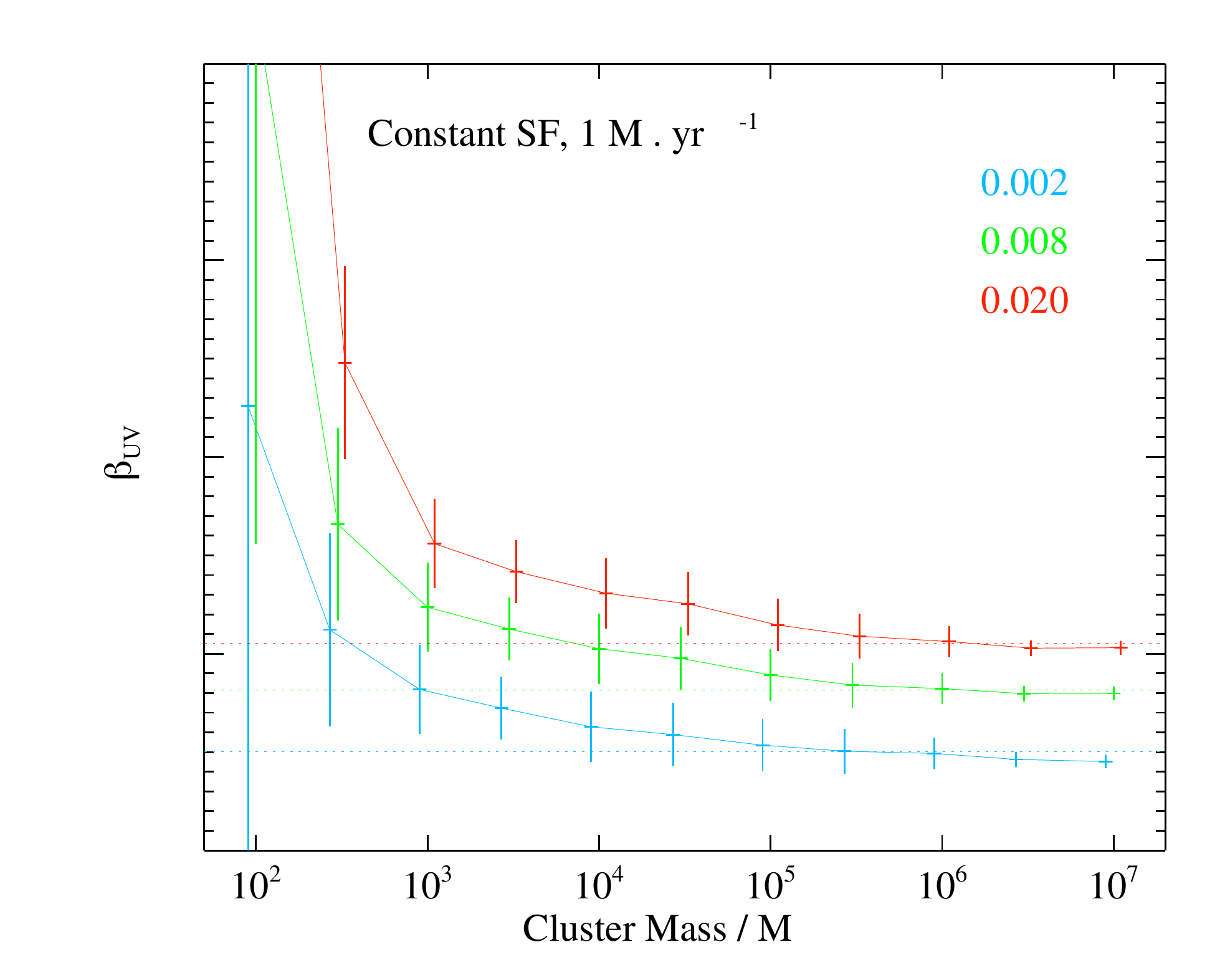}
      \caption{Average and standard deviation of rest-frame ultraviolet spectral slope from a continuous starburst as a function of metallicity and cluster mass. Dotted horizontal lines give the equivalent values from statistical sampling in BPASS v2.2.1.}
      \label{fig:uvslope}
  \end{figure}

 Moving longwards of the Lyman-break, the rest-frame ultraviolet slope of star forming systems is often used in the distant Universe to estimate their dust extinction, and can also be used in combination with other observables as an indicator of age and metallicity. In Figure \ref{fig:uvslope} we indicate how uncertainties on the ultraviolet spectral slope reflect stochastic sampling for a continuously star forming population. We define $F_\nu\propto\lambda^{\beta_\mathrm{UV}}$, and determine this using the ratio of luminosities at 1566 and 2266\,\AA\ (corresponding to the FUV and NUV Galex bands).  We calculate values assuming star formation at a constant rate for the previous 100\,Myr. We note that we do not account here for the reddening effect of nebular gas emission, but focus on uncertainties in the underlying stellar continuum.
 
 As was the case for the ionizing flux in Figure \ref{fig:ion_ind}, the statistical case is well reproduced by stochastic sampling in high mass clusters, and represents the bluest and hottest population (due to the presence of very high mass stars). Star forming clusters with lower \mcl\ typically show redder ultraviolet slopes at a given metallicity, due to the absence of massive stars, with the expected larger scatter on that slope. Numerical values are tabulated in the Appendix.

\begin{figure}
      \centering
      \includegraphics[width=\columnwidth]{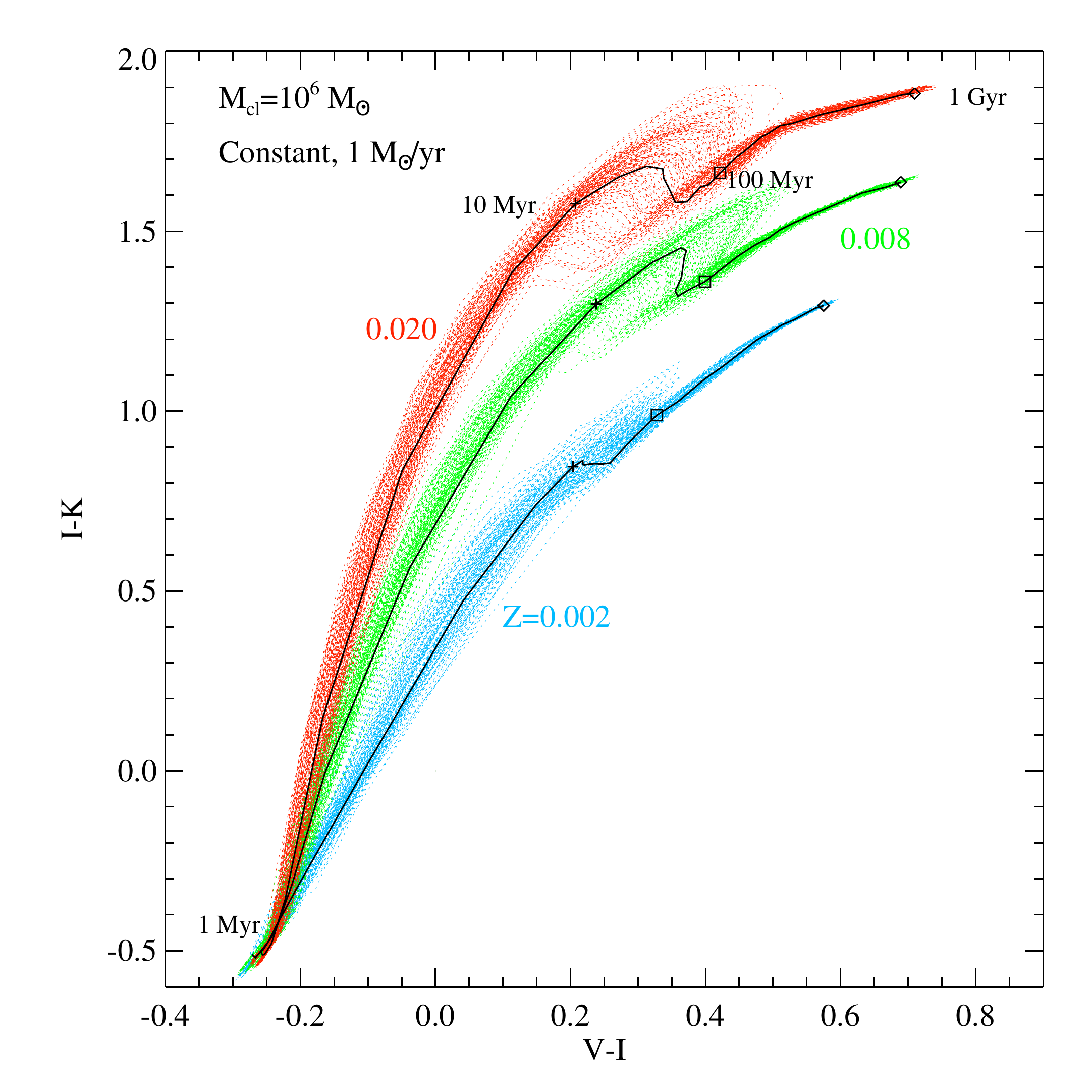}
      \caption{The evolution of a population with constant star formation in $V-I$ vs $I-K$ colour-colour space. Individual stochastic realisations for a cluster with \mcl$=10^6$\,M$_\odot$ are shown with dotted lines, coloured by metallicity. There is considerable uncertainty on the colour of star forming populations at ages 10-100\,Myr. }
      \label{fig:colcol}
  \end{figure}

 Moving redwards into the optical, in Figure \ref{fig:colcol} we consider the evolution of a stellar population forming stars at a constant rate in an optical-near-infrared colour-colour space, for a cluster with \mcl$=10^6$\,M$_\odot$. This is representative of the uncertainty on the fiducial BPASS $10^6$\,M$_\odot$ models. As the figure makes clear, the uncertainty on colour is generally small, with a scatter $<<0.1$\,Mag at most ages. By the time the population reaches 1 Gyr after the onset of star formation, the uncertainty on the colours arising from stochastic sampling at this mass is $\Delta(V-I)$=[0.014, 0.020, 0.023]\,mag and $\Delta(I-K)=$[0.012, 0.013, 0.015]\,mag at $Z=$[0.002, 0.008, 0.020] respectively. 
 
 However, in the interval between 10 and 100 Myr after the onset of star formation, the colour uncertainty arising from stochastic parameter sampling is considerable, with offsets of 0.2-0.3\,magnitudes in colour between different realisations common. This is likely a result of the combination of massive star IMF and binary parameter sampling - it is at these ages that primary stars with masses in the range 5-20\,M$_\odot$ are expanding off the main sequence. As a result, many stars in binaries are also undergoing mass transfer which can rejuvenate a population. These mid-sized stars are very much more common than the massive stars which dominate in the ultraviolet, but still show a high binary fraction. The presence or absence of stars in this mass range, and their individual mass ratios and periods as selected in each stochastic realisation, will thus have an impact on optical colour evolution at this time. It is perhaps a little surprising that the variation remains so large even with $10^6$\,M$_\odot$ samples, and this points to a need for caution when evaluating population properties at these moderate ages. Even in the most massive clusters, the number of very high mass stars is small.
  
 We note that for cluster masses \mcl$\sim10^3-10^5$\,M$_\odot$, and for ageing simple stellar populations, the colour evolution is far more erratic as individual realisations are strongly affected by small numbers of massive stars passing through luminous post-main sequence and mass transfer phases. At the lowest cluster masses, the scatter on photometric colour between stochastic realisations is large, but less time dependant due to the absence of stars with M$>10$\,M$_\odot$ in the majority of realisations.  The role of individual models is somewhat smoothed over time for the constant star formation case compared to that of simple stellar populations. Examples of the colour uncertainty for lower cluster masses are given in the Appendix (see e.g. figure \ref{fig:col_inst}).
 
\begin{figure*}
      \centering
      \includegraphics[width=0.32\textwidth]{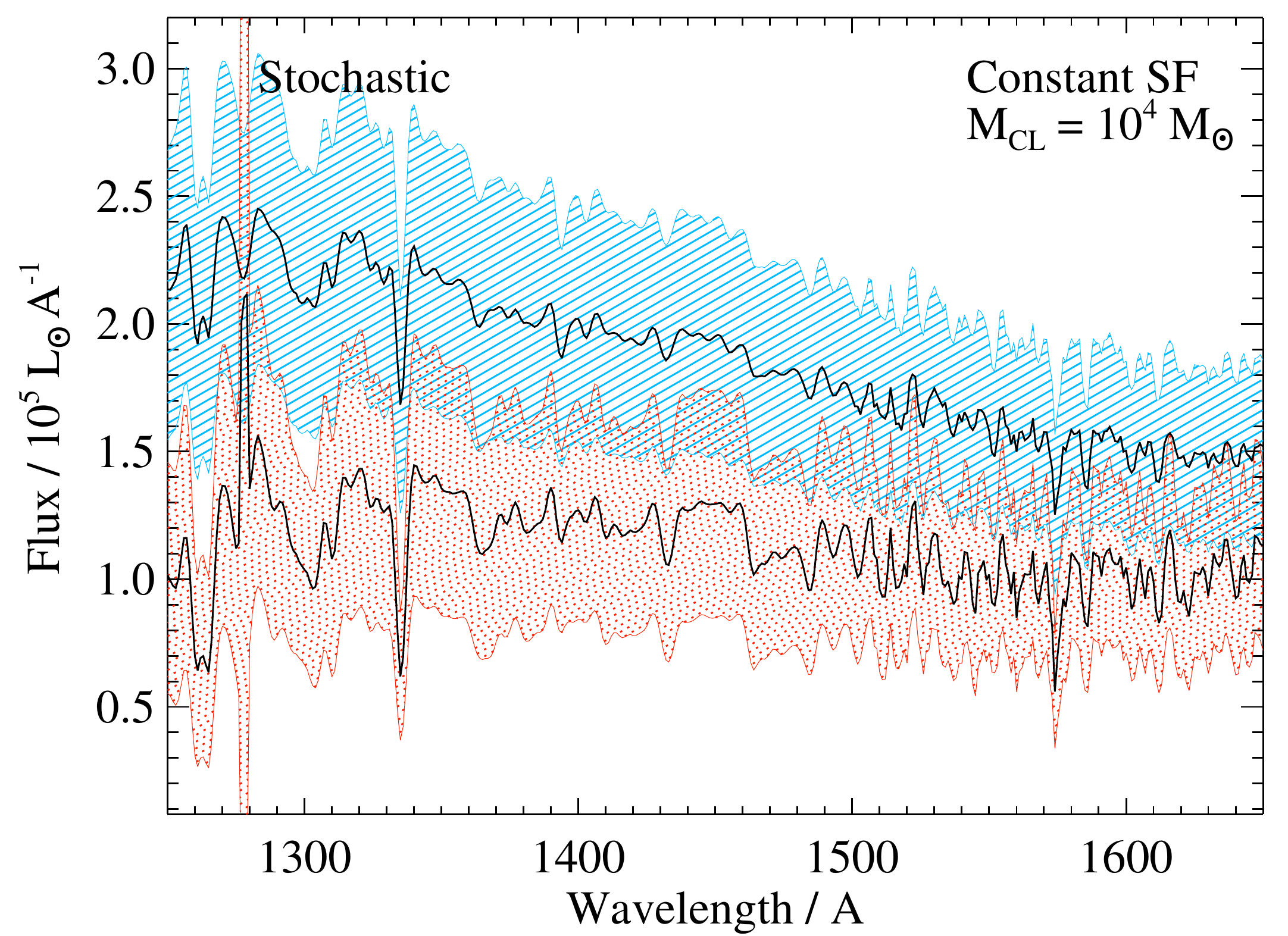}
      \includegraphics[width=0.32\textwidth]{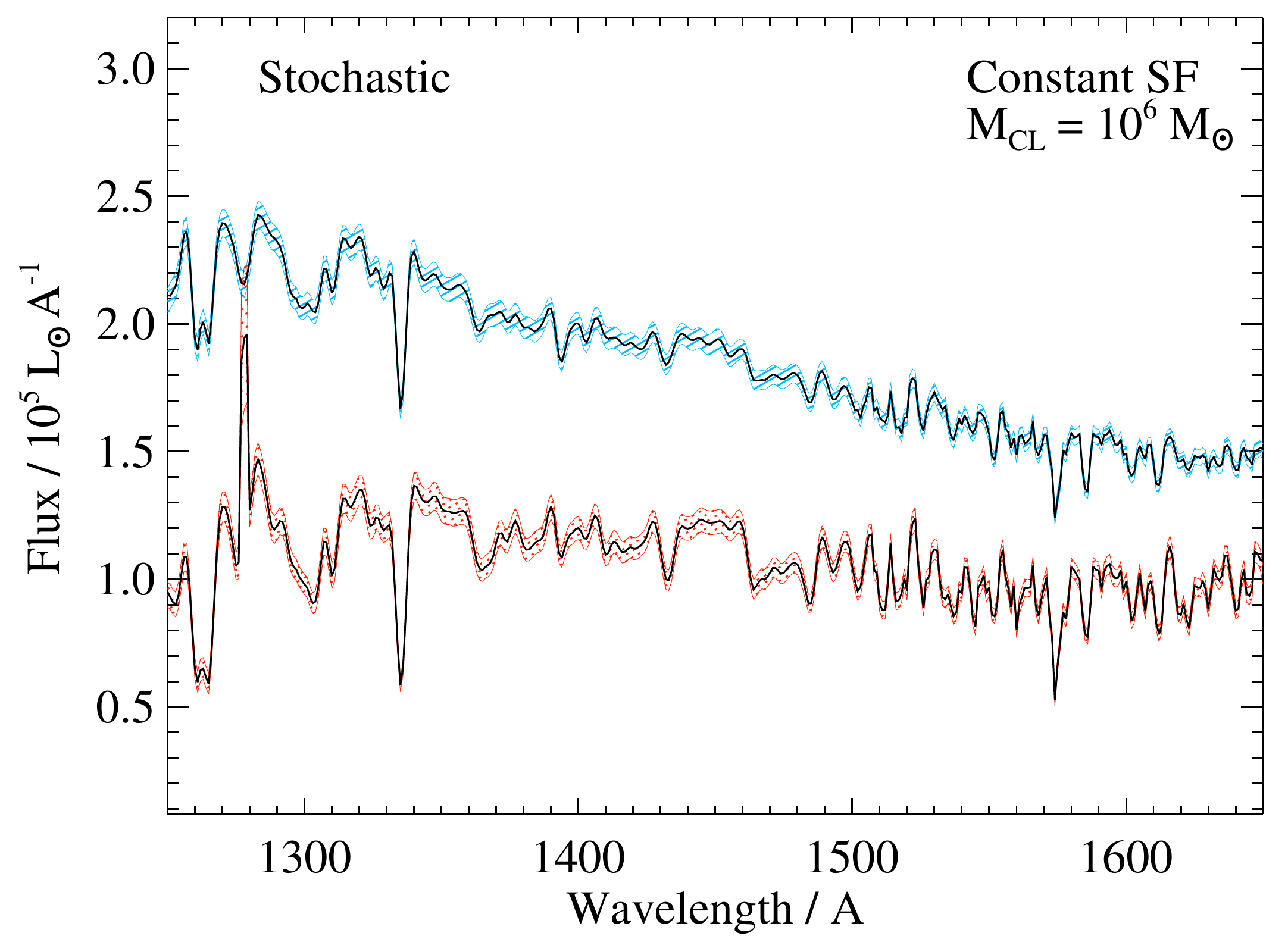}
      \includegraphics[width=0.32\textwidth]{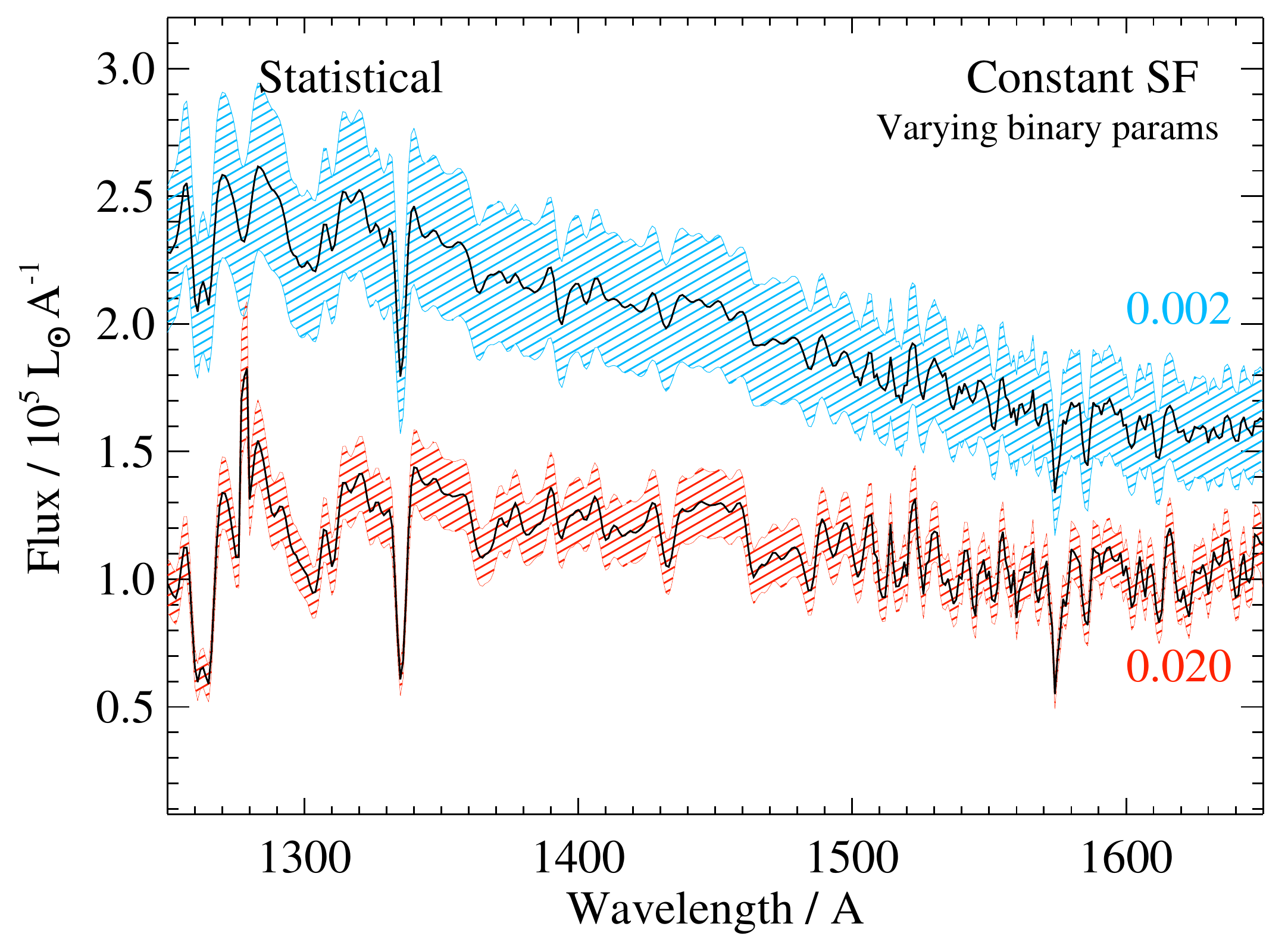}
      \caption{Examples of the uncertainty on the ultraviolet continuum due to the presence of stochastic sampling with \mcl$=10^4$\,M$_\odot$ and $10^6$\,M$_\odot$. Models with \mcl$=10^2$\,M$_\odot$ are not shown since the scatter on these exceeds the scale of the figure. Constant star formation is assumed to extend over 100\,Myr, giving a total stellar mass formed of $10^8$\,M$_\odot$ in each case. We show the standard deviation of the flux as a function of wavelength for $Z=0.002$ and $Z=0.020$ as shaded regions. For comparison, the left hand panel also shows the same measure for models which sample the IMF and binary parameters statistically but probe uncertainty in the underlying binary parameter distributions \citep{2020MNRAS.495.4605S}. }
      \label{fig:uvspec}
  \end{figure*}

Since the ultraviolet is of particular interest for many galaxy studies, Figure \ref{fig:uvspec} shows how the uncertainty on the rest-frame ultraviolet (1250-1650\AA) continuum and absorption lines depends on the cluster size when sampling stochastically, for the case of constant star formation. Shaded regions indicate the $\pm1\,\sigma$ range of the luminosity density estimates in  independent stochastic realisations, as a function of wavelength. The scatter is comparable in all three metallicities modelled, and we opt to show only $Z=0.002$ and 0.020 for clarity. The luminosity for \mcl$=10^4$\,M$_\odot$ clusters has been scaled to a total mass of $10^8$\,M$_\odot$ (formed by constant star formation at 1\,M$_\odot$\,yr$^{-1}$ over 100\,Myr) to allow direct comparison. For comparison, the right hand panel also shows the spectral uncertainty in the same wavelength range and metallicities which arises from uncertainty in the binary parameter distributions, using the models presented in \citet{2020MNRAS.495.4605S}. 

For clusters at our canonical mass of \mcl$=10^6$\,M$_\odot$, the flux uncertainty from stochastic sampling is only 2-4\,per\,cent in this wavelength range (depending on metallicity), compared to the 10-13\,per\,cent which arises from uncertainties in the binary parameter distribution. However, when the mass of individual clusters is reduced to $10^4$\,M$_\odot$, the uncertainty on the flux as a function of wavelength is increased to 24-36\,per\,cent. We note that the fractional uncertainty on the cluster flux (as shown in the left hand panel of the figure, and tabulated in Table \ref{tab:unc-spectra}) is perhaps a little misleading in instances where the flux has been scaled to higher masses; some of the stochastic variation will average out as individual clusters are summed to create a more massive population. Nonetheless, for clusters with individual masses of \mcl$\leq3\times10^4$\,M$_\odot$, the flux uncertainty from stochastic sampling will likely dominate over the binary parameter uncertainty. Similar behaviour is observing at optical wavelengths, where $\sigma F/F\sim3-4$\,per\,cent for \mcl$=10^6$\,M$_\odot$. For very low mass clusters (\mcl$\leq10^3$\,M$_\odot$), the flux uncertainty that results from stochastic sampling of a star-forming population approaches or exceeds 100\,per\,cent.

 \subsection{Supernova Rates}

The population synthesis stage of BPASS identifies stellar models which undergo a supernova at the end of their lifetimes. A star is deemed to undergo a core collapse supernova if it ends its life having formed a CO-core of at least 1.38\,M$_\odot$, has undergone core carbon burning and has a total mass exceeding 1.5\,M$_\odot$. While BPASS also tracks type Ia supernovae, these occur at significantly later times and are not considered further here.

\begin{figure}
      \centering
      \includegraphics[width=\columnwidth]{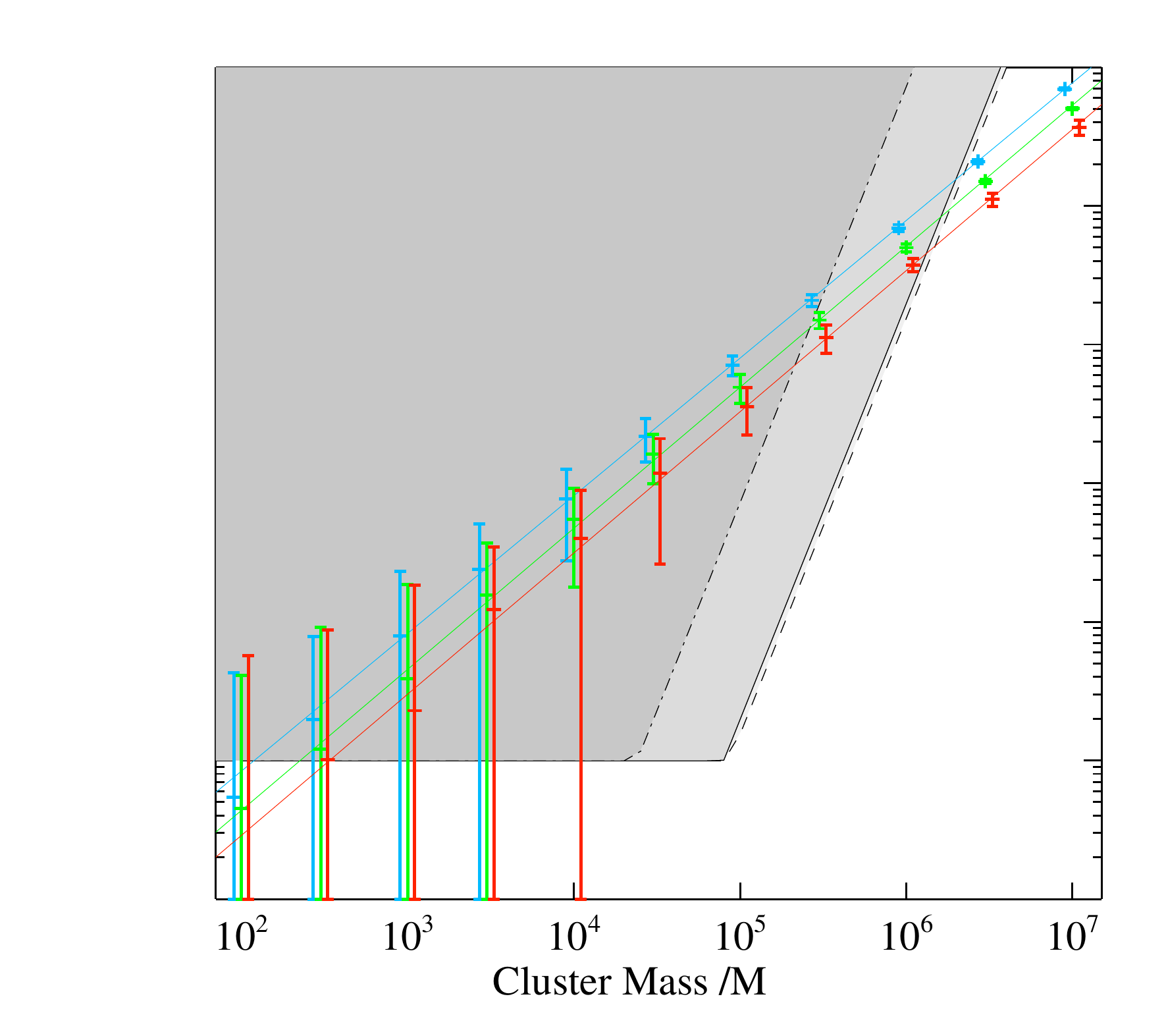}
      \caption{Average number of supernovae expected from the population by age 100 Myr, as a function of cluster mass, shown at three metallicities. The shaded region indicates energy input from SNe as discussed in Section \ref{sec:SN_clearance}.}
      \label{fig:mean_snrs}
  \end{figure}

In Figure \ref{fig:mean_snrs} we show how the total number of core-collapse supernovae in a population tracks the cluster mass of the stellar population. For each \mcl\ we indicate the mean number seen in realisations at that mass and the population standard deviation. As expected, the scatter on the supernova rate in the lowest mass clusters is considerable, with a significant number of \mcl$=10^4$\,M$_\odot$ clusters exhibiting no supernovae. The scatter decreases with increasing total mass in high mass clusters. The mean supernova rate closely tracks the total mass of stars formed, such that $N_\mathrm{SN} = N_0\,M_\mathrm{cl}^\alpha$,
where $N_0 = $ [2.1, 2.4, 2.5] and $\alpha = $ [0.99, 1.02, 1.02] at $Z=$0.002, 0.008 and 0.020 respectively. We tabulate the uncertainties on supernova rate in the Appendix.

It should be noted that low cluster masses do not necessarily preclude the occurrence of supernovae. While the mean number of SN in stochastic realisations of low mass clusters is consistent with expectations, $4.4\pm0.9$ per cent of 100\,M$_\odot$, and $17\pm2$ per cent of 300\,M$_\odot$  clusters at a metallicity of $Z=$0.002 experience one or more supernovae (see also Figure \ref{fig:fracsn}). 

\section{Discussion}\label{sec:disc}

\subsection{Stellar Population Uncertainties}\label{sec:uncertainties}

It comes as no surprise that stochastic sampling of the initial mass function and binary parameters can have a substantial impact on the properties of low stellar mass clusters. Indeed, the variation between realisations at \mcl$ = 100$\,M$_\odot$ and 300\,M$_\odot$ exceeds the variance in a given parameter with age or metallicity in most cases. Figure \ref{fig:ion_ind} in particular clearly indicates the role of the most massive stars in generating ionizing radiation, and also the dramatic impact of binary interactions in maintaining this at late times. Given the importance of massive stars, Figure \ref{fig:max_mass} suggests that the uncertainties calculated here may represent an upper limit. 

If clusters are indeed populated by top-down sampling, rather than purely stochastically, so that massive stars are more abundant than estimated in these populations, then low mass clusters would be expected to show less cluster-to-cluster variation than a purely stochastic model. The details of this will vary depending on the sampling prescription adopted, but this result follows from the key impact of massive stars in driving the stellar population properties. The analytic formalism of \citet{2007ApJ...671.1550P} indicates that no individual star in an \mcl=100\,M$_\odot$ cluster should exceed 9.1\,M$_\odot$ in mass. By contrast, 2 per cent of clusters in our sample populations have a star exceeding this mass, while almost 80 per cent have no single star exceeding 2.5\,M$_\odot$. As a result, the spectrum of the cluster (particularly in the ultraviolet) can be dominated by anything from a type O star to a cool F or G type star. While, in the latter case, a handful of such stars may form, the highly non-linear dependence of luminosity and temperature on mass means that the latter case will have much lower ionizing photon production rates. Requiring that each cluster first forms a mid B-star (removing a tenth of its total mass reservoir) and then fragments into smaller mass components according to an initial mass function ensures a more similar distribution in stellar masses than pure stochastic sampling, and ensures that every cluster's UV spectrum is dominated by a star with a similar temperature.

\begin{figure}
    \centering
    \includegraphics[width=\columnwidth]{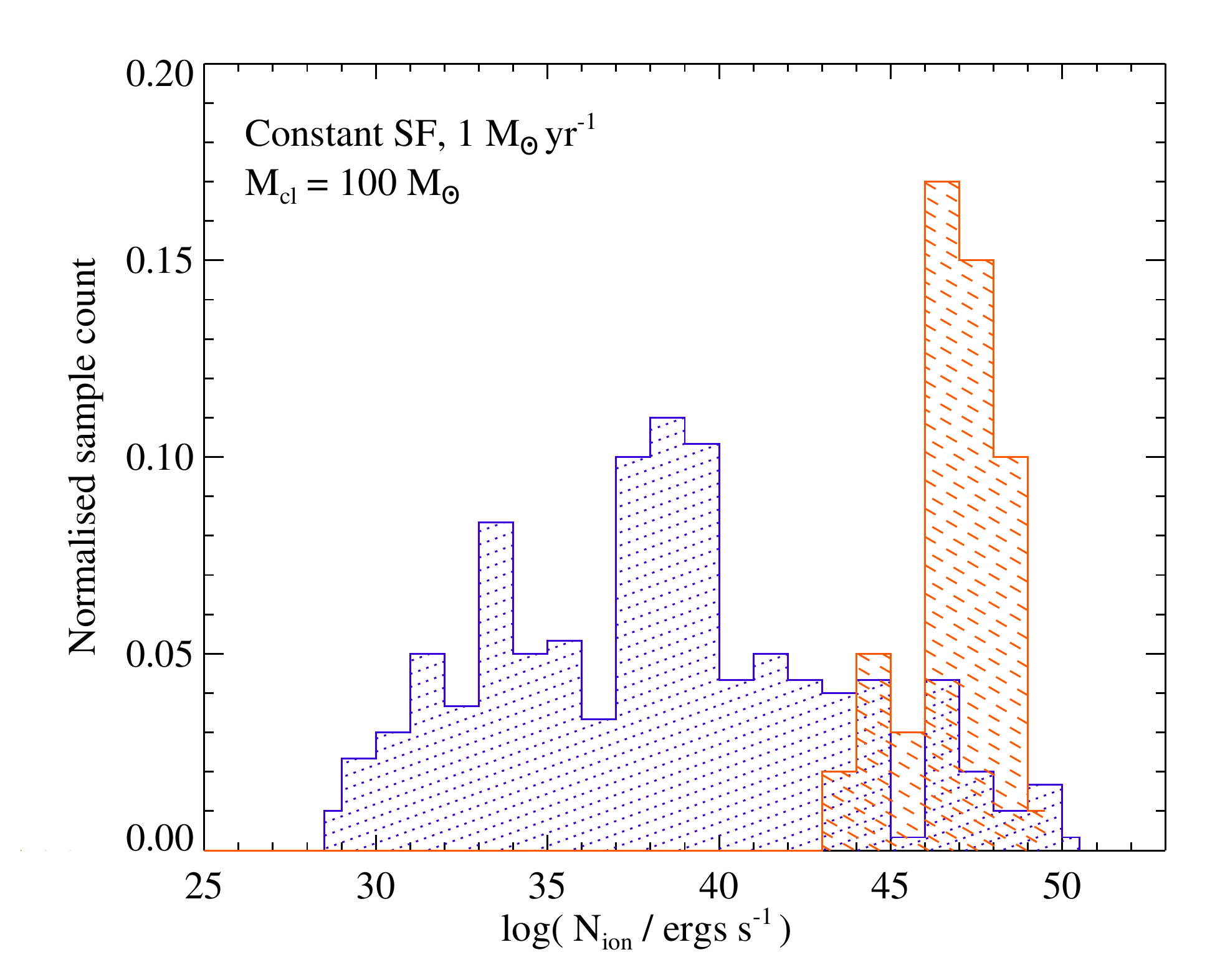}
    \caption{The range of ionizing photon production rates determined from stochastically sampled clusters with \mcl=100\,M$_\odot$ (blue, dotted region), compared to the narrower range measured in a small test grid which implements top-down sampling of the IMF (red, dashed region). See section \ref{sec:uncertainties} for details.
    \label{fig:topdown}}
\end{figure}

To demonstrate this, we compare the distribution of ionizing photon production rates, N$_\mathrm{ion}$, for our \mcl=100\,M$_\odot$ clusters with a small grid of \mcl=100\,M$_\odot$ models in which we have required the most massive star to be within 20 per cent of the expected mass proposed by \citet{2007ApJ...671.1550P}, i.e. in the range $\sim$7.3-10.9\,M$_\odot$. All remaining stars are sampled stochastically from an IMF stochastically sampled entirely below this mass, and we perform 100 iterations to explore the resulting variation. As Fig.~\ref{fig:topdown} demonstrates, the ionizing photon production rate, integrated over the first 100 Myrs falls into a far narrower range when top-down sampling is implemented.

However, in clusters of this mass the dynamical and evolutionary timescales are extremely short. The shallow gravity well of these systems permits stars to escape rapidly from their birth clusters, while the short dynamical timescales permit dynamical interactions to modify the stellar populations on timescales of a few Myrs \citep[see e.g.][]{2000ASPC..211...12G}. Such dynamical interactions are neglected in our models, but are likely responsible for the disruption of some stellar multiple systems, and the formation of others. This will modify the binary fraction, mass ratio distributions and initial period distributions in a manner not accounted for in BPASS. Such interactions provide a pathway for the formation of massive binaries (or higher multiples), and the mergers of such binaries (again on short timescales) has the potential to repopulate the upper end of a stellar mass function depopulated by stochastic sampling effects \citep{2002MNRAS.336..705B}. If the most massive star in these low mass clusters typically arises from such effects, rather than sampling of the IMF during stellar formation, then these will be underestimated by our model approach, and the impact of stochastic sampling somewhat mitigated. The effect of dynamical interactions will extend to higher mass clusters but will have the most profound impacts in these low mass samples. 

For populations of \mcl$ = 10^3-10^4$\,M$_\odot$, the effects of stochastic sampling are smaller, but still likely dominate the stellar population uncertainties, particularly for young or bursty star formation. In continuously star-forming populations the scatter on output properties is substantially lower. The effect of stochastic sampling is often to favour stars of a slightly higher or lower mass, and to force binary interactions to affect one or a few time bins. Constant star formation, by contributing stars with a range of ages, smooths over much of this mass-dependant event timing. For clusters with masses \mcl$ = 10^5$\,M$_\odot$ or higher, the mean properties of stochastically-sampled populations converge with the statistical values in all measurements considered here, except in the mass of the largest star, although the scale of variations between samples continues to depend on the cluster mass up to \mcl$ = 10^7$\,M$_\odot$, the highest mass population considered in this work.

Given that the canonical, statistically-sampled stellar population in BPASS is scaled to a total stellar mass of $10^6$\,M$_\odot$, it is informative to consider the magnitude of stochastic sampling uncertainties at this cluster mass, and compare these to other sources of uncertainty evaluated in previous work.

  \begin{figure*}
      \centering
      \includegraphics[width=1.6\columnwidth]{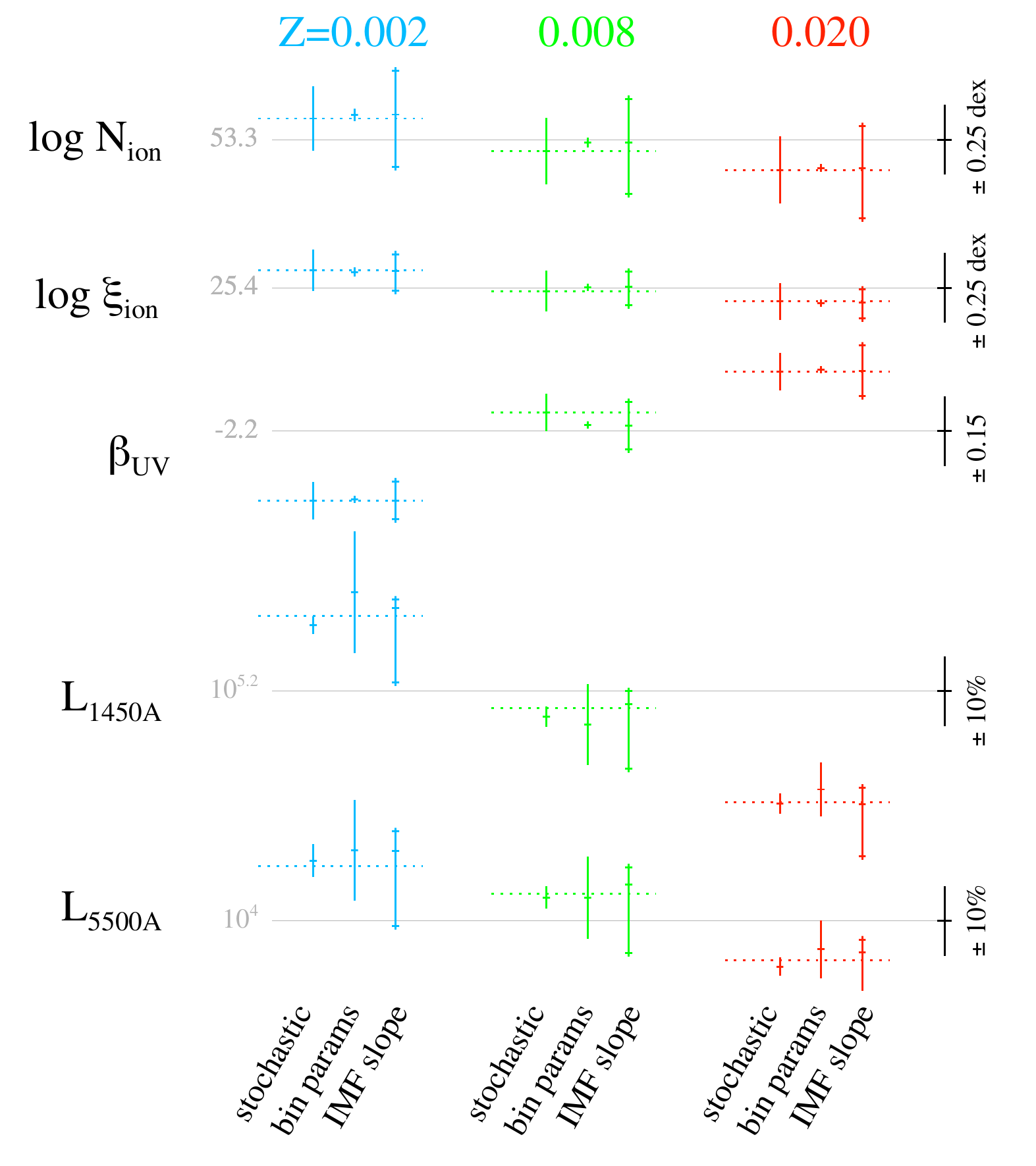}
      \caption{A visualisation of the relative scale of uncertainties on spectral output as a result of metallicity and stellar population uncertainties due to stochastic sampling (this work), initial binary fraction, mass ratio and separation uncertainties \citep{2020MNRAS.495.4605S} and initial mass function slope uncertainty \citep{2019A&A...621A.105S}. In each case the dotted horizontal line indicates the value in the statistically sampled BPASS v2.2.1 models. 
       See section \ref{sec:uncertainties} for details.}
      \label{fig:uncertainties}
  \end{figure*}

In figure \ref{fig:uncertainties} we visualise the variation in key outputs for a stellar population with ongoing constant star formation, as a result of stellar population parameter uncertainties. For each output, predictions are shown on a defined scale, such that relative position indicates the variation in that quantity as a result of variations in metallicity and population assumptions. Dotted horizontal lines indicate the values predicted at each metallicity in the statistically-sampled, fiducial BPASS v2.2.1 models \citep{2018MNRAS.479...75S}. The first data point associated with each line indicates the scatter which arises from stochastic sampling in a star formation event with \mcl$ = 10^6$\,M$_\odot$, with an indication of the absolute value on the far left. The middle data point on each line indicates the scatter arising from uncertainties in the initial distribution of binary parameters (binary fraction, mass ratio, separation) as evaluated by \citet{2020MNRAS.495.4605S}. That study calculated models at $Z=0.002$, 0.010 and 0.020; we supplement these with new models calculated at $Z=0.008$. The right hand estimate on each line indicates the scatter due to variation in the IMF upper mass slope $\alpha_u$ with predictions at -2.10, -2.35 and -2.60 indicated as connected points. This is actually a fairly conservative range in IMF slope compared to variation observed in the local Universe \citep{2018PASA...35...39H}. These uncertainties were explored by \citet{2019A&A...621A.105S}, and that work is supplemented here with new models at $Z=0.002$ and $0.008$. In each case offsets in measurements and uncertainty ranges are shown on a constant scale for any given measured quantity, allowing direct visual comparison between different variants and metallicities.

Of the quantities explored, N$_\mathrm{ion}$ and $\xi_\mathrm{ion}$ are most sensitive to the most massive stars in the stellar population. For these stars the binary fraction and initial parameter distributions are both very high and comparatively well-constrained. The remaining uncertainties in binary parameters act to slightly delay or advance the onset of binary interactions, and thus have little effect on a continuously star-forming stellar population, to which stars of all ages contribute. As a result, in these quantities the uncertainties arising from stochastic sampling dominate over those from binary parameter distributions, even for \mcl$ = 10^6$\,M$_\odot$. The presence or absence of very massive stars ($M>100$\,M$_\odot$) has a dramatic effect on the ionizing photon output. The scatter due to stochastic sampling is comparable to the effects of a variation in the initial mass function upper slope (i.e. that applied between 0.5 and 300\,M$_\odot$) from -2.1 to -2.6. We caution that the majority of IMF slope estimates published to date have been derived by comparison of data to stellar population synthesis models that neglect the effect of binary interactions  \citep{2018PASA...35...39H}. This analysis suggests that stochastic population of the IMF and binary parameters may also complicate such estimates.

The far-ultraviolet spectral slope, $\beta_\mathrm{UV}$, is a probe of the luminosity-weighted temperature of the dominant stellar population. It is thus sensitive to a slightly larger range of stellar masses than the ionizing flux, but is less sensitive to the presence of large quantities of fainter, lower mass stars than the integrated population luminosity at fixed mass. As Figure \ref{fig:uncertainties} demonstrates, uncertainties in $\beta_\mathrm{UV}$ follow a similar trend to those seen in the ionizing photon flux. By contrast, the continuum luminosity at both ultraviolet and optical wavelengths is only weakly sensitive to stochastic sampling in an \mcl$ = 10^6$\,M$_\odot$ stellar population. Both the mass and the continuum normalisation are dominated by the low mass stellar population. In these quantities, uncertainty arising from binary parameter distribution sampling dominates, although varying the slope of the IMF (again affecting the number of low mass stars and thus the luminosity normalisation at fixed mass) has a comparable impact.

Given their importance in interpretation of galaxies in the distant Universe, it should be noted that realisations of both the ionizing photon production rate and its efficiency continue to show scatter exceeding 0.2\,dex even in galaxy-wide starbursts forming $10^6$\,M$_\odot$ of stars. Given that very few, if any, single star formation episodes exceed this mass, this is likely to be an unavoidable source of uncertainty in the interpretation of young stellar populations.

\subsection{Supernovae and Birth Cloud Clearance}\label{sec:SN_clearance}

A key point in cosmic history at which these uncertainties become critical to our understanding is during the epoch of reionization. At this epoch the total stellar mass of a typical galaxy was $<10^{10}$\,M$_\odot$ \citep{2021ApJ...922...29S} and the majority of galaxies were likely undergoing or had just undergone a star formation event. These faint, low mass galaxies are believed to be the primary source of ionizing photons powering the reionization process. However a key question that has arisen in modelling this epoch is the fraction of such photons that are able to escape their original galaxies in order to reach and ionize the intergalactic medium. While this escape fraction appears to be small when considered for typical sightlines, it can approach unity if a channel through the interstellar medium has been cleared by supernovae or other feedback processes.

As a result, it is informative to explore the impact of star forming cluster mass on the possibility that mechanical feedback from supernovae is sufficient to clear gas from the stellar birth cloud. To do so, we need to consider two main parameters: the amount of kinetic energy injected by each supernova to its surroundings and the strength of the gravitational potential that must be overcome. 

Previous theoretical work has shown that each supernova in a population can be modelled as injecting a mean amount of momentum into the surrounding gas.  This will, of course, depend on a large number of parameters, including the density and opacity of that gas and the efficiency with which electromagnetic energy is converted to mechanical energy through shocks and other processes. In the interests of making a simple approximation, we assume that each supernova that occurs injects a total momentum of $p_\mathrm{SN}=2.7\times10^4$\,M$_\odot$\,km\,s$^{-1}$ integrated over its lifetime \citep{2015ApJ...802...99K}. 

We assume that this momentum is distributed across the residual gas in a cluster after the formation of the stellar population, in the form of kinetic energy. Given these assumptions, we calculate the critical number of supernovae such that the resulting kinetic energy in the cloud exceeds the gravitational potential energy required to unbind the cloud: 

\begin{equation}
\indent N_\mathrm{SN}\,\left(\frac{p_\mathrm{SN}^2}{2 M_g}\right) > \frac{3\,G}{5\,R}\left(M_\mathrm{total}^2 - M_\mathrm{cl}^2\right),
\end{equation}

where $M_g$ is the cluster gas mass, $M_\mathrm{cl}$ is the cluster stellar mass, $M_\mathrm{total} = M_g + M_\mathrm{cl}$ and $R$ is the physical size of the cluster, which we take as 100\,pc. We note that this is an overestimate for most of our clusters. The typical size of individual star forming clusters in local galaxies is typical of order $\sim3-10$\,pc \citep{2017ApJ...841...92R}. However such clusters often form embedded within much larger giant molecular clouds and star formation complexes, with extreme examples such as 30 Doradus in the Large Magellanic Cloud which has a clumpy structure extending over a kiloparsec \citep{1993ASPC...48..588M} or NGC\,604 which has a scale of order 100\,pc, while many other regions have scales of tens of parsecs \citep[see][]{2010ARA&A..48..431P}. Hence we err on the side of a conservative estimate and require the gas to be cleared to well beyond the core radius of a typical cluster.

The gas mass appropriate for each stellar cluster is determined by a star formation efficiency parameter $\epsilon$, i.e. the fraction of total original mass of the cluster which has been turned into stars before the gas is cleared, such that $M_\mathrm{cl} = \epsilon M_\mathrm{total}$ and $M_g = (1-\epsilon)\,M_\mathrm{total}$. We do not include the effects of radiation pressure feedback or energy injection from stellar winds at this point. Earlier studies have shown that over the evolution of a starburst, supernova feedback dominates the energy injection, particularly at low metallicity, although stellar wind feedback likely dominates in the first 10\,Myr  \citep[e.g.][]{1992ApJ...401..596L,2005ARA&A..43..769V}.

For an initial momentum injection $p_\mathrm{SN}$, the critical number of supernovae required to clear a natal gas cloud is thus given by,
\begin{equation}
\indent N_\mathrm{SN,crit} = \frac{3\,G\,M_\mathrm{cl}^2}{5\,R}\left(\frac{1}{\epsilon^2}-1\right)\,\frac{2\,M_\mathrm{cl}}{p_\mathrm{SN}^2}\,\left(\frac{1-\epsilon}{\epsilon}\right)
\end{equation}

The dependence of critical supernova number on cluster mass and star formation efficiency is shown in Figure \ref{fig:mean_snrs}, in which shaded regions exceed this number. As the figure demonstrates, star forming clusters with stellar masses between $10^4$ and $10^6$\,M$_\odot$ are likely to generate sufficient supernovae to clear their natal clouds of gas through mechanical feedback, given reasonable assumptions for star formation efficiency. At higher cluster masses, the cluster's self-gravity is too high for supernova feedback alone to overcome. At very low masses the cluster self-gravity is low, but at least one supernova must occur before the gas cloud is dispersed. 

\begin{figure}
      \centering
      \includegraphics[width=\columnwidth]{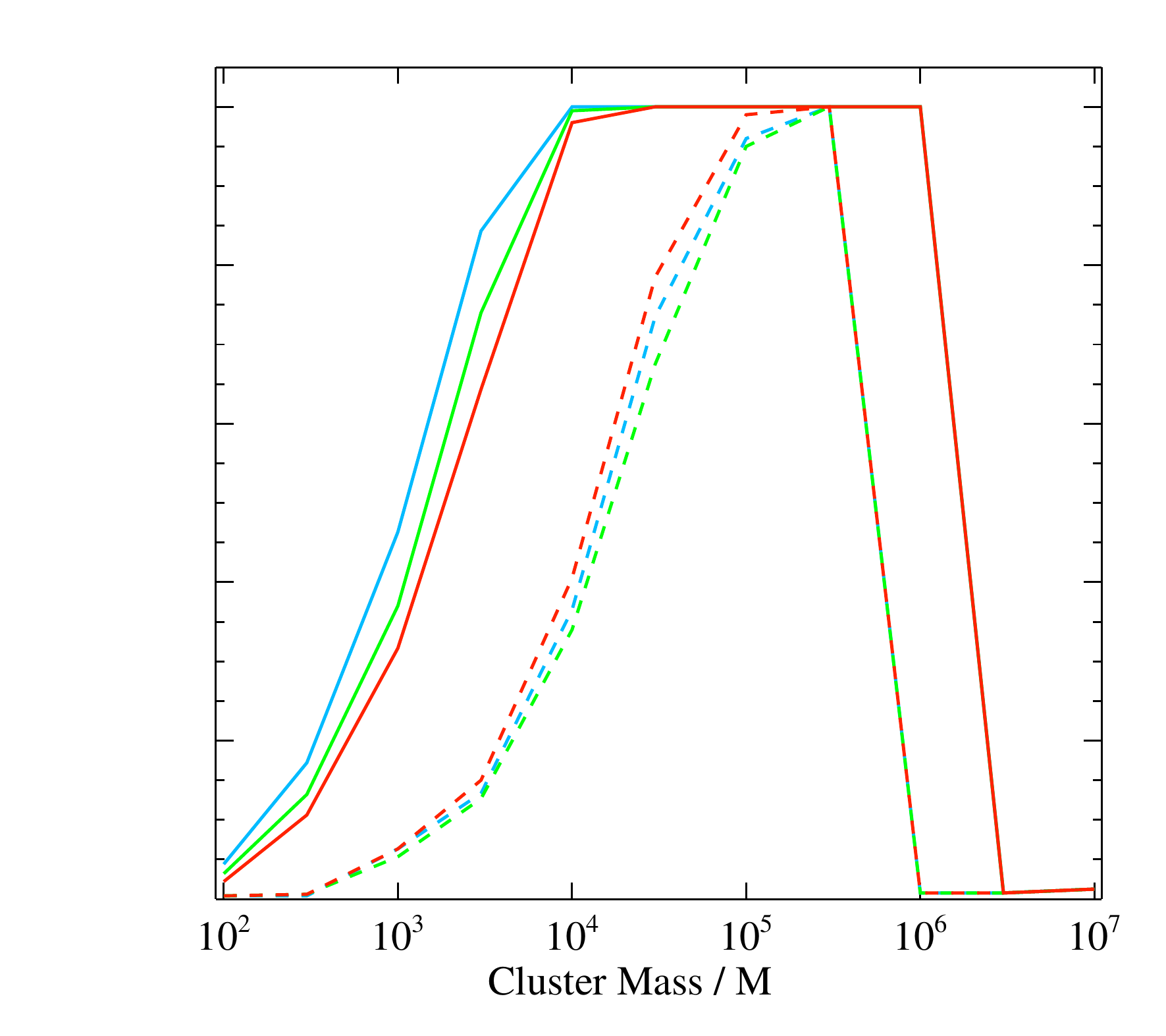}
      \caption{The fraction of clusters which inject enough momentum from supernovae to clear their natal gas clouds within 10 Myr (dashed lines) and 100 Myr (solid lines), assuming a star formation efficiency of 0.5.}
      \label{fig:sn_clearance}
  \end{figure}

However of equal interest is the timescale on which such clearances can occur. In order to contribute significantly to reionization, gas must be cleared from the interstellar medium before the ionizing photon production rate of the cluster drops significantly. Binary stellar populations have an advantage here over populations constructed using only single star evolution, since they remain hot, blue and ionizing for longer \citep{2016MNRAS.456..485S}.  In Figure \ref{fig:sn_clearance}, we explore the effect of stochastic sampling on this issue by calculating the fraction of star forming clusters capable of clearing their natal clouds within two defined intervals: 10\,Myr (appropriate for the hot single star evolution) and 100\,Myr (allowing for rejuvenated and stripped binary products). 

This Figure demonstrates a weak metallicity effect and a strong effect with age and cluster mass. 
At low masses, only a small fraction of clusters can clear their immediate environs of gas within 10\,Myr, with around 60\,per\,cent of \mcl$=3\times10^4$\,M$_\odot$ clusters able to do so.  By contrast,  
the clearing of higher mass clusters becomes more efficient, with virtually all realisations at \mcl$=1-3\times10^5$\,M$_\odot$ exceeding the critical number of supernovae.
If we permit longer timescales,
more than 90\,per\,cent of clusters with $10^4<M_\mathrm{cl}<3\times10^5$\,M$_\odot$ exceed the threshold number of supernovae within 100\,Myr at $Z=0.020$, with slightly lower mass clusters cleared more efficiently at lower metallicities. On these longer timescales, a significantly non-zero fraction of low mass clusters exceed the threshold supernova number. About 25-40\,per\,cent of 1000\,M$_\odot$ clusters have sufficient supernova kinetic energy injection to eject the gas from their gravitational self-potential, despite their low mean number of supernovae.

Note that we have focussed here on the properties of the Type Ib, Ic and II supernovae that dominate the number statistics of explosive transients in young stellar populations. However our models consider initial stellar masses extending to 300\,M$_\odot$. At solar and near-solar metallicities, a combination of strong stellar winds and binary stripping substantially reduce the mass of such a star to of order 10\,M$_\odot$ before explosion. However at lower metallicities, some very massive stars may retain a substantial fraction of their initial mass, resulting in more exotic transients such as pair-instability and pulsational pair-instability supernovae \citep[PISNe and PPISNe, see e.g.][]{2003ApJ...591..288H}. These have been explored using BPASS in the context of binary black hole formation by \citet{2023MNRAS.520.5724B}. However they are also relevant in this context since, while these events are rare, they are very luminous and any that occur will inject substantial amounts of energy and momentum into their surroundings. Since they arise from the most massive stars, their rates are highly subject to stochastic sampling uncertainties. Following \citet{2003ApJ...591..288H}  and \citet{2023MNRAS.520.5724B}, we consider the end state of each star in our \mcl=$10^6$\,M$_\odot$ populations and flag it as likely to undergo a PISN if its helium core mass at collapse is in the range 64-133\,M$_\odot$ and a PPISN if its carbon-oxygen core is in the range 38-60\,M$_\odot$. No stars met this condition in our $Z=0.020$ models. At $Z=0.002$, the two conditions are met by a small subset of models with initial masses greater than 100 and 50\,M$_\odot$ respectively. The PISNe require very massive stars with little or no binary interaction - unusual since most massive stars are born in close binaries \citep{2013A&A...550A.107S}. The PPISNe are found in post-merger systems, again a subset of the already-rare massive stars \citet{2023MNRAS.520.5724B}. As Fig \ref{fig:max_mass} demonstrated, these initial masses are not populated for clusters with \mcl$<10^3$\,M$_\odot$, and are well-populated only above \mcl$=10^6$\,M$_\odot$. Thus (P)PISNe are expected to be a significant source of cluster-to-cluster variability in supernova energy injection for clusters between these mass limits.

The prescription for supernova feedback adopted in the analysis here is, of course, a toy model. Star forming clusters do not form in isolation, but will be embedded within a much deeper gravitational potential associated with galaxy-scale dark matter haloes. They will be affected by mechanical feedback from supernovae in neighbouring clusters, as well as their own stellar population. We have also neglected the effect of mechanical feedback from stellar winds, and the physics of shocks and gas-dust interactions in a clumpy interstellar medium. Full hydrodynamic simulations are needed to follow these issues in detail \citep[e.g.][]{2021MNRAS.506.2199G}. Often even these must resort to subgrid prescriptions for supernova rates and energy injection \citep[e.g.][]{2009ApJ...704..137J,2016MNRAS.459.3460K}, and these prescriptions are based on insights from single star evolution models. However we note that recent simulations agree with our conclusion that massive giant molecular clouds (\mcl$ \sim 10^6$\,M$_\odot$) are unlikely to be completely disrupted by supernova feedback alone, but that regions on 10\,pc scales may well be evacuated by supernovae \citep{2016MNRAS.459.3460K}. 

An increasing number of simulations are also beginning to consider the impacts of stochastic sampling. \citet{2020MNRAS.492....8A}, for instance, considered the impact of discrete star sampling on the supernova rate in their dwarf galaxy simulations, showing that this makes feedback burstier, leading to a enhanced cooling and a boost in the star formation rate relative to the statistical case in galactic haloes as large as $10^{8.5}$\,M$_\odot$.
\citet{2021MNRAS.502.5417S} also looked at the impact of this approach in dwarf galaxy simulations and suggested that the impact on photoionization was more significant than the impact on supernova feedback.  They suggest that stochastic sampling is necessary when \mcl$ < 500$\,M$_\odot$, a rather weaker constraint than that suggested by our models in which the ionizing photon production rate begins to differ significantly from that of a statistically sampled population at between \mcl$ = 10^4$ and $10^5$\,M$_\odot$, and the uncertainty on the supernova rate rises sharply in the same mass range.

Despite the simplicity of our approach, clearing gas from the circumstellar environment is a necessary first step for creating channels through the ISM, and so this analysis is indicative of the scale of the impact of stochastic parameter sampling (or equivalently expected astrophysical cluster-to-cluster variation) on the potential sources of mechanical feedback. It demonstrates that while the occurrence star formation in very low mass clusters may restrict the effectiveness of supernova feedback in facilitating the escape of ionizing photons from their birth cloud, it will not eliminate it entirely.

\section{Conclusions}\label{sec:conc}

In this paper we have investigated the impacts of simultaneous stochastic sampling of the stellar initial mass function and of stellar binary mass ratio and period distributions on the properties of stellar populations. Our principle conclusions are as follows: 

\begin{enumerate}
    \item The most massive star expected in a stellar cluster, and its binary fraction, are both highly sensitive to the mass of the cluster being modelled, and show large scatter for low mass clusters.
    
    \item For cluster masses \mcl$<10^6$\,M$_\odot$, statistical sampling overpredicts the ionizing photon production rate and production efficiency, compared to typical clusters formed by stochastic sampling. Stellar clusters above this mass are well described by stochastic sampling of the IMF and binary parameters, but continue to show substantial variation between stochastic realisation. The uncertainties for an \mcl$=10^6$\,M$_\odot$ constantly star-forming population are $\sigma$log(N$_\mathrm{ion}$)=[0.20, 0.38, 0.53]\,dex and
    $\sigma$log($\xi_\mathrm{ion}$) =[0.16, 0.30, 0.1]\,dex at metallicities $Z=$[0.002, 0.008, 0.020] respectively.
    
    \item For populations with cluster masses \mcl$<10^6$\,M$_\odot$, the random variation in ultraviolet flux and optical colours due to stochastic randomness in the stellar population exceeds uncertainties due to stellar population age or metallicity in most cases. However in constantly star forming populations with \mcl$>10^6$\,M$_\odot$, the variation in ultraviolet luminosity or optical colour are reduced to just a few per cent. Uncertainty in the spectrum due to stochastic sampling is largest at ages of 10-100\,Myr, corresponding to the evolutionary timescales of primary stars with masses in the range 5-20\,M$_\odot$.
    
    \item The number of supernovae expected in a stellar population scales strongly with \mcl. Stochastic sampling results in supernovae occurring in lower mass stellar clusters than might be expected; $4.4\pm0.9$ per cent of 100\,M$_\odot$, and $17\pm2$ per cent of 300\,M$_\odot$  clusters at a metallicity of $Z=$0.002 experience one or more supernovae.
    
    \item Supernovae in low mass clusters could be important for clearing channels for ionizing photon escape during the epoch of reionization. We evaluate the fraction of cluster realisations which generate sufficient mechanical feedback through supernovae to clear their natal clouds within 10 (100) Myr of the onset of star formation. This fraction exceeds 80 per cent for clusters in the range $3\times10^4-3\times10^5$ ($3\times10^3-10^6$) M$_\odot$.
    
    \item For properties dominated by the most massive stars (e.g. $N_\mathrm{ion}$, $\xi_\mathrm{ion}$, $\beta_\mathrm{UV}$), the uncertainties on the output of a $10^6$\,M$_\odot$ stellar cluster are more likely to be dominated by stochastic sampling than by uncertainties in the initial binary parameter distribution, and produce a comparable uncertainty to a change in the initial stellar mass function slope by $\pm0.3$. For properties sensitive to less massive stars (e.g. the UV-continuum or optical luminosity), binary distribution parameter uncertainties dominate over the effects of stochastic sampling, and again are comparable to the effects of changing the initial mass function slope by $\pm0.3$. 
    
\end{enumerate}

The impact of stochastic sampling of the initial mass function on the physical properties of massive star-dominated populations thus exceeds other sources of uncertainty. It highlights the danger of over-interpreting properties such as the ionizing photon production rate in samples of one or a few extreme galaxies, which may not be well-reproduced by statistically-sampled stellar population synthesis models, or by any single realisation of a stochastically sampled cluster.

\section*{Acknowledgements}

The authors thank past and present BPASS team members for fruitful collaboration over many years. JJE acknowledges support by the University of Auckland and funding from the Royal Society Te Apar\={a}ngi of New Zealand Marsden Grant Scheme. ERS acknowledges support by the University of Warwick and funding from the UK Science and Technology Facilities Council (STFC) through Consolidated Grant ST/T000406/1.  We thank the anonymous reviewer for their constructive input.

\section*{Data Availability}

The data required to reproduce figures in this paper will be made available on the BPASS websites at www.warwick.ac.uk/bpass and www.bpass.auckland.ac.uk. The underlying population and spectral synthesis model realisations represent a significant data volume but will be made available on reasonable request to the first author.





\bsp	
\label{lastpage}
\appendix
\clearpage
\setcounter{page}{0}
    \pagenumbering{roman}
    \setcounter{page}{1}

\section{Supplementary and Numerical Results}

\bigskip

In this Appendix we provide supplementary figures and numerical values for the uncertainties in key parameters as a function of cluster mass. Where appropriate these have been scaled to the same total stellar mass for ease of comparison.

\bigskip

In Table \ref{tab:unc-nionconst} we give the standard deviation for our stochastic realisations at each \mcl\ in log($N_\mathrm{ion}$) and in the ionizing photon production rate log($\xi_\mathrm{ion}$). Table \ref{tab:unc-uvsope} indicates the uncertainties on the ultraviolet spectral slope, measured between 1566 and 2266\,\AA, with values at each metallicity split over two lines for clarity. 

\bigskip

In Table \ref{tab:unc-spectra} we give the fractional uncertainties on the continuum flux, $\sigma F/F$ in two wavelength ranges, 1250-1650\AA\ in the ultraviolet and 5000-6000\AA\ in the ultraviolet. For comparison, we also show the same values calculated from a set of models generated through statistical sampling of a range of plausible binary parameters, as discussed in \citet{2020MNRAS.495.4605S}.

\bigskip

In Table \ref{tab:unc-SNe} we give the mean supernova rate as a function of cluster mass, and again present the scatter in this value expressed as the standard deviation of values in stochastic realisations.

\bigskip

\begin{figure*}
      \centering
      \includegraphics[width=0.32\textwidth]{Figures/ionizing_comp_z020_1e2.pdf}
      \includegraphics[width=0.32\textwidth]{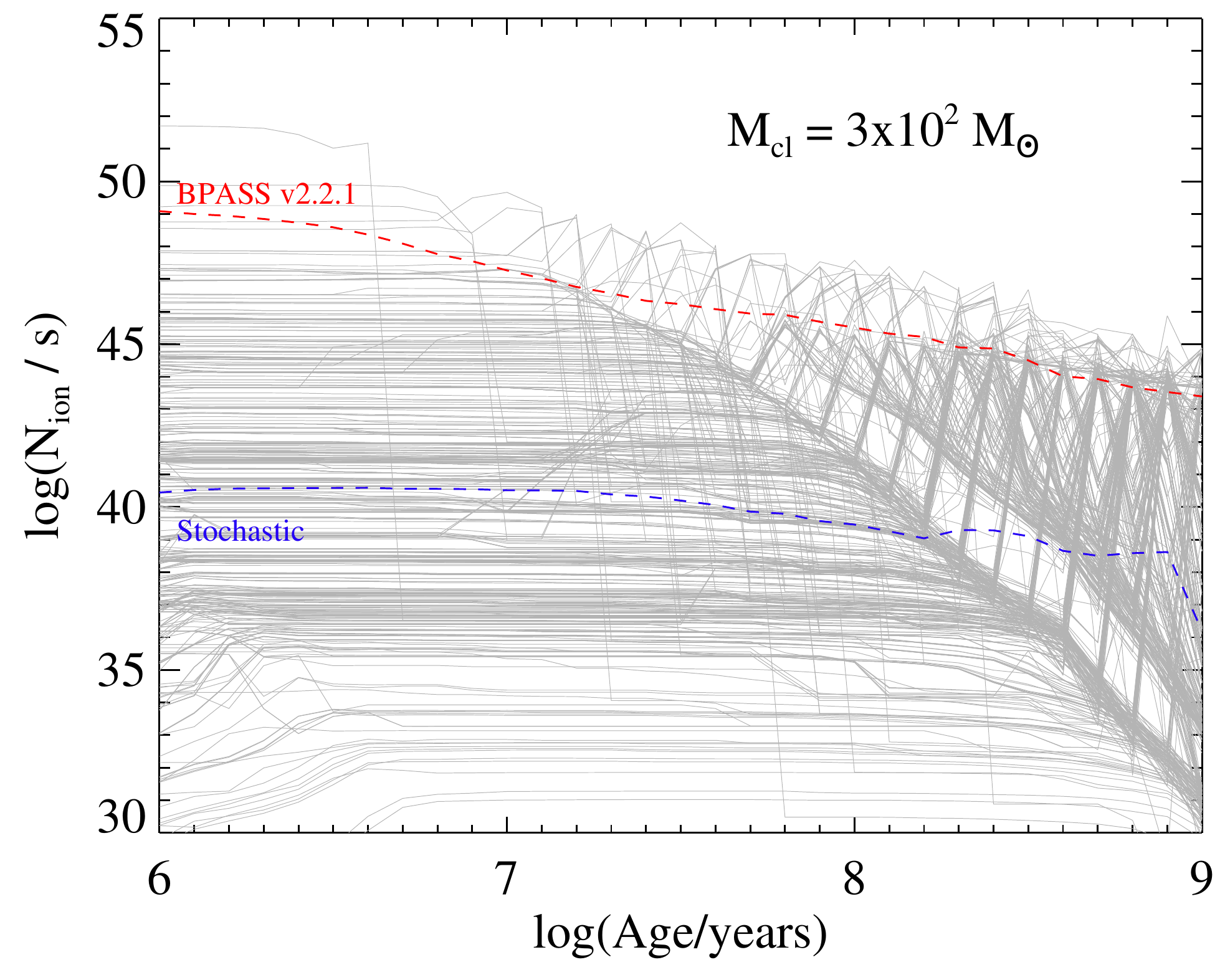}
      \includegraphics[width=0.32\textwidth]{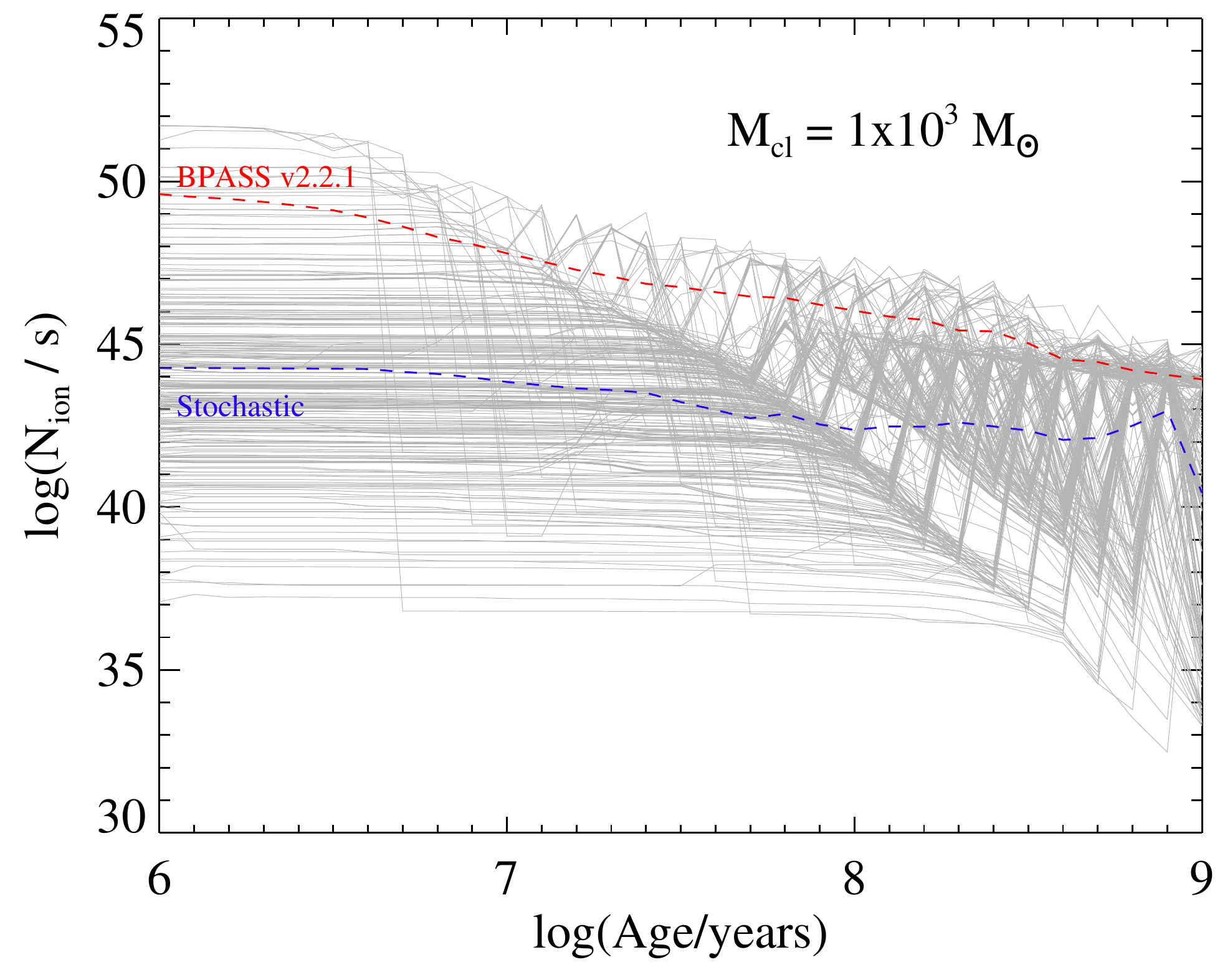}
      \includegraphics[width=0.32\textwidth]{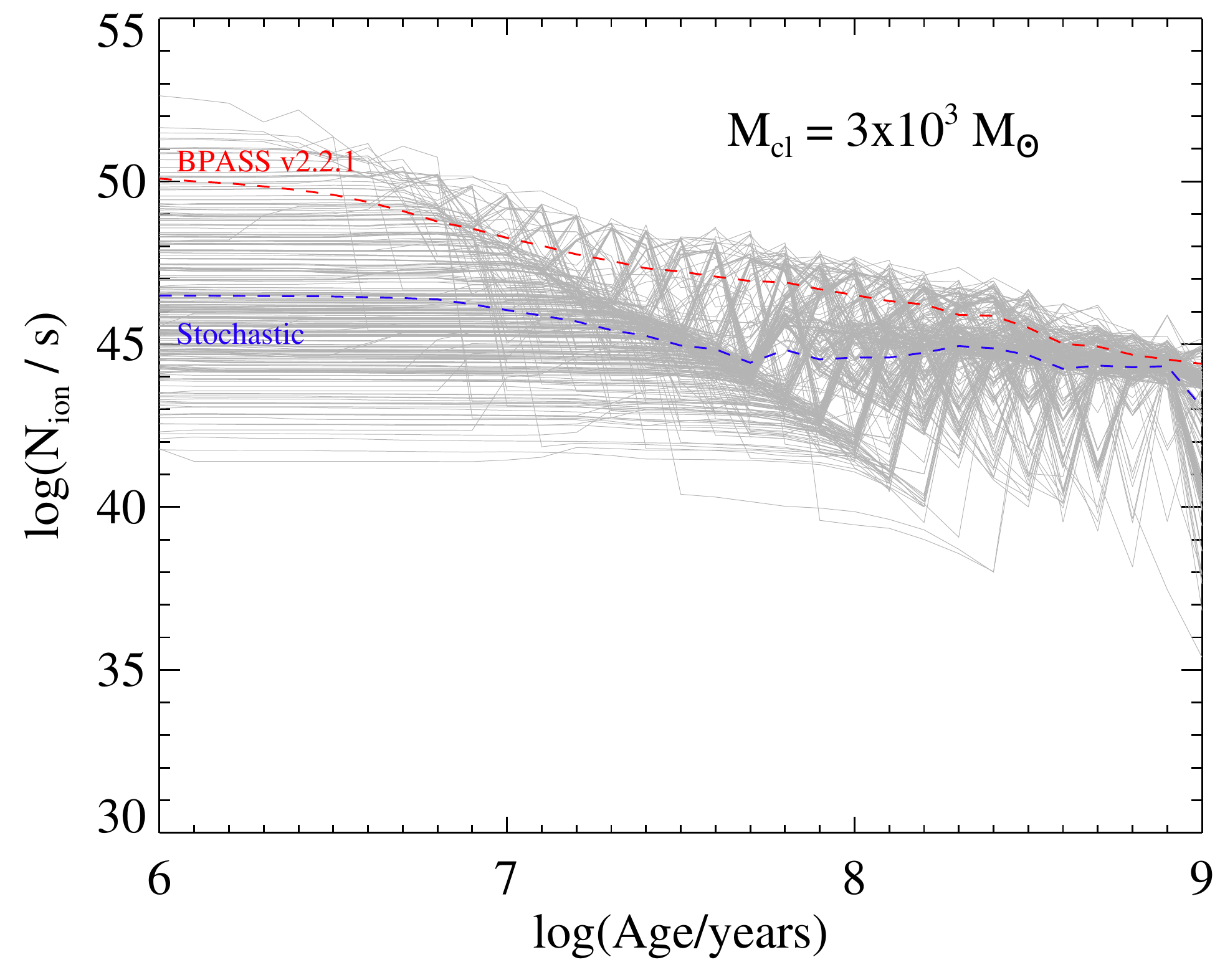}
      \includegraphics[width=0.32\textwidth]{Figures/ionizing_comp_z020_1e4.pdf}
      \includegraphics[width=0.32\textwidth]{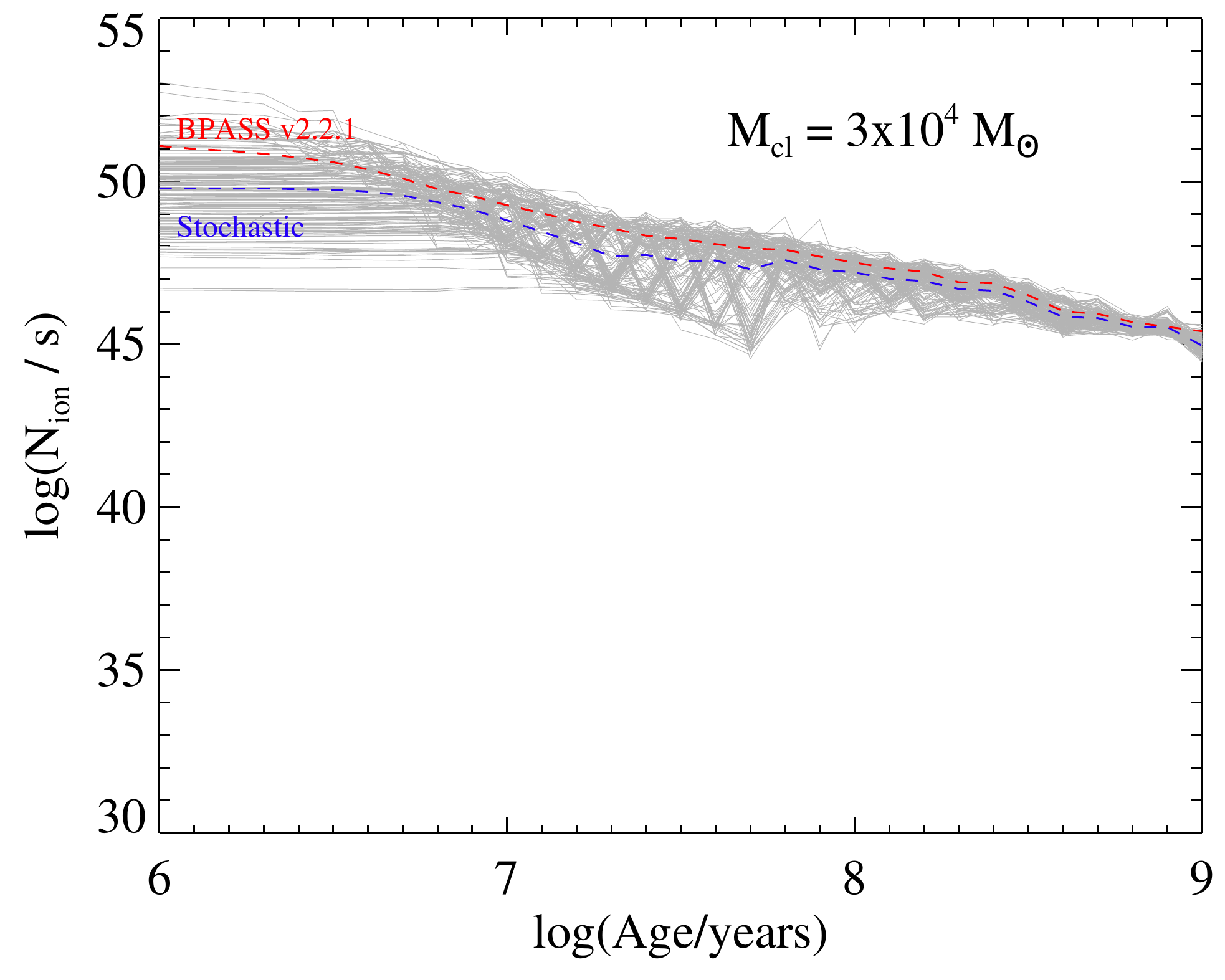}
      \includegraphics[width=0.32\textwidth]{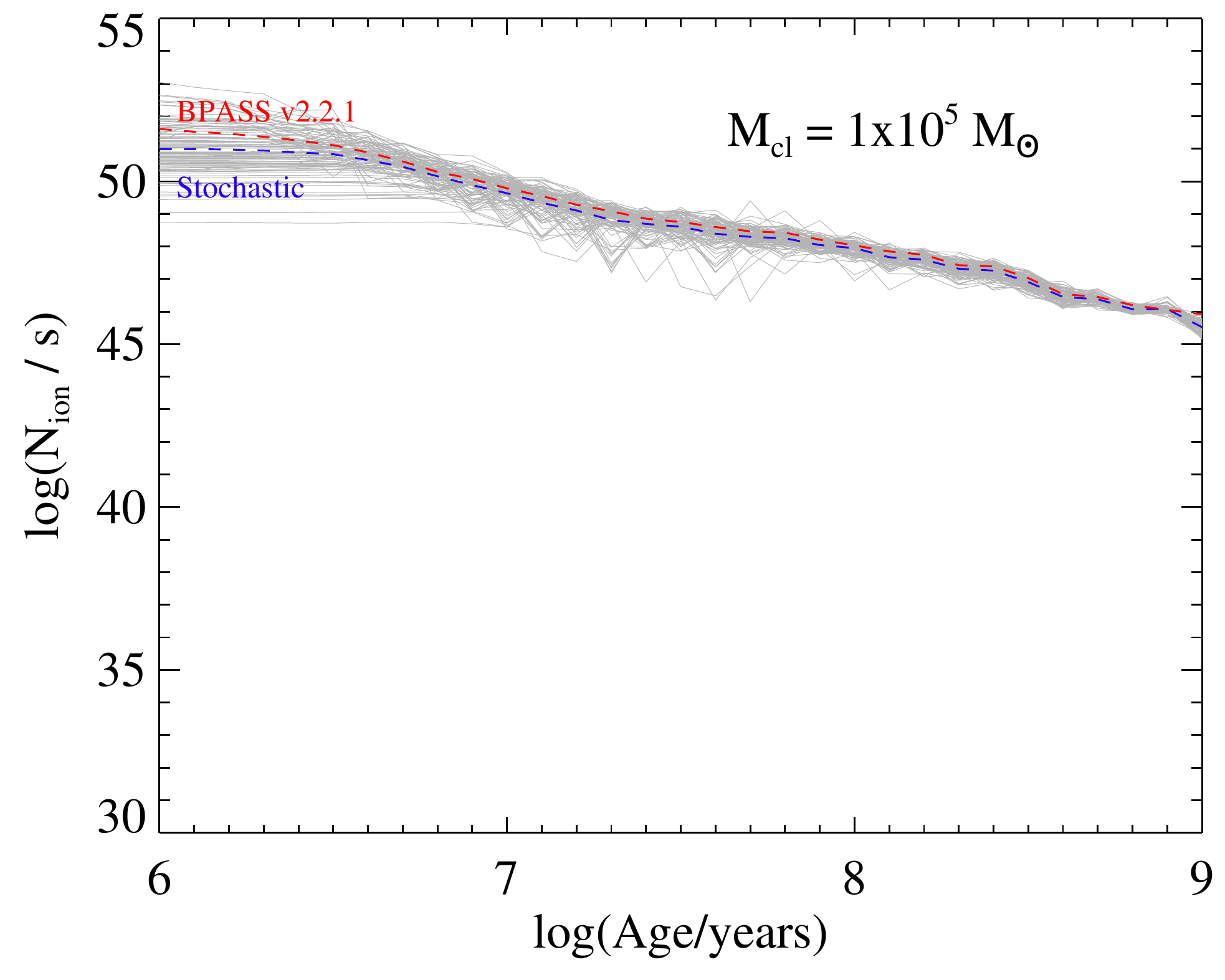}
      \includegraphics[width=0.32\textwidth]{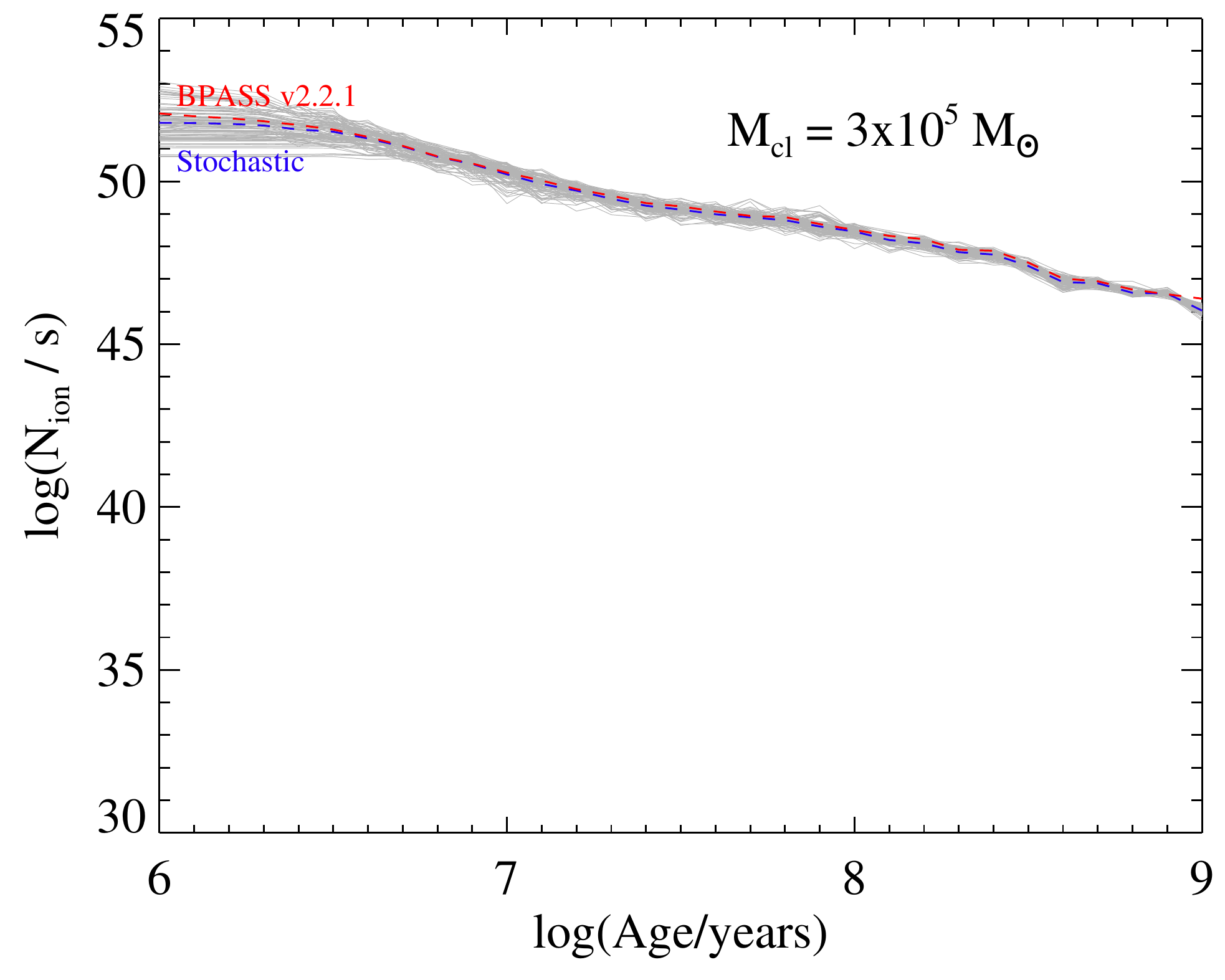}
      \includegraphics[width=0.32\textwidth]{Figures/ionizing_comp_z020_1e6.pdf}
      \includegraphics[width=0.32\textwidth]{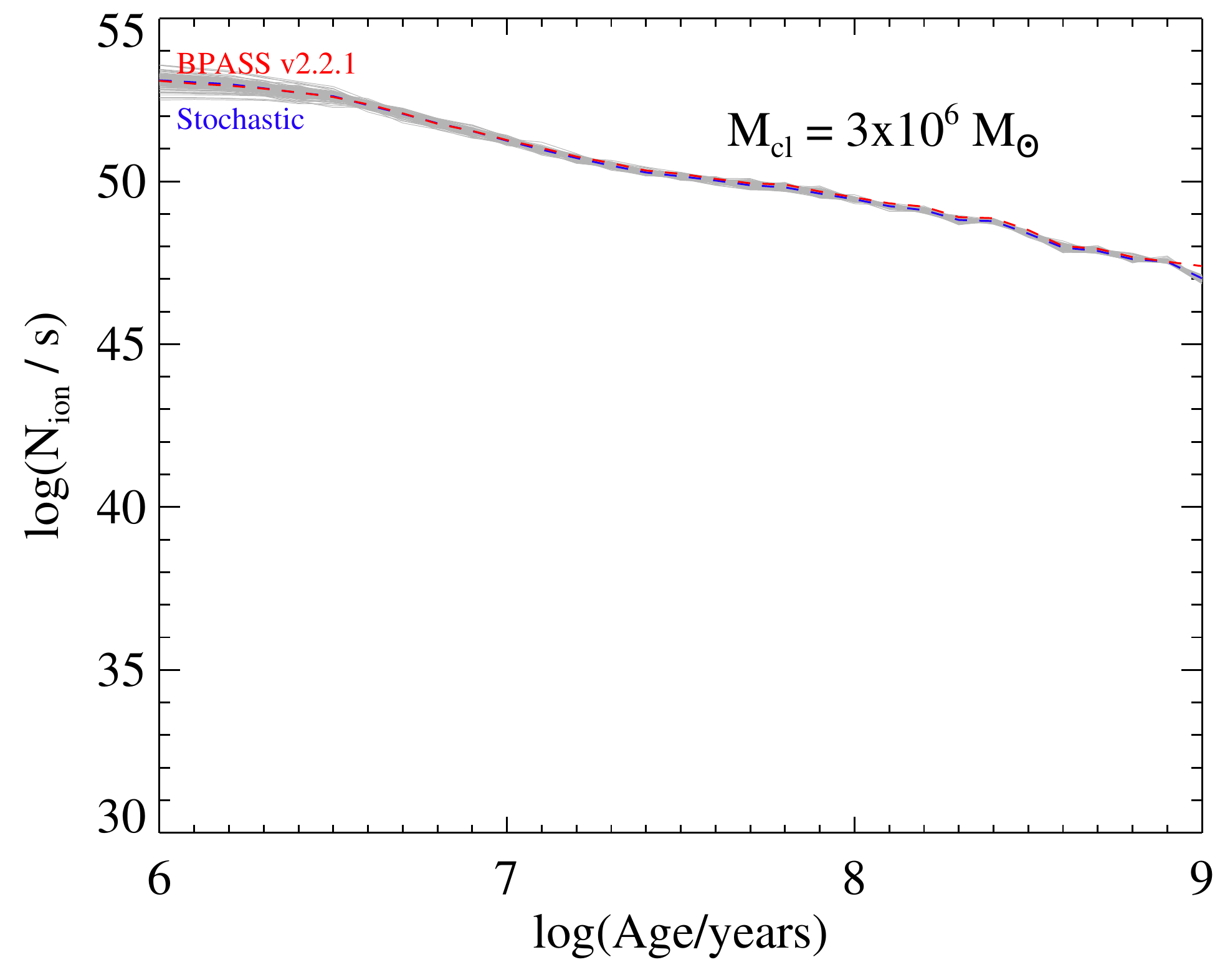}
      \includegraphics[width=0.32\textwidth]{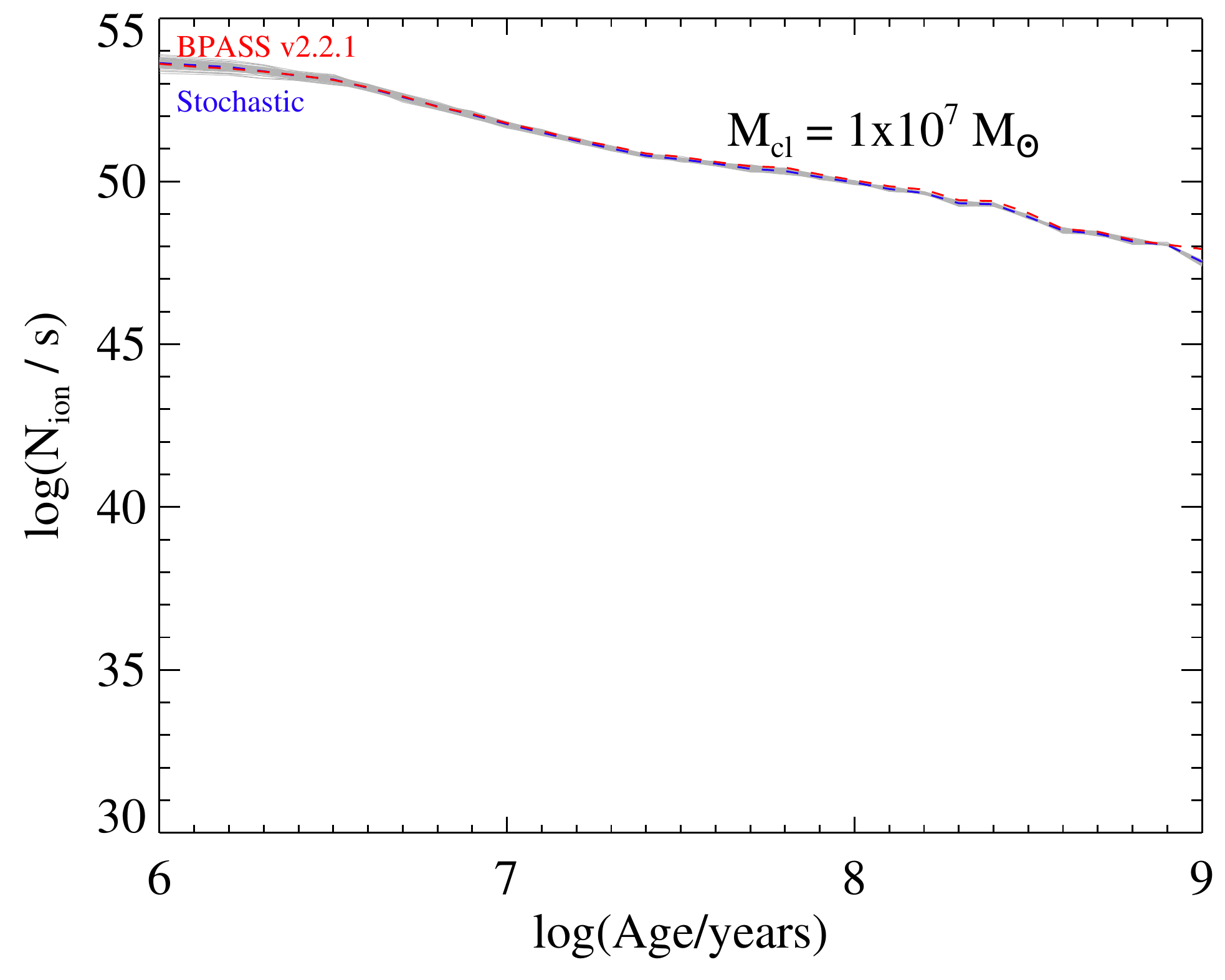}
      
      \caption{As in Figure \ref{fig:ion_ind}, but now for all values of \mcl.}
      \label{fig:appendix_ind}
  \end{figure*}

Figure \ref{fig:appendix_ind} shows the variation between different stochastic realisations of ionizing photon production rate, and its dependence on age of the stellar population. This reproduces Figure \ref{fig:ion_ind}, but shows the scatter for all \mcl\ values considered in this work.

\bigskip

\begin{figure*}
      \centering
      \includegraphics[width=0.32\textwidth]{Figures/plot_bfracm_1e2.pdf}
      \includegraphics[width=0.32\textwidth]{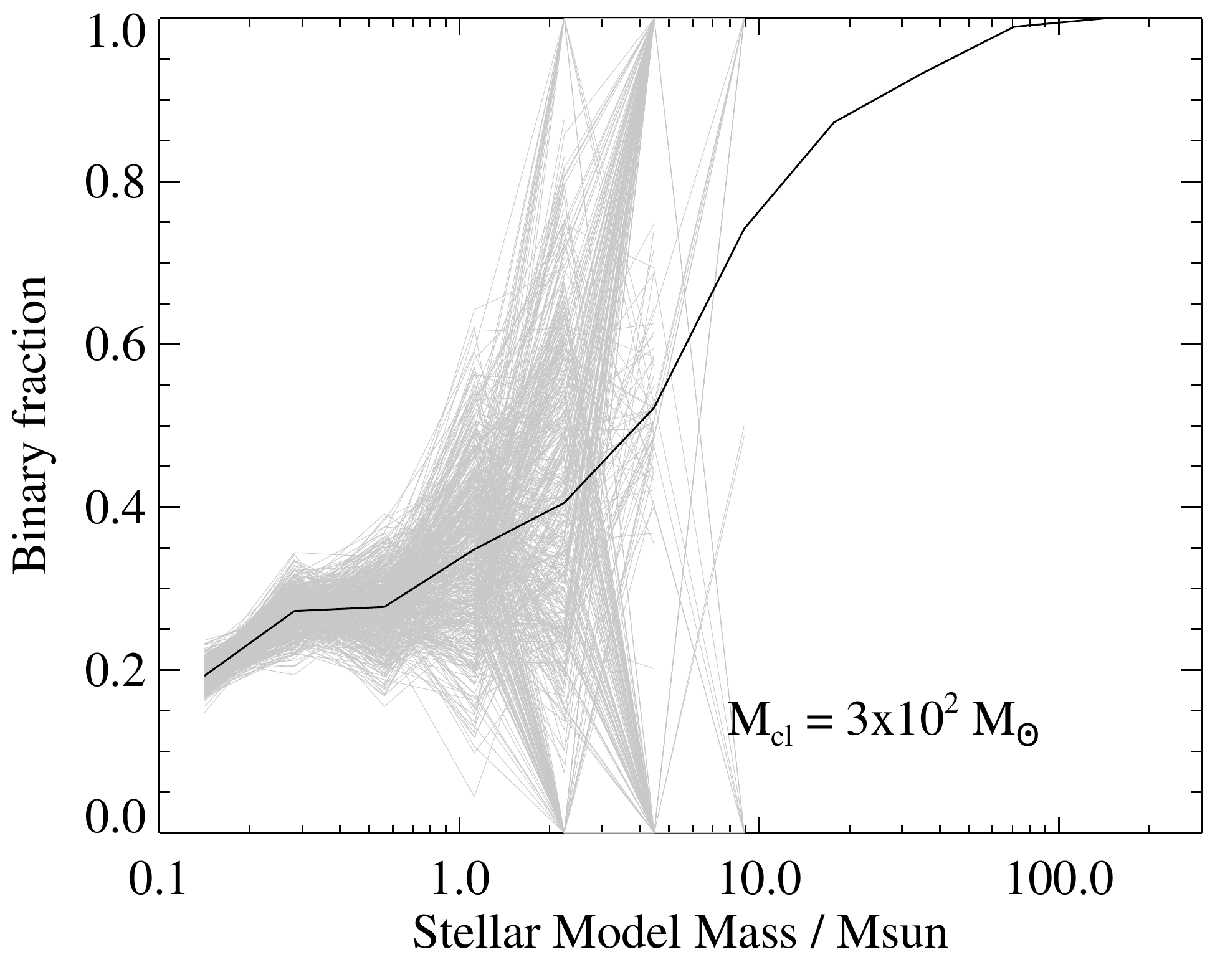}
      \includegraphics[width=0.32\textwidth]{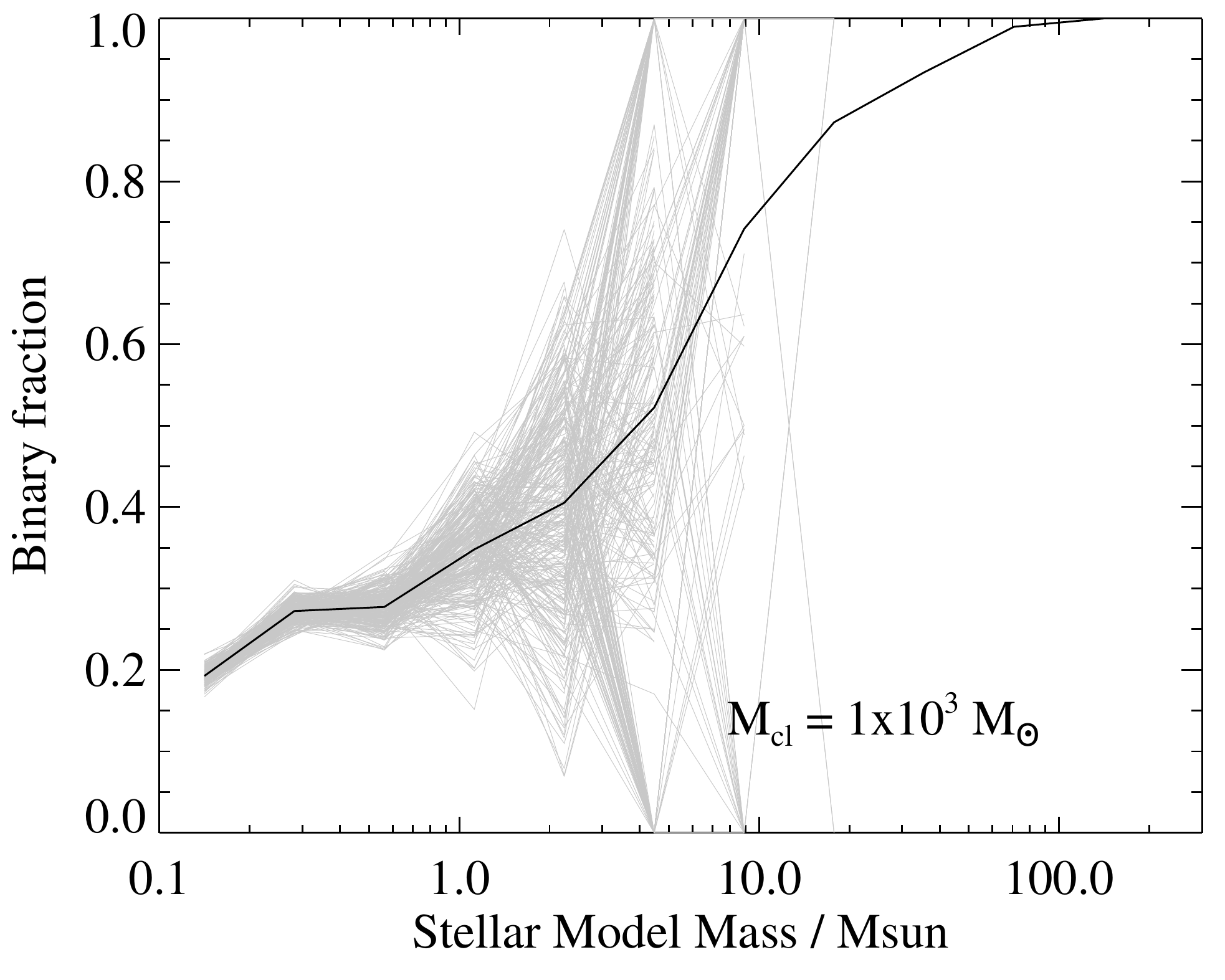}
      \includegraphics[width=0.32\textwidth]{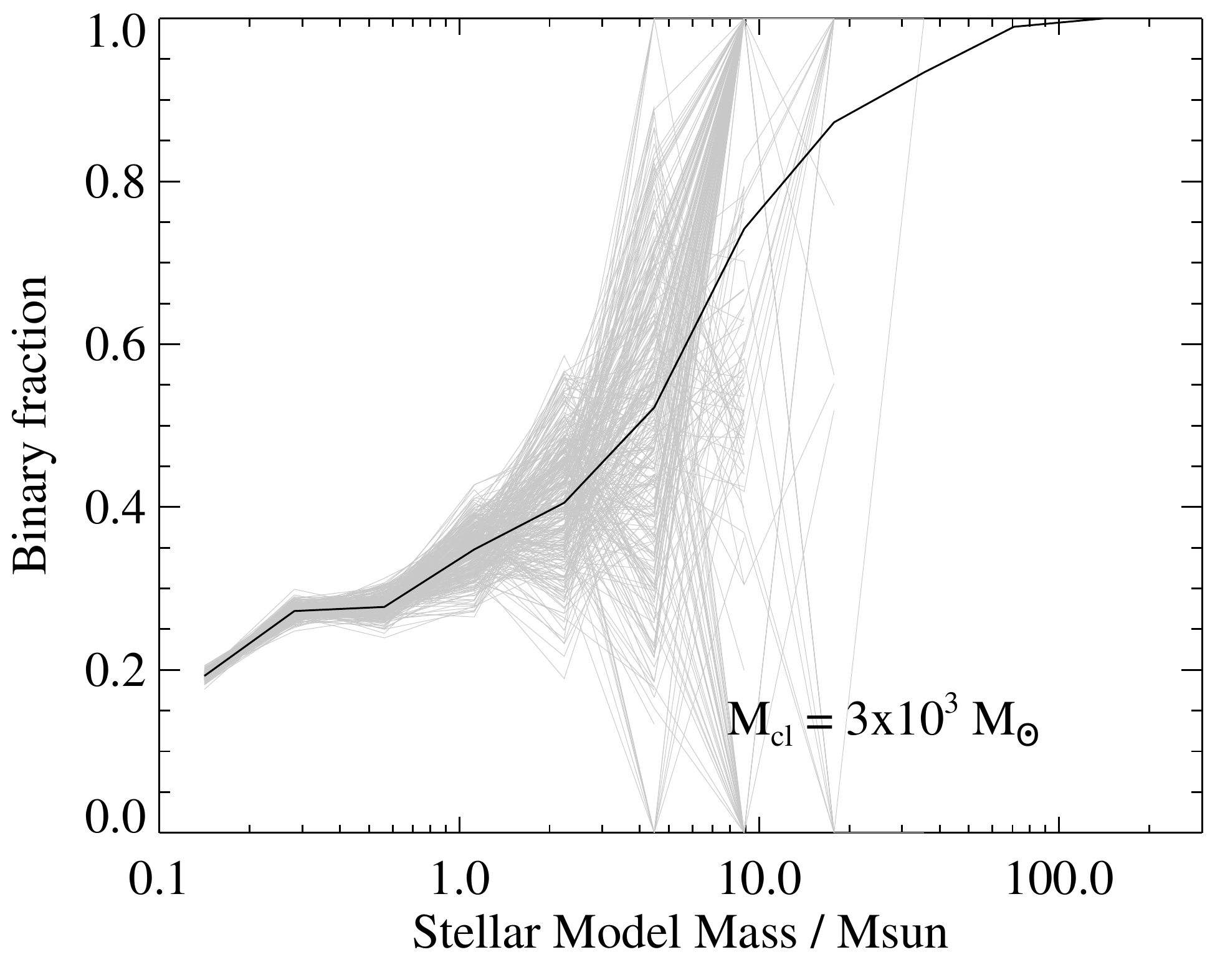}
      \includegraphics[width=0.32\textwidth]{Figures/plot_bfracm_1e4.pdf}
      \includegraphics[width=0.32\textwidth]{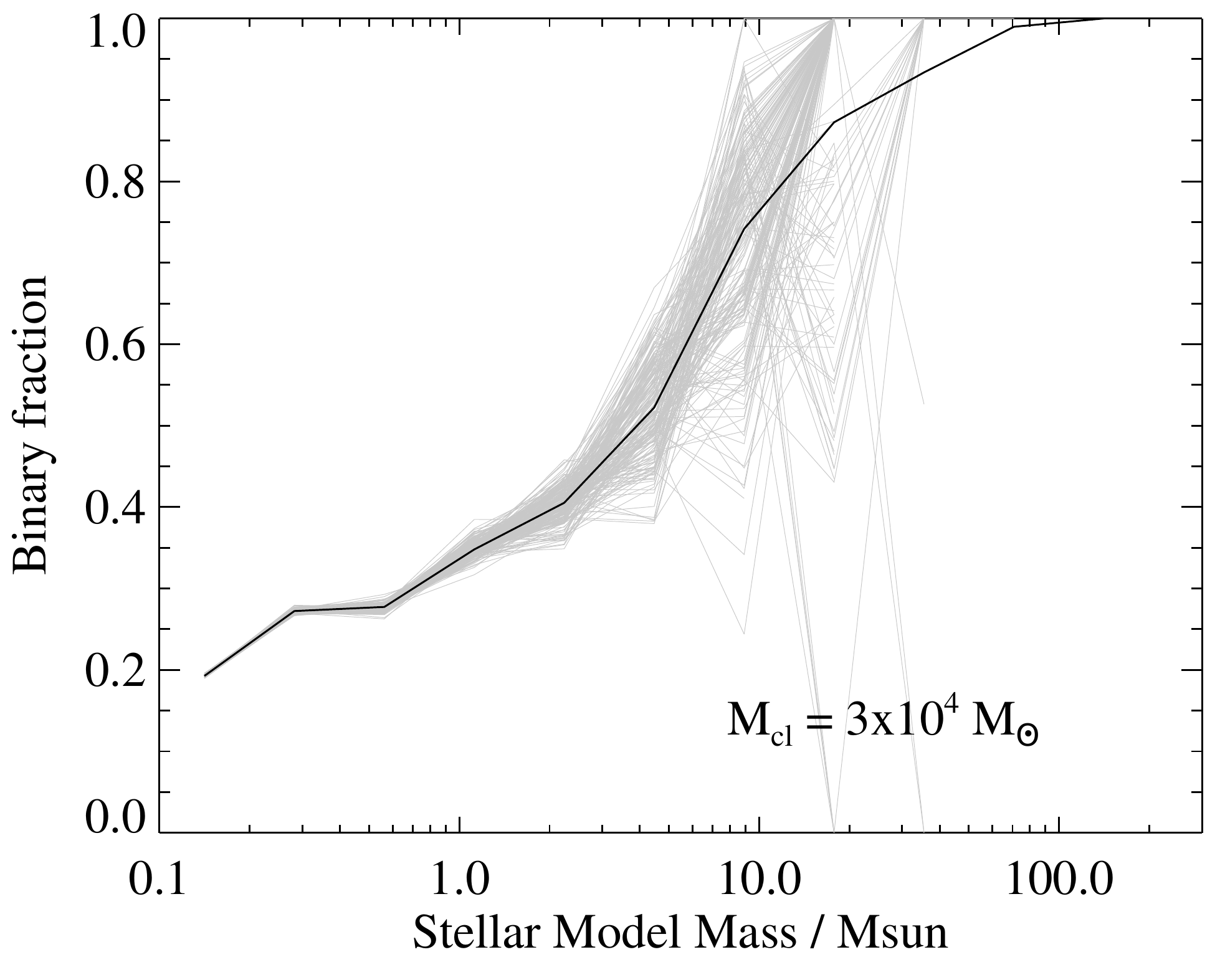}
      \includegraphics[width=0.32\textwidth]{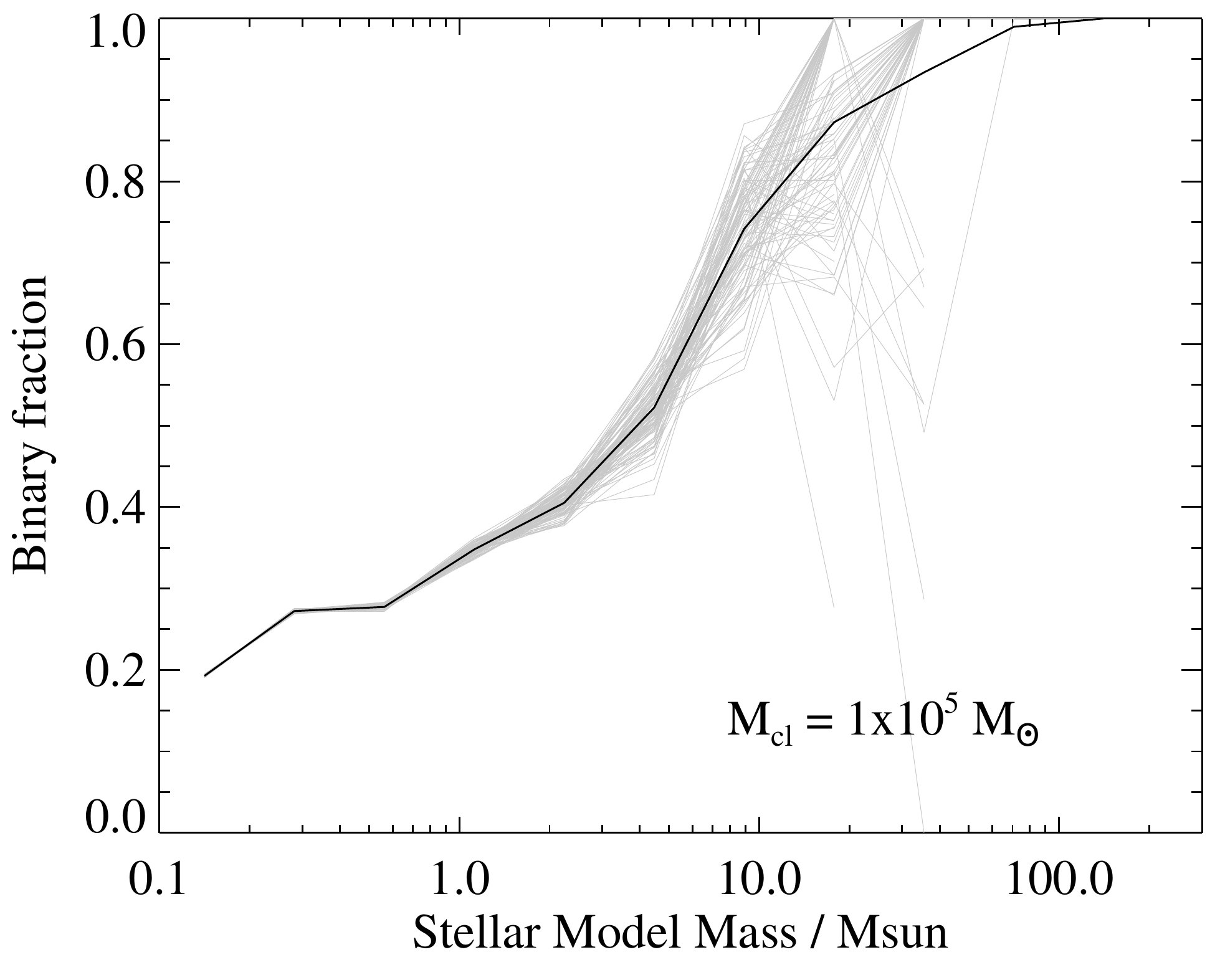}
      \includegraphics[width=0.32\textwidth]{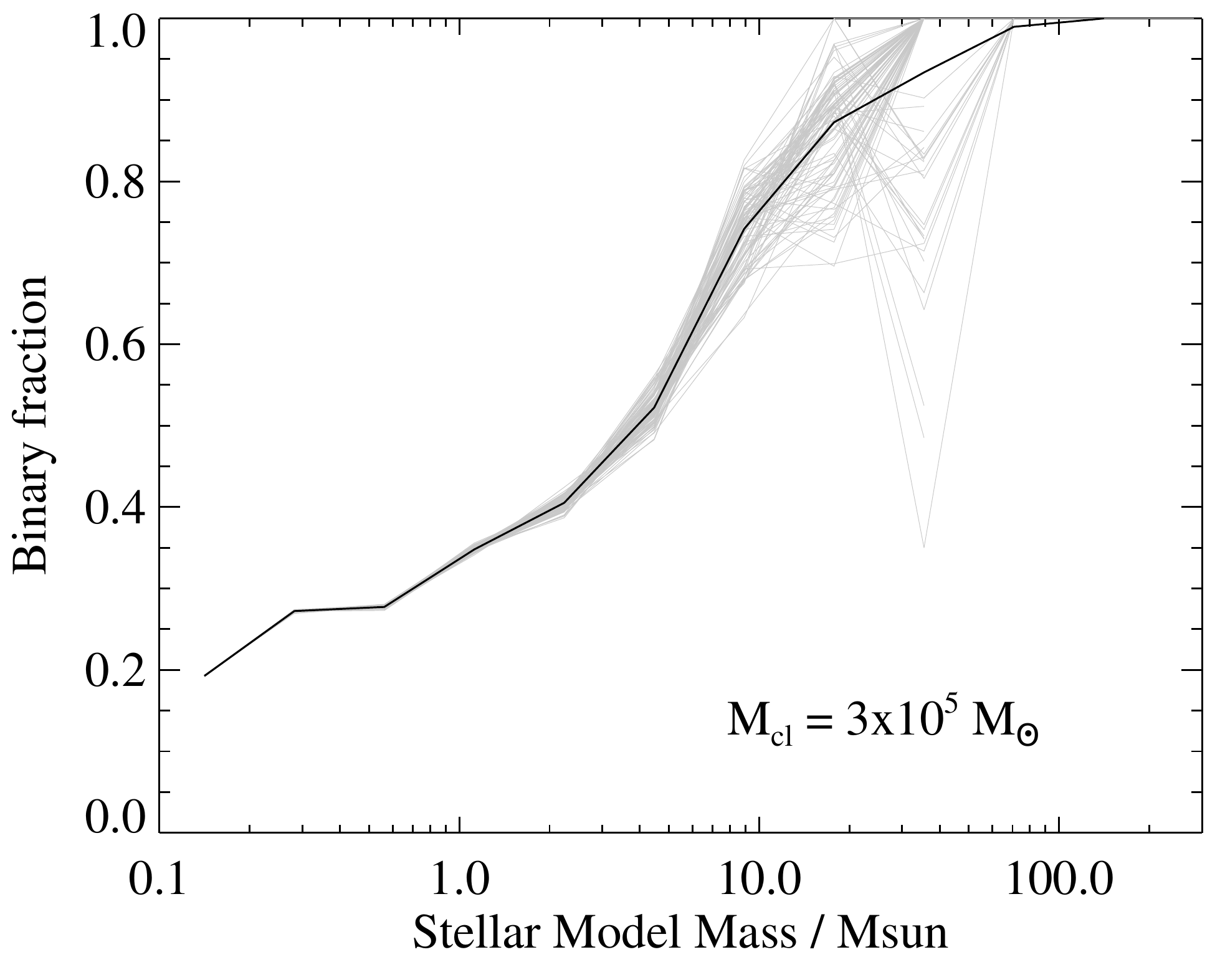}
      \includegraphics[width=0.32\textwidth]{Figures/plot_bfracm_1e6.pdf}
      \includegraphics[width=0.32\textwidth]{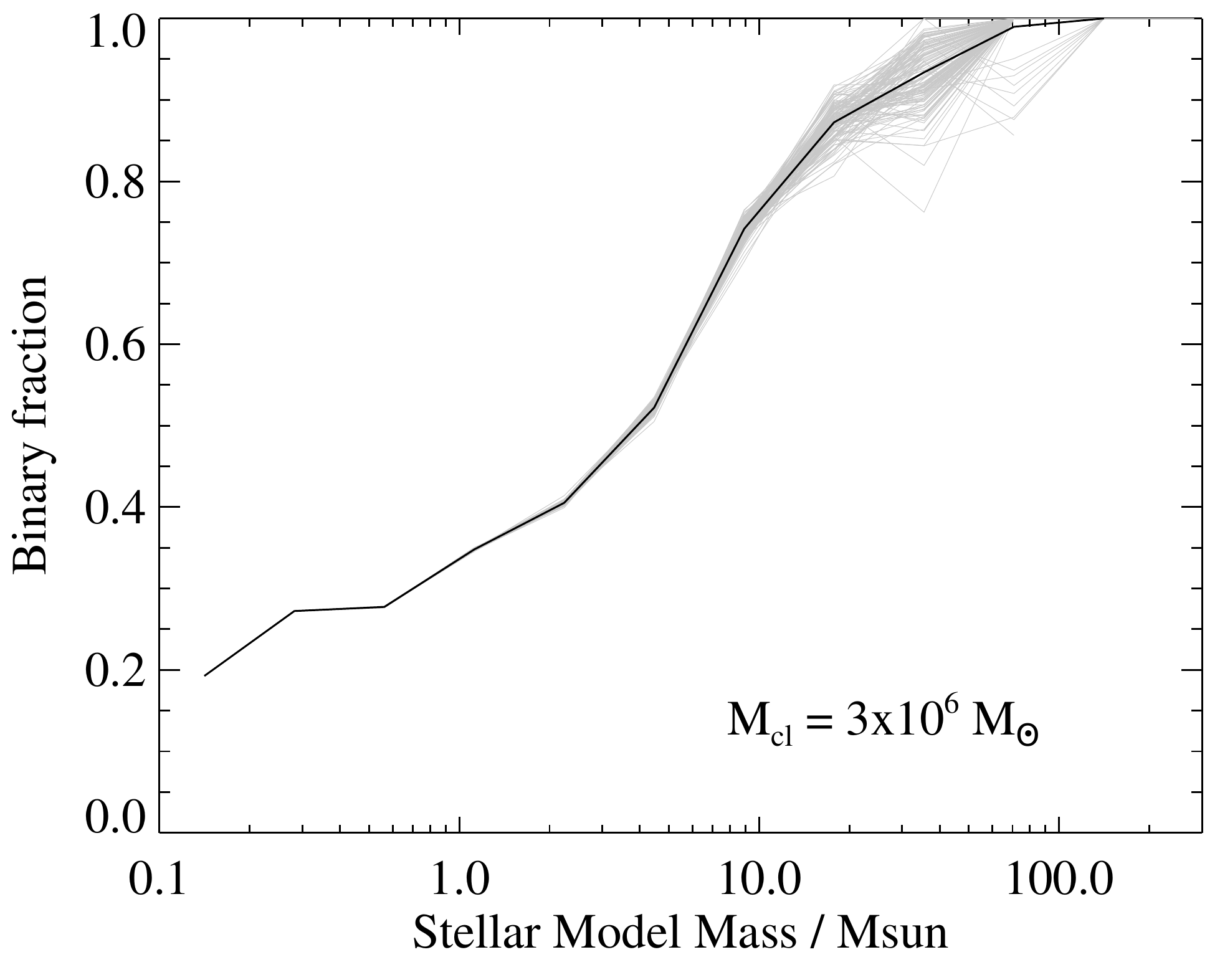}
      \includegraphics[width=0.32\textwidth]{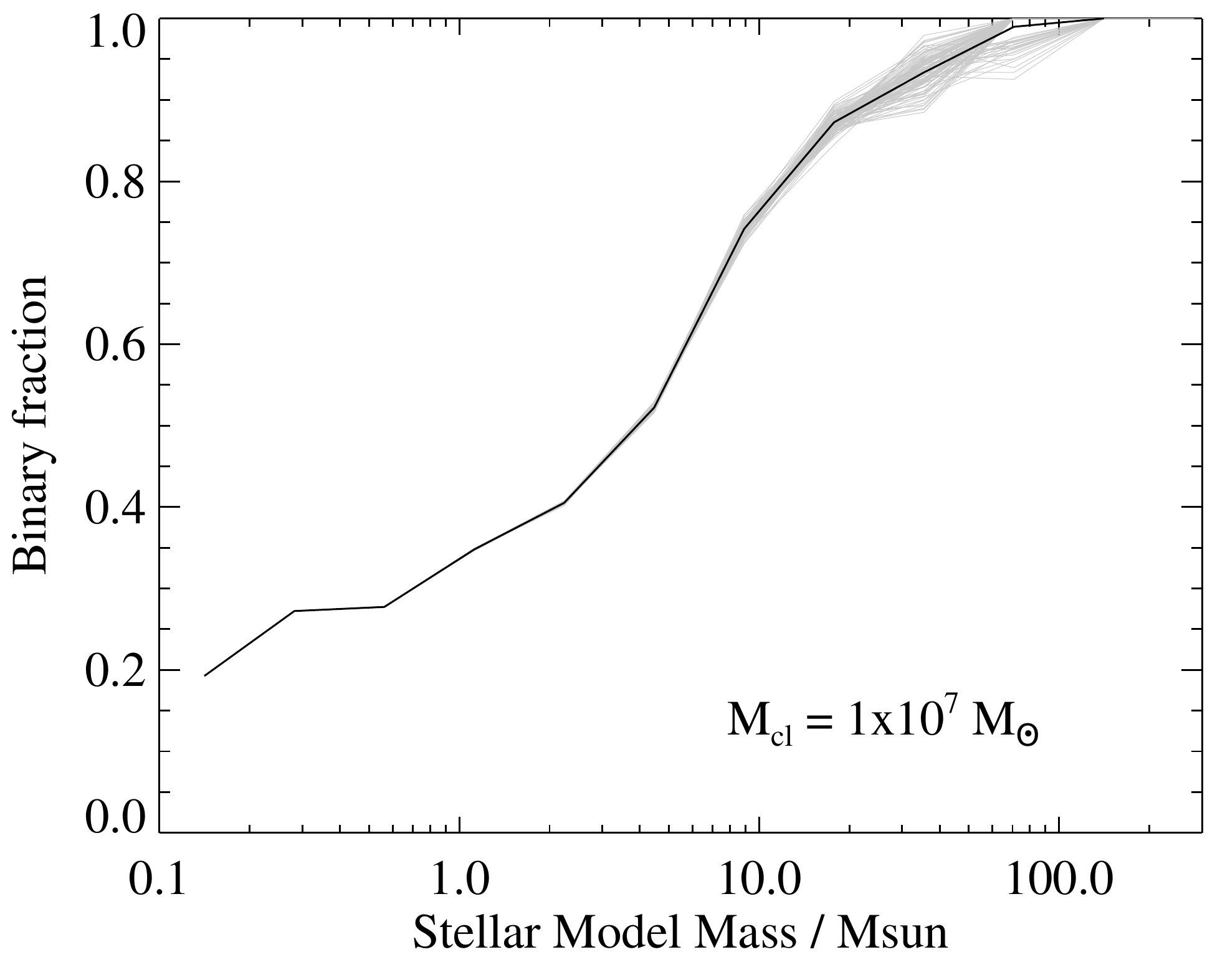}
      
      \caption{As in Figure \ref{fig:bfrac}, but now for all values of \mcl.}
      \label{fig:appendix_bfrac}
  \end{figure*}

Figure \ref{fig:appendix_bfrac} shows the variation in binary fraction as a function of mass between different stochastic realisations. The stellar models have been divided into twelve bins, evenly separated in logarithmic initial mass, and the contribution of binary models by mass calculated for each bin. This reproduces Figure \ref{fig:bfrac}, but shows the scatter for all \mcl\ values considered in this work.

\bigskip

\begin{figure*}
    \centering
    \includegraphics[width=0.46\textwidth]{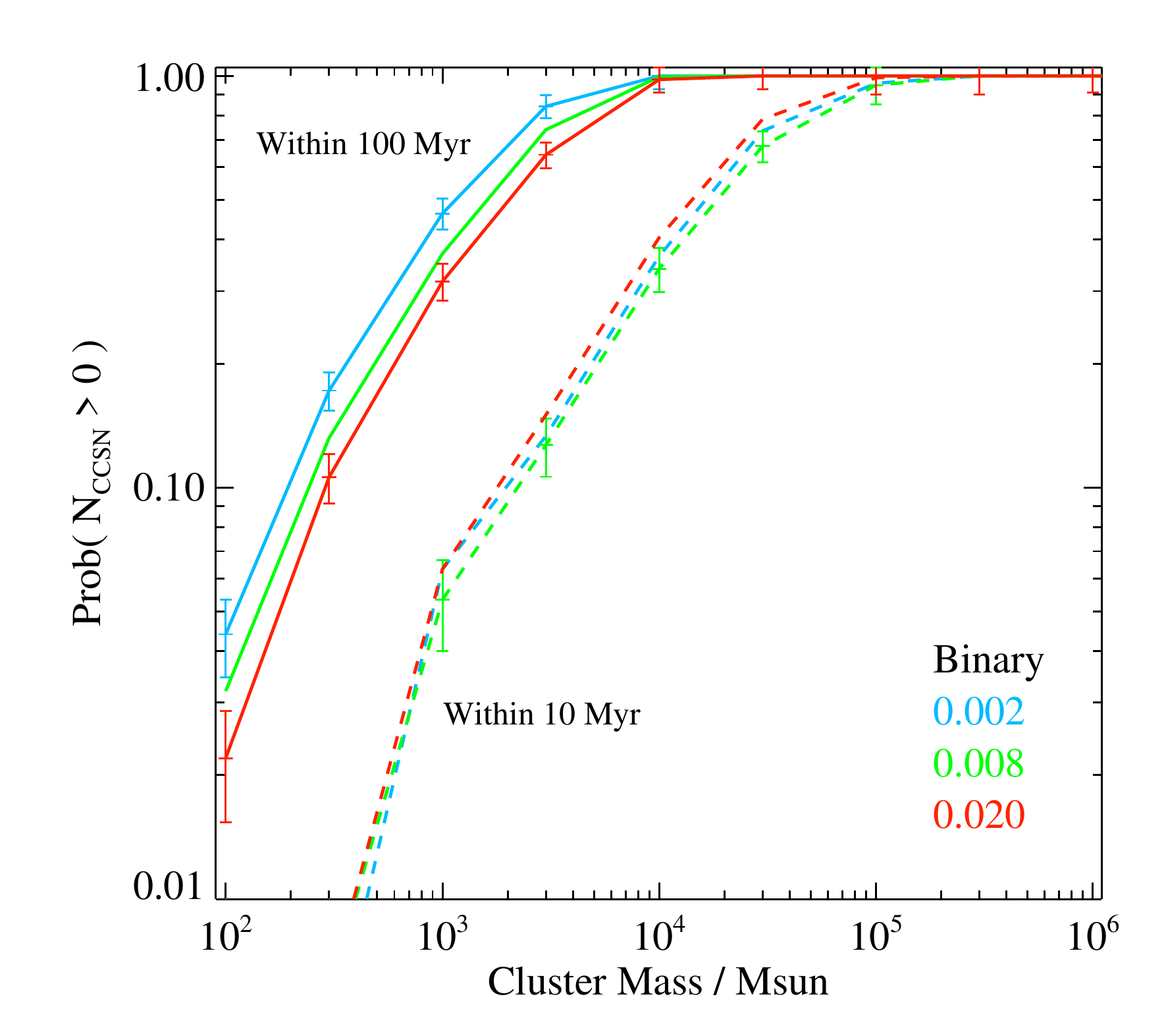}
    \includegraphics[width=0.46\textwidth]{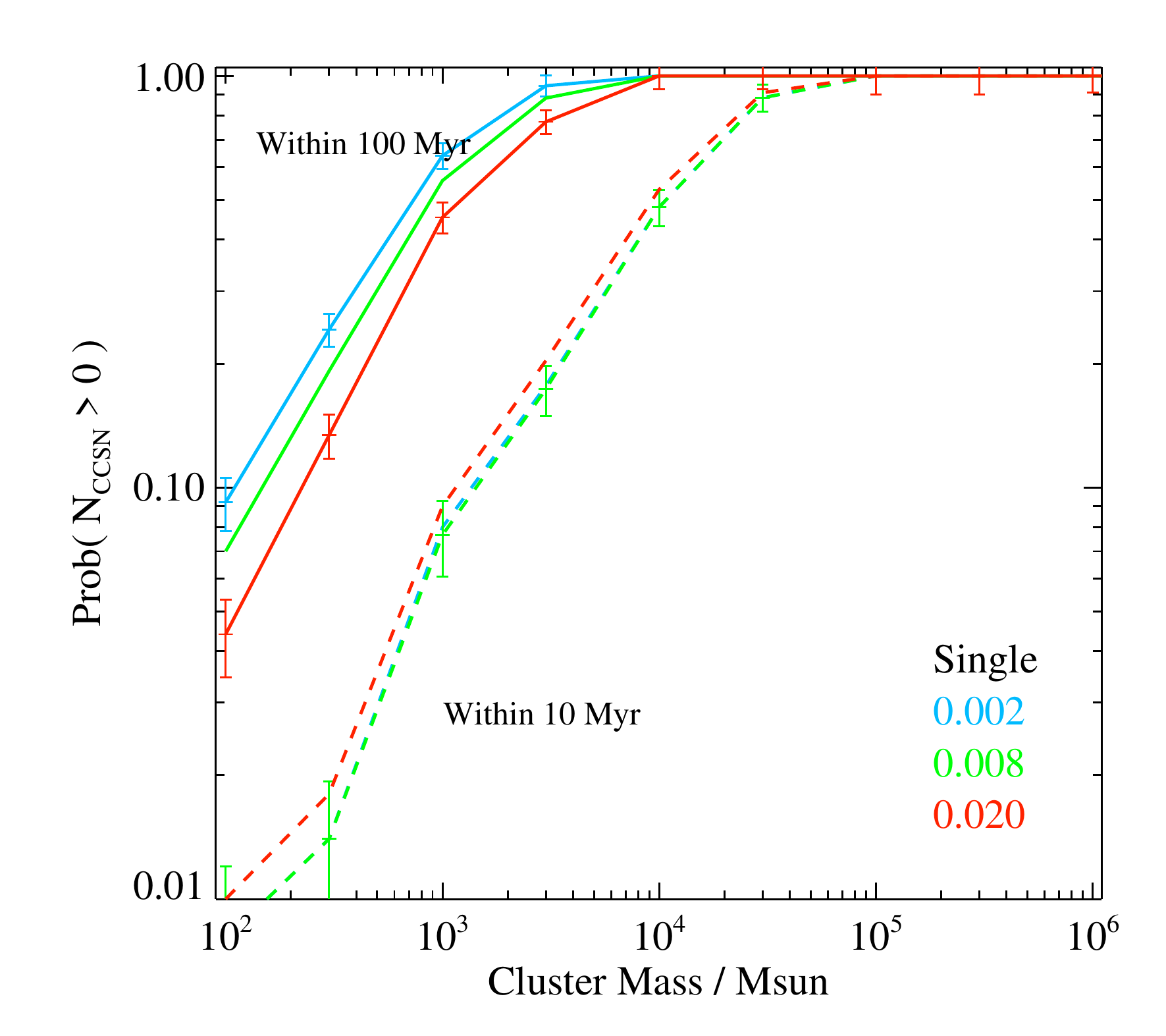}
    \caption{The fraction of stochastically-clusters hosting at least one core collapse supernova within 10 or 100\,Myr, as a function of cluster stellar mass and metallicity. Error ranges are determined by Poisson uncertainties, given the number of realisations undertaken at each cluster mass. Some error bars are omitted for clarity. Right: binary population, left: single stars only.}
    \label{fig:fracsn}
\end{figure*}

Figure \ref{fig:fracsn} shows the probability of clusters at a given cluster stellar mass hosting one or more core collapse supernovae, as a function of stellar population age and metallicity. Probabilities are evaluated as the fraction of stochastic realisations which satisfy the criteria in a binary stellar population, with associated Poissonian uncertainties. 

\bigskip

\begin{figure}
    \centering
    \includegraphics[width=0.5\textwidth]{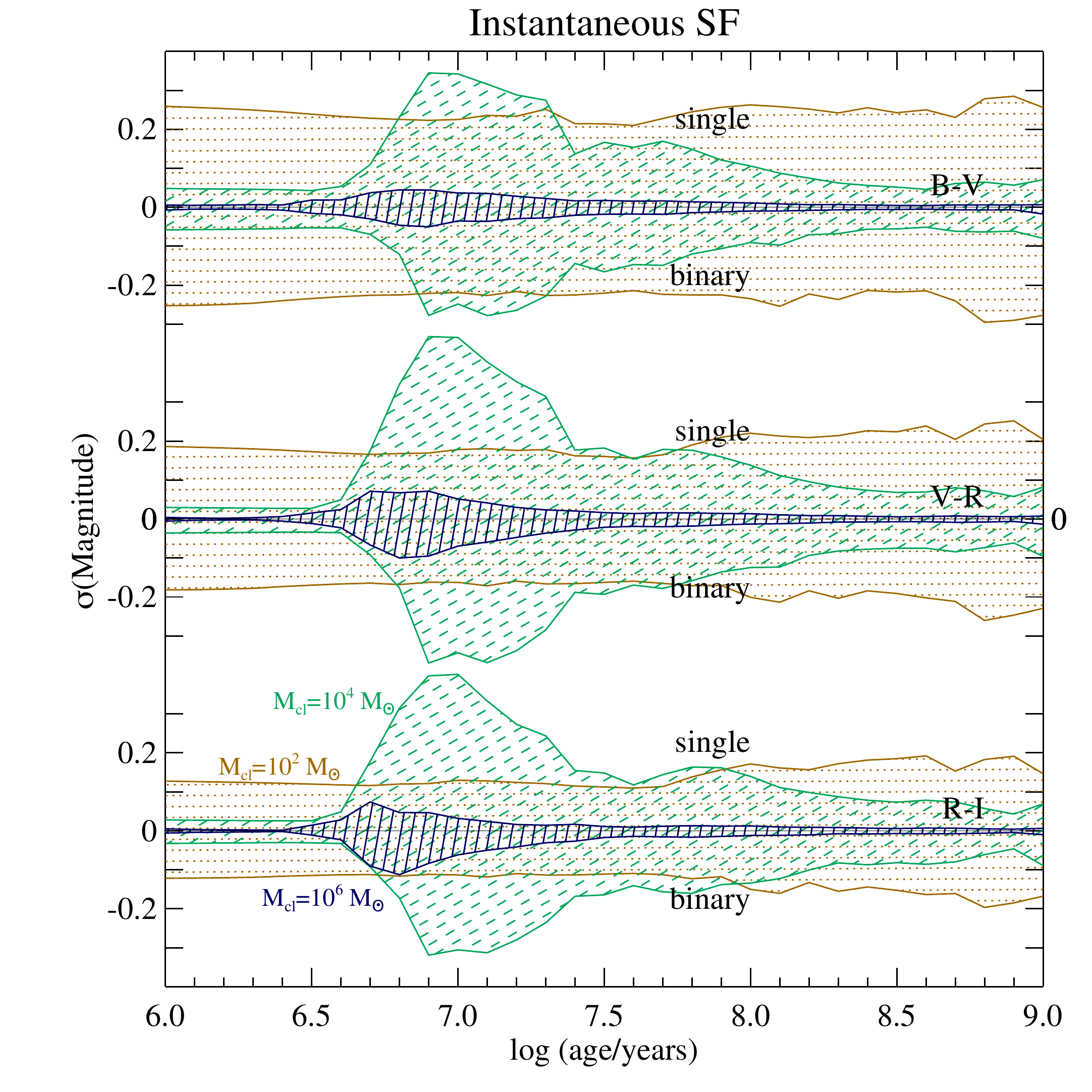}
    \caption{Colour uncertainties with age for instantaneous star formation (i.e. a simple stellar population). The standard deviation of the colour in magnitudes in binary and single star populations are shown either side of the centre line, in the case of \mcl$=10^2$, $10^4$ and $10^6$\,M$_\odot$ populations (shown as orange dotted, green dashed, and blue hashed regions respectively). All models are shown at $Z=0.020$.}
    \label{fig:col_inst}
\end{figure}

\begin{figure}
    \centering
    \includegraphics[width=0.5\textwidth]{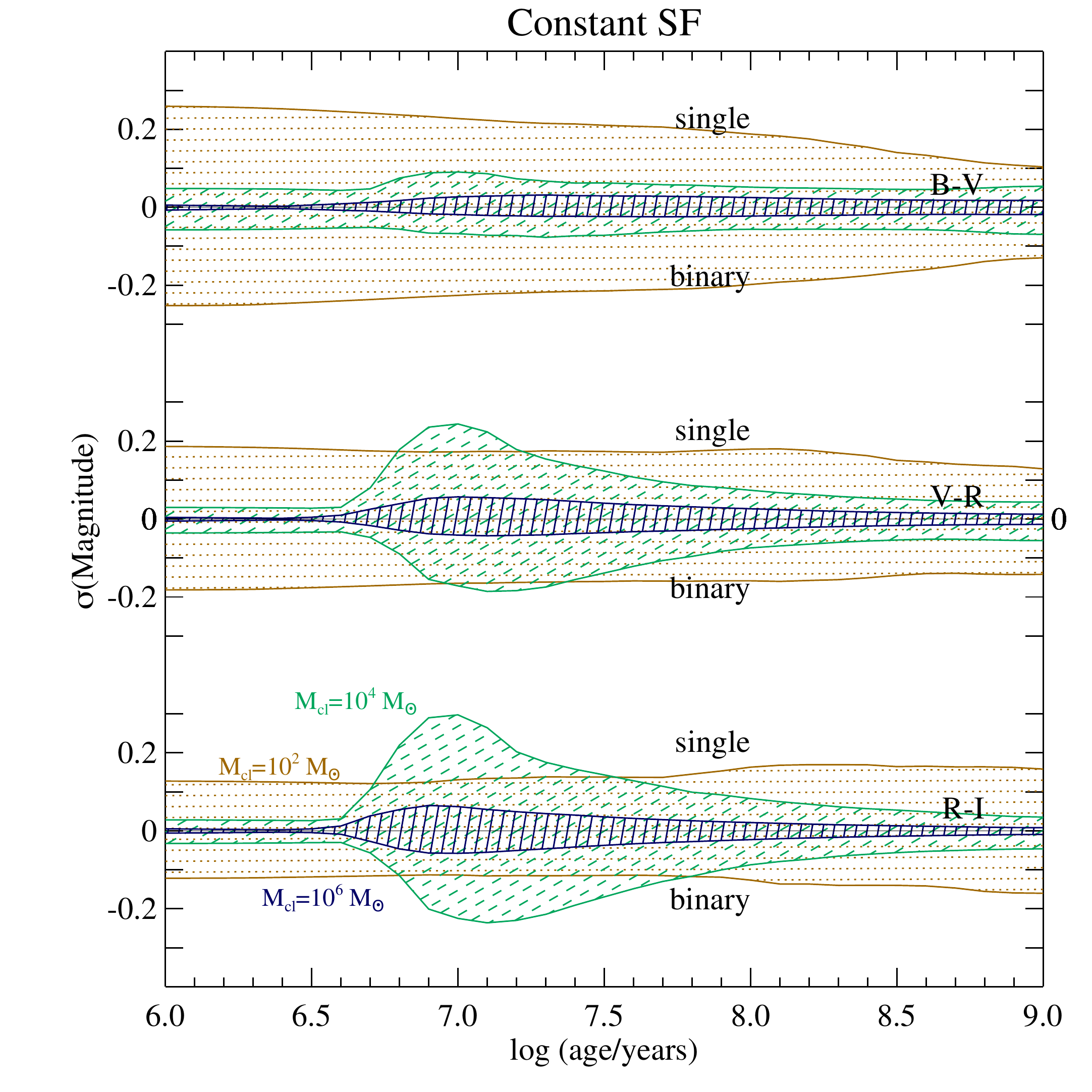}
    \caption{As in Figure \ref{fig:col_inst}, but now showing colour uncertainties with age for constant star formation. The magnitude of uncertainties in binary and single star populations are shown either side of the centre line in the case of \mcl$=10^2$, $10^4$ and $10^6$\,M$_\odot$ populations (shown as orange dotted, green dashed, and blue hashed regions respectively).}
    \label{fig:col_const}
\end{figure}

Figure \ref{fig:colcol} demonstrated the age-dependence of scatter in the photometric colours of an \mcl$=10^6$\,M$_\odot$ cluster undergoing constant star formation. In Figures \ref{fig:col_inst} and \ref{fig:col_const} we extend this analysis by comparing the 1\,$\sigma$ scatter in the $B-V$, $V-R$ and $R-I$ photometric colours of stellar populations stochastically-sampled at three different cluster masses. For clusters of $10^2$\,M$_\odot$, the scatter in the photometric colours is large (typically $\sigma(V-R)\sim0.2$\,mag) at all ages. However it shows little age dependence: relatively few stars with M$>5$\,M$_\odot$ are selected and so the colour of each cluster remains relatively stable on Gyr timescales. Clusters with $10^4$\,M$_\odot$ of stars show the largest variation, with photometric uncertainty peaking at ages around 10\,Myr. These clusters typically host stars with M$>20$\,M$_\odot$, but only in small numbers. Stochastic sampling can significantly change the number and timing of individual massive stars entering luminous post-main sequence phases. By contrast, clusters with $10^6$\,M$_\odot$ of stars select substantially larger numbers of these massive stars, reducing the variation between individual stochastic samples compared to the lower mass cluster.

Note that we do not recommend the use of single star synthetic populations as these are demonstrably incompatible with the binary fraction in observed stellar populations. Nonetheless, in these two figures we also provide a comparison between populations built using binary models and those constructed entirely from single star models in order to evaluate whether stochastically  sampling the mass or binary parameter distributions has the larger effect. At most ages and cluster masses, the uncertainty in colour from single and binary populations is similar but shows a different time structure, particularly in the redder colours, hinting that both sampling effects are important. The inclusion of binary models does not significantly change the average number of massive stars selected in each cluster (which dominates the scatter), but the presence of binary interactions introduces a sample-dependent delay in the colour evolution, shifting the epoch of greatest colour uncertainty in the population to a later time step. 

\bigskip

\begin{figure}
    \centering
    \includegraphics[width=0.5\textwidth]{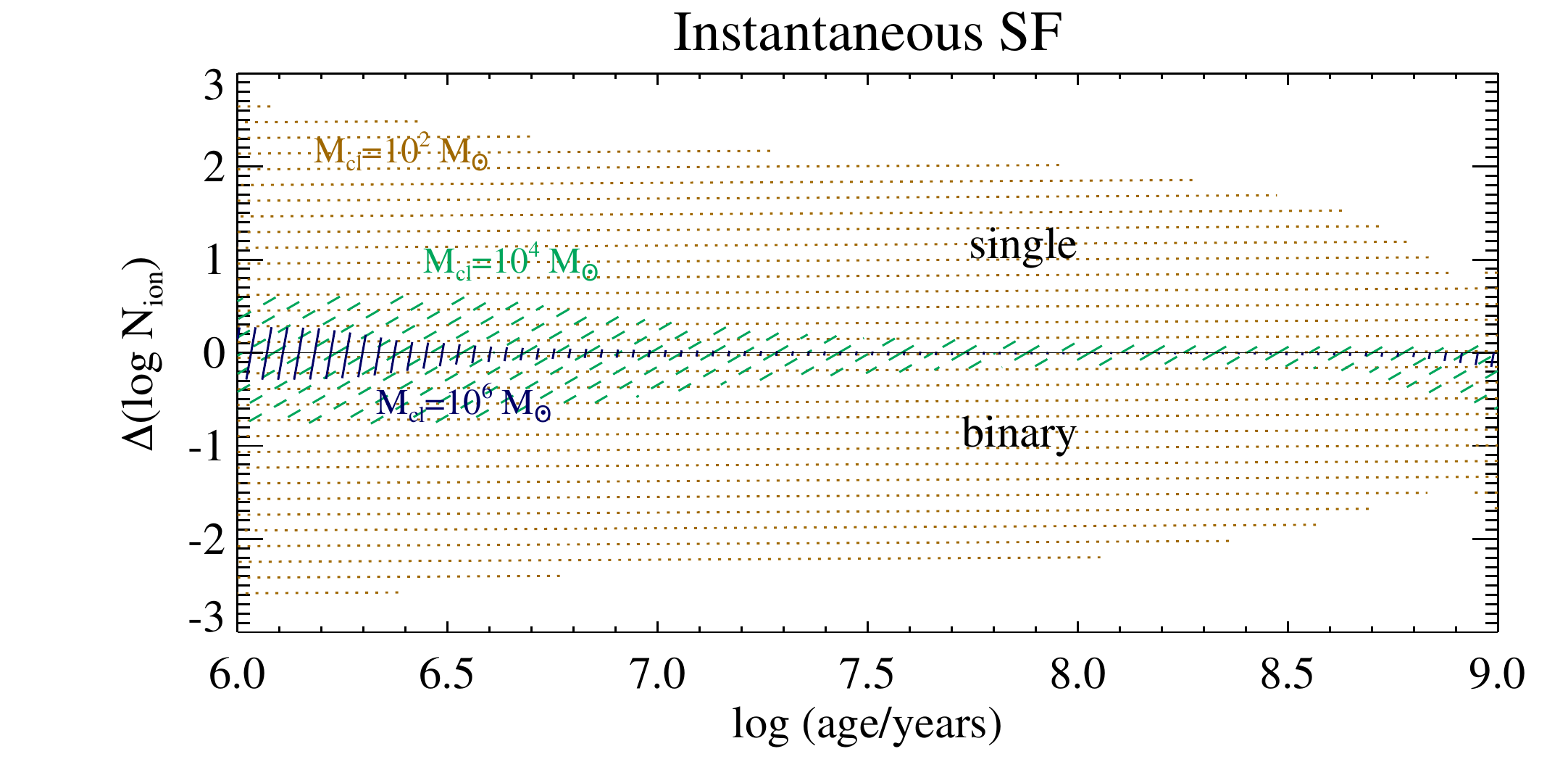}
    \caption{As in Figure \ref{fig:col_inst}, but now showing uncertainties with age on the ionizing photon production rate for instantaneous star formation. The magnitude of uncertainties in binary and single star populations are shown either side of the centre line.}
    \label{fig:ion_inst}
\end{figure}

\begin{figure}
    \centering
    \includegraphics[width=0.5\textwidth]{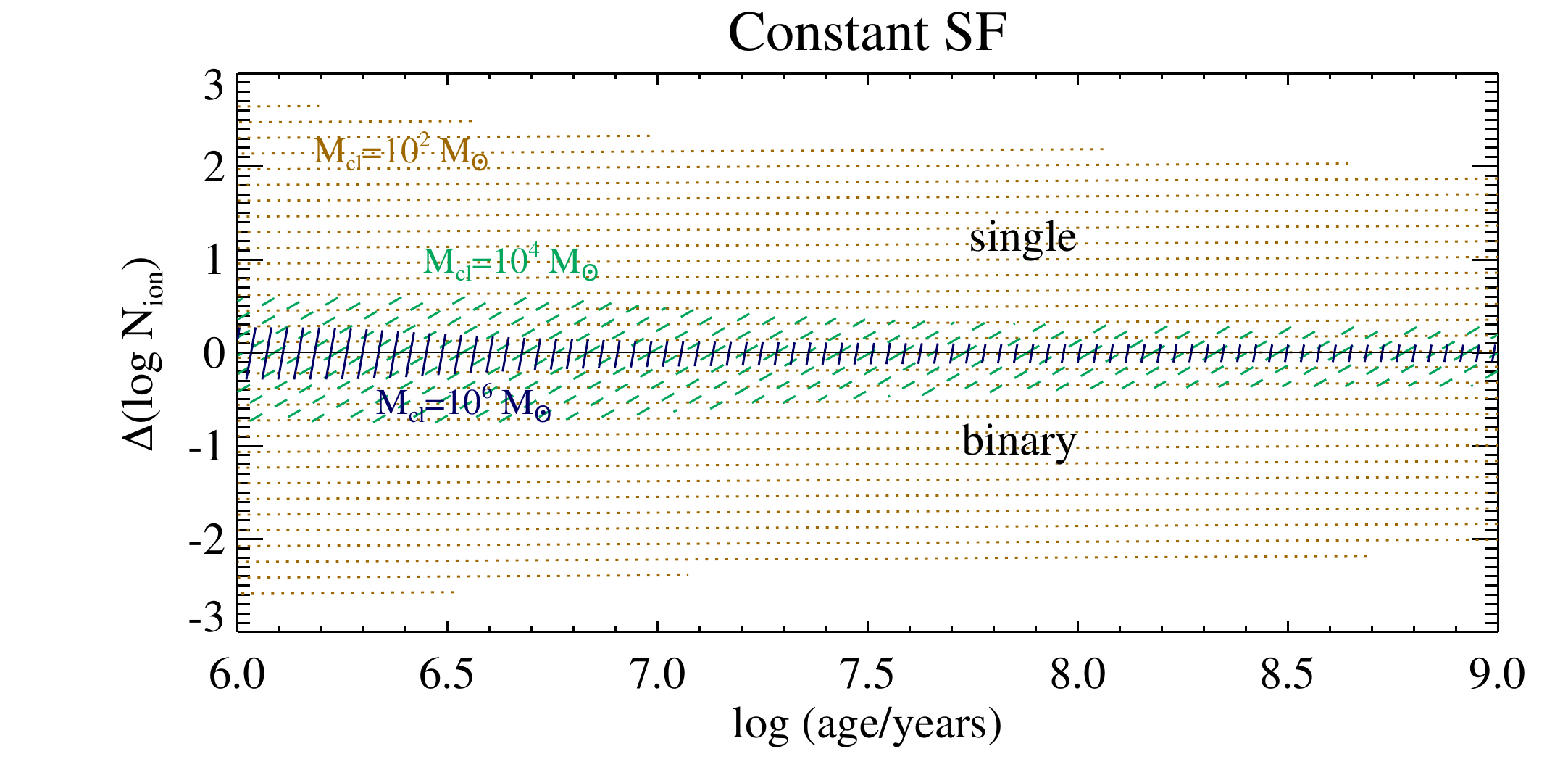}
    \caption{As in Figure \ref{fig:col_inst}, but now showing uncertainties with age on the ionizing photon production rate for constant star formation. The magnitude of uncertainties in binary and single star populations are shown either side of the centre line in each case.}
    \label{fig:ion_cont}
\end{figure}

In Figures \ref{fig:ion_inst} and \ref{fig:ion_cont} we show the 1\,$\sigma$ scatter in the ionizing photon production rate of $Z=0.020$ stellar populations stochastically-sampled at the same three different cluster masses. Since this emission is dominated by very young massive stars, there is relatively little time-dependence in the statistical scatter in the stochastic sample. There is also little difference in the scatter between our single and binary star models.

\bigskip

\begin{figure*}
      \centering
      \includegraphics[width=0.85\columnwidth]{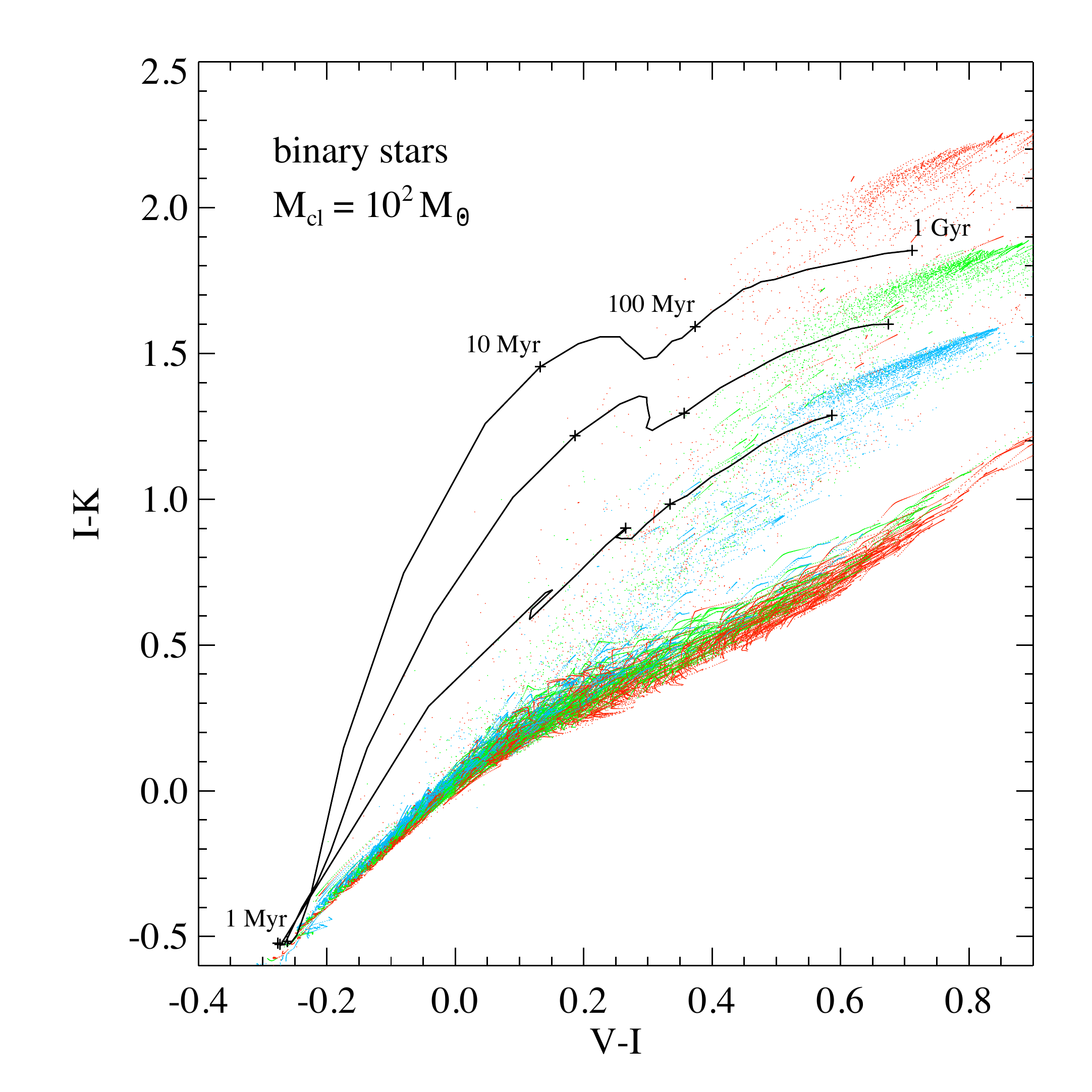}
      \includegraphics[width=0.85\columnwidth]{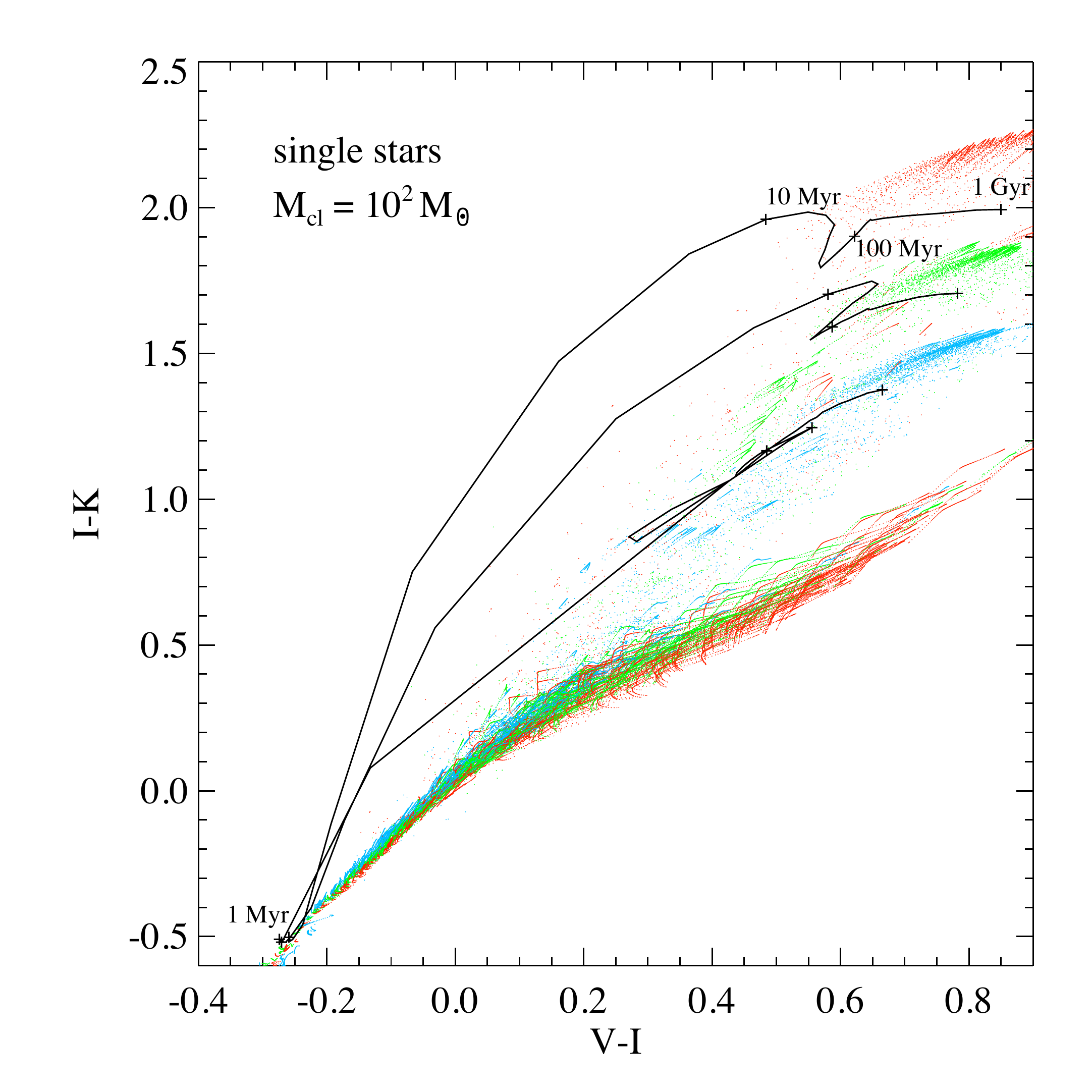}
      \includegraphics[width=0.85\columnwidth]{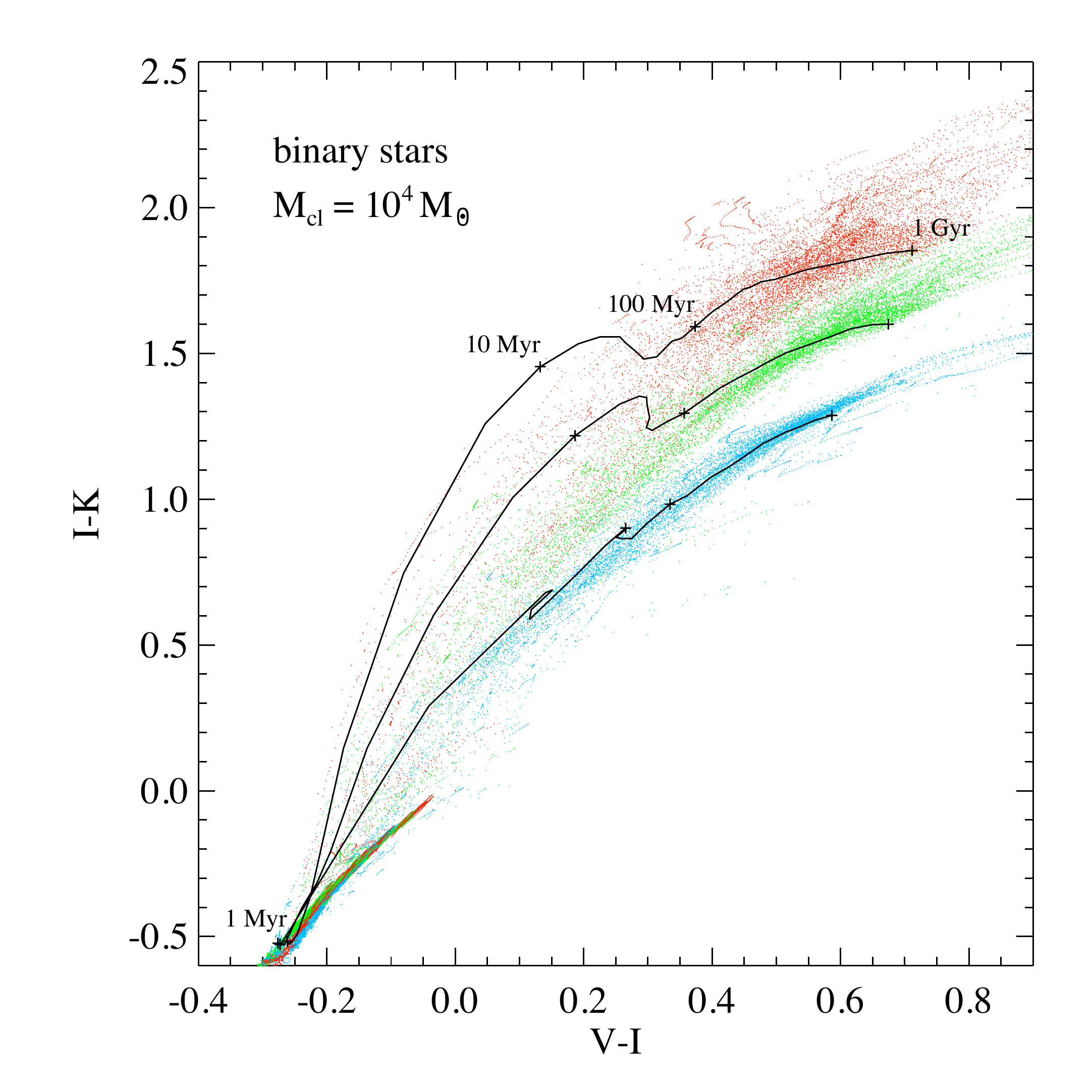}
      \includegraphics[width=0.85\columnwidth]{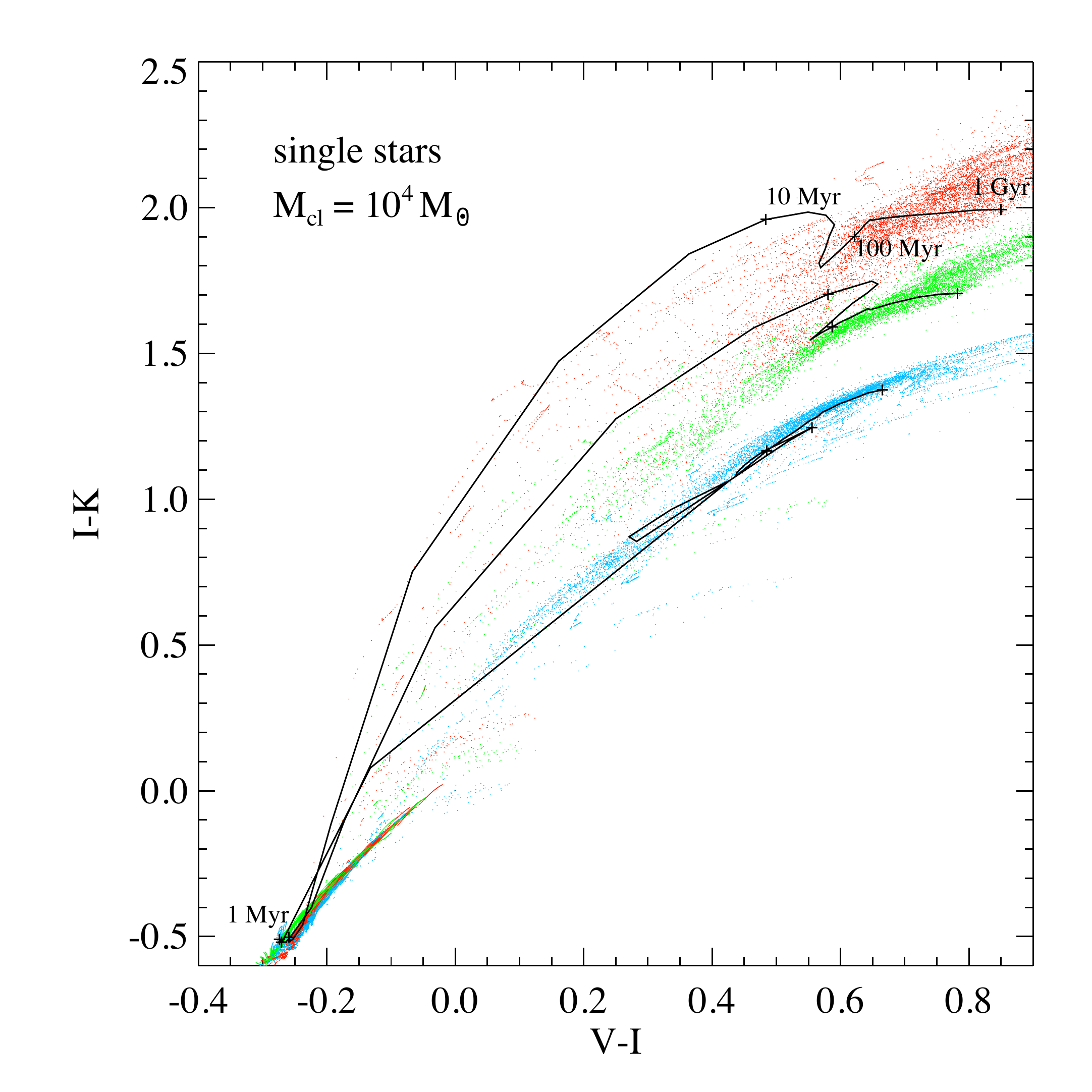}
      \includegraphics[width=0.85\columnwidth]{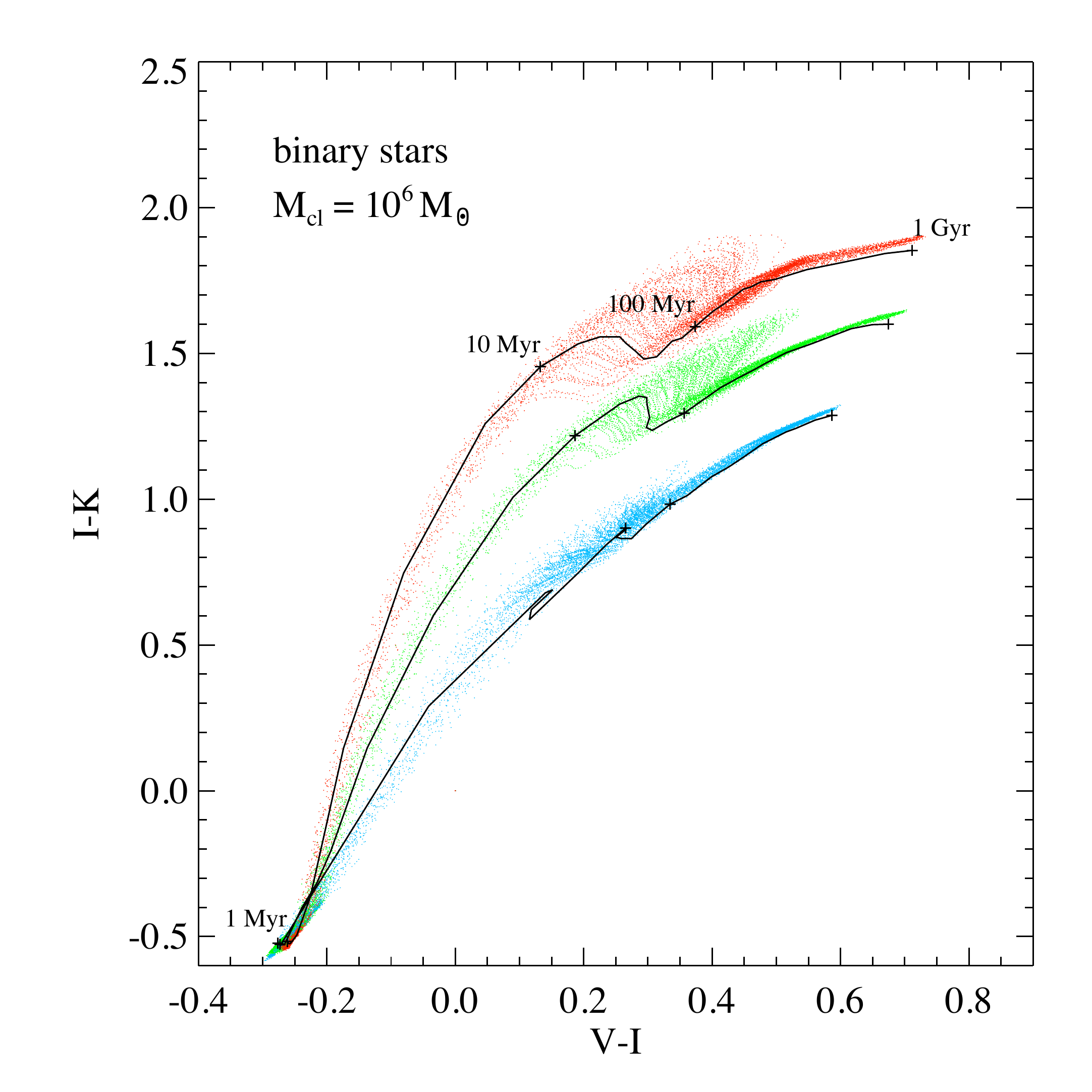}
      \includegraphics[width=0.85\columnwidth]{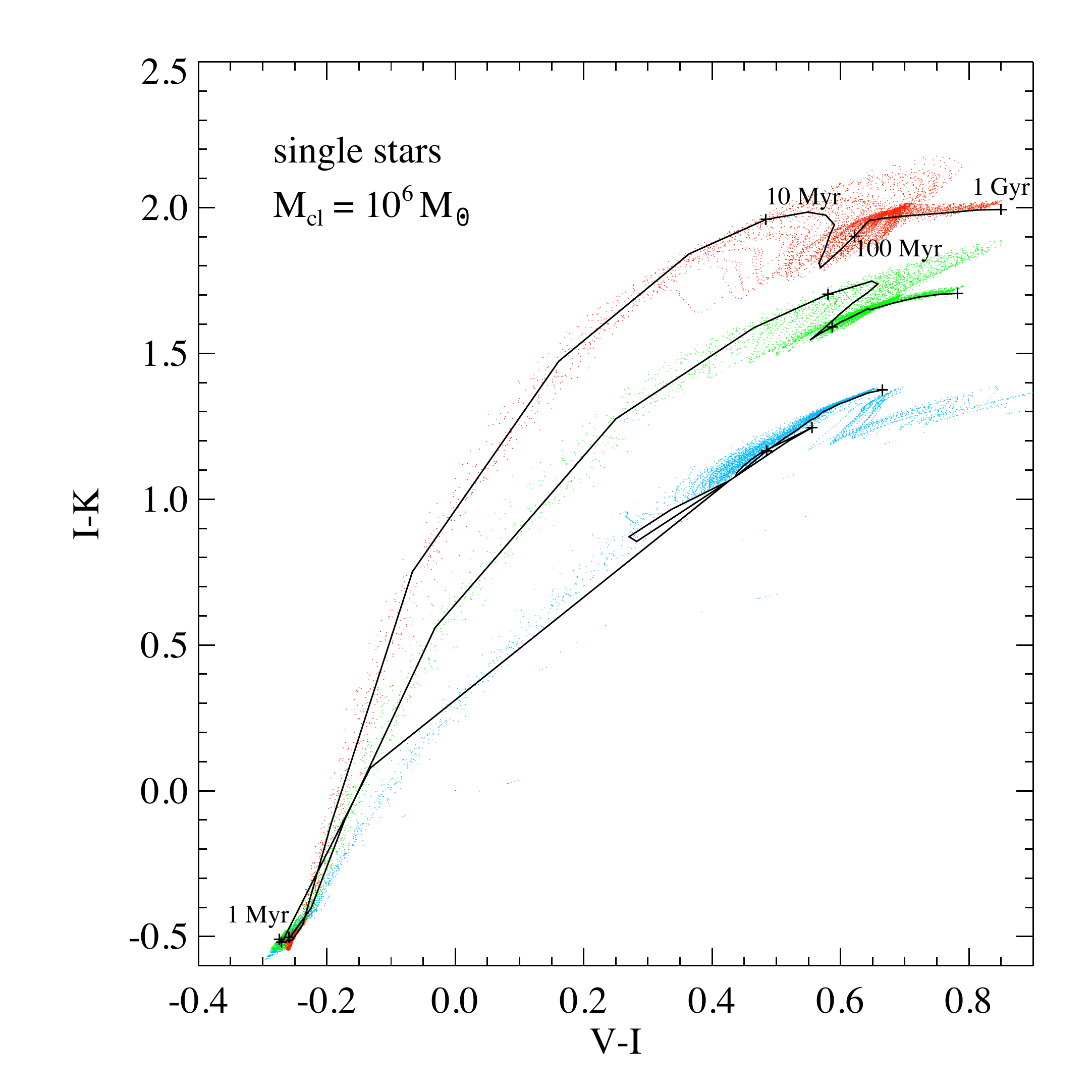}
      \caption{The $(V-I)$ vs $(I-K)$ colour evolution of stellar populations, coloured by metallicity. As in Figure \ref{fig:colcol}, but now comparing the impact of stochastic sampling on populations of different masses, and those including binary pathways to those involving only single stars. Solid lines indicate the time evolution of BPASS v2.2.1 statistically-sampled models at each metallicity}
      \label{fig:colcols}
  \end{figure*}

Figure \ref{fig:colcols} extends Figure \ref{fig:colcol} to consider the photometric colour evolution of continuously star forming stellar populations with different cluster masses and including or excluding binary stellar evolution pathways.  Stellar populations in low mass clusters are redder than assumed by statistical sampling. In the lowest masses clusters, the colour distribution becomes bimodal at ages above about 10 Myr. The colours of those stochastic realisations which include moderately massive stars become dominated by the red giant and supergiant population, while those which do not continue to show colours consistent with main sequence stars. The presence of binary interactions has a relatively small effect, although the fraction of realisations on the upper branch of the colour-colour diagram is slightly higher. At high cluster masses, both single and binary populations are well described by the statistically sampled BPASS v2.2.1 model colour evolution.

\bigskip

\begin{figure*}
      \centering
      \includegraphics[width=1.8\columnwidth]{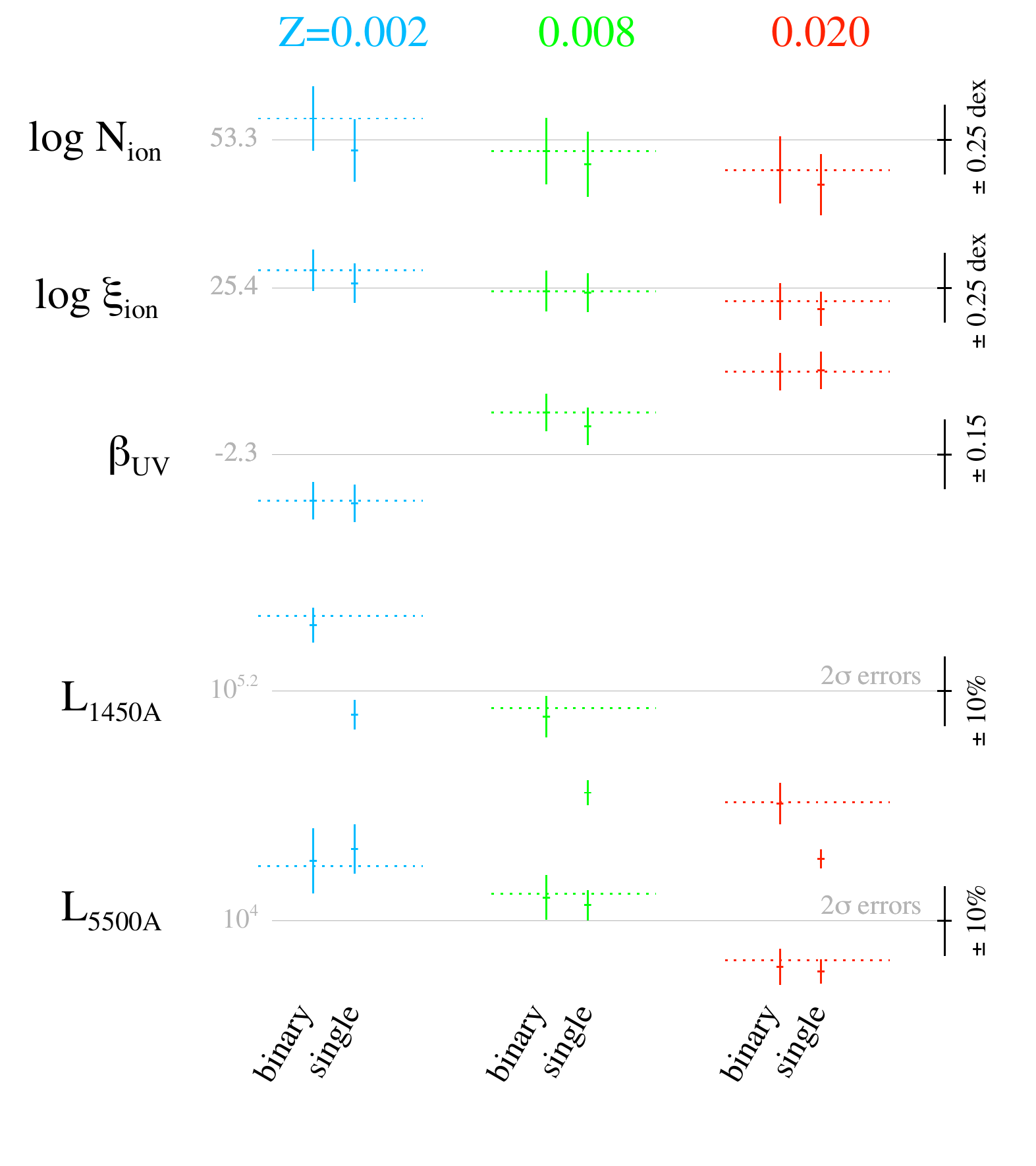}
      \caption{As in Figure \ref{fig:uncertainties} but now comparing the impact of stochastic sampling on populations including binary pathways to those involving only single stars. Uncertainties indicate 1\,$\sigma$ scatter in the stochastic realisations except for the continuum luminosity densities which are shown with $2\,\sigma$ error bars for clarity.}
      \label{fig:sinuncertainties}
  \end{figure*}

Finally, in Figure \ref{fig:sinuncertainties} we provide a comparison of the magnitude of uncertainties arising from stochastic sampling of a stellar population involving binary pathways (i.e. drawing randomly from distributions in mass, binary mass ratio, binary period) and those with only single pathways (i.e. sampling in mass alone). The magnitude of the uncertainties are similar in all cases considered here.

\bigskip

A machine readable version of the data tabulated here (including Table \ref{tab:maxmasses}) can be downloaded from the BPASS project website at \hyperlink{bpass.auckland.ac.uk}{bpass.auckland.ac.uk} or \hyperlink{warwick.ac.uk/bpass}{warwick.ac.uk/bpass}

\begin{table*}
\begin{tabular}{lccccccccccc}
   & \multicolumn{11}{|c|}{Uncertainty on log($N_\mathrm{ion}$/s) as a function of cluster mass, \mcl/M$_\odot$} \\
 Z\ \ \textbackslash\ \ \mcl  & 1$\times10^2$ & 3$\times10^2$ & 1$\times10^3$ & 3$\times10^3$ & 1$\times10^4$ & 3$\times10^4$ & 1$\times10^5$ & 3$\times10^5$ & 1$\times10^6$ & 3$\times10^6$ & 1$\times10^7$  \\
\hline\hline
0.002 & 1.47 &  1.51 &  1.69 &  1.37 &  0.98 &  0.71 &  0.52 &  0.36 &  0.23 &  0.14 &  0.08\\
0.008 & 1.51 &  1.46 &  1.70 &  1.43 &  0.96 &  0.70 &  0.55 &  0.37 &  0.24 &  0.13 &  0.08\\
0.020 & 1.43 &  1.25 &  1.54 &  1.35 &  0.96 &  0.67 &  0.53 &  0.36 &  0.24 &  0.13 &  0.08\\
\hline
\\
   & \multicolumn{11}{|c|}{Uncertainty on log($\xi_\mathrm{ion}$/s) as a function of cluster mass, \mcl/M$_\odot$} \\
 Z\ \ \textbackslash\ \ \mcl  & 1$\times10^2$ & 3$\times10^2$ & 1$\times10^3$ & 3$\times10^3$ & 1$\times10^4$ & 3$\times10^4$ & 1$\times10^5$ & 3$\times10^5$ & 1$\times10^6$ & 3$\times10^6$ & 1$\times10^7$  \\
\hline\hline
0.002 &  2.4 &  1.8 &  1.4 &  1.0 &  0.68 &  0.50 &  0.34 &  0.23 &  0.15 &  0.09 &  0.05\\
0.008 &  2.6 &  2.3 &  1.7 &  1.1 &  0.63 &  0.44 &  0.35 &  0.22 &  0.14 &  0.08 &  0.05\\
0.020 &  2.8 &  2.7 &  2.1 &  1.2 &  0.61 &  0.40 &  0.31 &  0.19 &  0.13 &  0.07 &  0.04\\
\hline

\end{tabular}
\caption{Uncertainties on the ionizing photon production rate and efficiency of a stellar population forming 1\,M$_\odot$\,yr$^{-1}$ in stellar mass continuously for 100\,Myr. The first column gives the metallicity.  Remaining columns give the 1\,$\sigma$  scatter on the photon rate or efficiency in individual stochastic realisations, as a function of \mcl. All values are in dex. }\label{tab:unc-nionconst}
\end{table*}

\begin{table*}
\begin{tabular}{lccccccccccc|c}
   & \multicolumn{7}{|c|}{$\beta_\mathrm{UV}$ as a function of cluster mass, \mcl/M$_\odot$} & \\
 Z\ \ \textbackslash\ \ \mcl  & 1$\times10^2$ & 3$\times10^2$ & 1$\times10^3$ & 3$\times10^3$ & 1$\times10^4$ & 3$\times10^4$ & 1$\times10^5$ & 3$\times10^5$ \\
 &&&&& 1$\times10^6$ & 3$\times10^6$ & 1$\times10^7$ & Statistical\\
\hline\hline
0.002 &  -0.7 $\pm$ 2.7 &   -1.88 $\pm$  0.49 &   -2.18 $\pm$   0.22 &  -2.28  $\pm$ 0.16 &   -2.37  $\pm$ 0.18 &   -2.41  $\pm$ 0.16 &   -2.47  $\pm$ 0.13 &   -2.50 $\pm$  0.11 \\
&&&&&   -2.51  $\pm$ 0.08 &   -2.54  $\pm$ 0.04 &   -2.55  $\pm$  0.01 & -2.50\\
0.008 & 
   1.2  $\pm$  2.7 &  -1.34 $\pm$  0.49 &   -1.76 $\pm$  0.23 &   -1.88  $\pm$ 0.16 &   -1.98  $\pm$ 0.18 &   -2.02 $\pm$  0.16 &   -2.11 $\pm$  0.13 &   -2.16 $\pm$  0.11 \\
   &&&&&   -2.18  $\pm$  0.08 &  -2.20 $\pm$  0.04 &   -2.20 $\pm$  0.01 & -2.19 \\
0.020 &
   4.6  $\pm$ 2.7 &   -0.52 $\pm$  0.49 &   -1.44 $\pm$  0.23 &   -1.58  $\pm$ 0.16 &   -1.69  $\pm$ 0.18 &   -1.75  $\pm$ 0.16  &   -1.85 $\pm$  0.13 &   -1.91 $\pm$  0.11 \\
   &&&&&   -1.94  $\pm$  0.08 &  -1.97 $\pm$  0.04 &    -1.97  $\pm$ 0.01 & -1.95\\
\hline

\end{tabular}
\caption{Uncertainties on the 1566-2266\AA\ ultraviolet spectral slope of a stellar population forming 1\,M$_\odot$\,yr$^{-1}$ in stellar mass continuously for 100\,Myr. The first column gives the metallicity. The final column gives the canonical value for the parameter from the statistical sampling method used in BPASS v2.2.1. }\label{tab:unc-uvsope}
\end{table*}

\begin{table*}
\begin{tabular}{lcccccccccccc}
   & \multicolumn{11}{|c|}{Mean $\sigma F/F$(UV) as a function of cluster mass, \mcl/M$_\odot$} & S20\\
 Z\ \ \textbackslash\ \ \mcl  & 1$\times10^2$ & 3$\times10^2$ & 1$\times10^3$ & 3$\times10^3$ & 1$\times10^4$ & 3$\times10^4$ & 1$\times10^5$ & 3$\times10^5$ & 1$\times10^6$ & 3$\times10^6$ & 1$\times10^7$  \\
\hline\hline
0.002 & 2.20 & 1.3 & 0.73 &  0.40 &  0.24 &   0.14 & 0.08 &  0.04 &  0.02 & 0.01 & 0.01 & 0.13\\
0.008 & 3.1  & 1.5  & 0.86 & 0.45 &  0.28 &   0.15 & 0.09 &  0.054 & 0.03 & 0.02 & 0.01 & 0.12 \\ 
0.020 & 4.2 & 2.1 & 1.1 &  0.60 &  0.36 &   0.19 & 0.13 &   0.07 & 0.04 & 0.03 & 0.03 &  0.10\\
\hline
\\
   & \multicolumn{11}{|c|}{Mean $\sigma F/F$(Opt) as a function of cluster mass, \mcl/M$_\odot$} & S20\\
 Z\ \ \textbackslash\ \ \mcl  & 1$\times10^2$ & 3$\times10^2$ & 1$\times10^3$ & 3$\times10^3$ & 1$\times10^4$ & 3$\times10^4$ & 1$\times10^5$ & 3$\times10^5$ & 1$\times10^6$ & 3$\times10^6$ & 1$\times10^7$  \\
\hline\hline
0.002 & 3.3  & 1.9  & 1.1  & 0.70  & 0.39  & 0.21  & 0.13  &  0.06  & 0.04  & 0.02  & 0.01  & 0.12   \\
0.008 & 3.1  & 1.6  & 0.89 & 0.58  & 0.34  & 0.19  & 0.11  &  0.05  & 0.03  & 0.02  & 0.01  & 0.11   \\
0.020 & 2.7  & 1.4  & 0.75 & 0.43  & 0.26  & 0.15  & 0.09  &  0.04  & 0.03  & 0.02  & 0.01  & 0.09  \\
\hline

\end{tabular}
\caption{Uncertainties on the continuum flux of a stellar population forming 1\,M$_\odot$\,yr$^{-1}$ in stellar mass continuously for 100\,Myr. The first column gives the metallicity. The final column provides a comparison value for scatter due to binary parameter uncertainties from \citet{2020MNRAS.495.4605S}. Remaining columns give the 1\,$\sigma$  scatter on the flux in individual stochastic realisations, normalised by the continuum luminosity. We give the typical scatter in in the UV at 1250-1650\AA, and in the optical at 5000-6000\AA. }\label{tab:unc-spectra}
\end{table*}

\begin{table*}
\begin{tabular}{lccccccccccc|c}
   & \multicolumn{11}{|c|}{Mean $N_\mathrm{SN}$ per $10^3$\,M$_\odot$ formed as a function of cluster mass, \mcl/M$_\odot$} & Statistical\\
 Z\ \ \textbackslash\ \ \mcl  & 1$\times10^2$ & 3$\times10^2$ & 1$\times10^3$ & 3$\times10^3$ & 1$\times10^4$ & 3$\times10^4$ & 1$\times10^5$ & 3$\times10^5$ & 1$\times10^6$ & 3$\times10^6$ & 1$\times10^7$ & \\
\hline\hline
 0.002 & 5.40 &   6.63 &   7.94 &   8.05 &   7.72 &   7.27 &   7.13 &   6.99 &   6.97 &   6.99 &   6.99 & 6.14\\
 0.008 &    4.51 &   4.01 &   3.94 &   5.24 &   5.49 &   5.41 &   4.95 &   5.04 &   5.03 &   5.05 &   5.06 & 4.53\\
 0.020 & 0.00 &   3.36 &   2.30 &   4.10 &   3.99 &   3.93 &   3.56 &   3.75 &   3.77 &   3.72 &   3.71 & 3.63\\
\hline\\
\\
  & \multicolumn{11}{|c|}{Standard Deviation of $N_\mathrm{SN}$ per M$_\odot$ formed as a function of cluster mass, \mcl/M$_\odot$}\\
 Z\ \ \textbackslash\ \ \mcl  & 1$\times10^2$ & 3$\times10^2$ & 1$\times10^3$ & 3$\times10^3$ & 1$\times10^4$ & 3$\times10^4$ & 1$\times10^5$ & 3$\times10^5$ & 1$\times10^6$ & 3$\times10^6$ & 1$\times10^7$ & \\
\hline\hline
0.002 &  37.7 &   19.4 &   15.1 &   8.90 &   4.95 &   2.48 &   1.14 &   0.69 &   0.39 &   0.21 &   0.12 & \\
0.008 &  36.7 &   26.3 &   14.6 &   7.07 &   3.67 &   2.10 &   1.15 &   0.66 &   0.33 &   0.16 &   0.097 & \\
0.020 &  57.0 &   25.8 &   16.1 &   7.42 &   4.92 &   3.05 &   1.34 &   0.86 &   0.40 &   0.42 &   0.46 & \\

\hline
\end{tabular}
\caption{Uncertainties on the supernova number counts from clusters, normalised by the mass of stars formed. The first column gives the metallicity. Remaining columns give the 1\,$\sigma$  scatter on the supernova number per unit mass in the cluster in individual stochastic realisations, as a function of \mcl. }\label{tab:unc-SNe}
\end{table*}

\end{document}